%% file: Beambook_20190717.tex
\documentclass{ieeeaccess}
\usepackage{cite}

\usepackage{amsmath,amssymb,amsfonts}
\usepackage{algorithmic}
\usepackage{graphicx}
\usepackage{textcomp}
\usepackage{hhline}

\newlength{\xfigwd}
\def\BibTeX{{\rm B\kern-.05em{\sc i\kern-.025em b}\kern-.08em
		T\kern-.1667em\lower.7ex\hbox{E}\kern-.125emX}}

\usepackage{caption}
\DeclareCaptionFont{ieeeblue}{\color{accessblue}}
\DeclareCaptionLabelFormat{myformat}{\figcapfont{\textbf{#1} \textbf{#2}}}
\input{input.tex}
\DeclareMathOperator*{\argmax}{argmax} 

\newcommand{\revision}[1]{{\leavevmode\color{black}#1}}

\begin{document}
	%
	
	
	\history{Date of publication xxxx 00, 0000, date of current version xxxx 00, 0000.}
	\doi{10.1109/ACCESS.2019.2930224}
	
	\title{Beam Codebook Design for 5G mmWave Terminals}
	\author{\uppercase{Jianhua Mo}\authorrefmark{1}, \uppercase{Boon Loong Ng}\authorrefmark{1}, \uppercase{SangHyun Chang}\authorrefmark{3}, \uppercase{Pengda Huang}\authorrefmark{1}, \uppercase{Mandar Kulkarni}\authorrefmark{1},  \uppercase{Ahmad Alammouri}\authorrefmark{2}, \uppercase{Jianzhong Charlie Zhang}\authorrefmark{1}, \uppercase{Jeongheum Lee}\authorrefmark{3}, \uppercase{Won-Joon Choi}\authorrefmark{3}}
	\address[1]{Samsung Research America, Plano, TX 75023, USA (Email: \{jianhua.m, b.ng,  p.huang, mandar.kulkarni, jianzhong.z\}@samsung.com)}
	\address[2]{Wireless Networking and Communications Group (WNCG), The University of Texas at Austin, Austin, TX 78712, USA (Email: alammouri@utexas.edu)}
	\address[3]{Samsung Electronics Co., Ltd., Suwon 16677, Korea (Email: \{s29.chang, jh0413.lee, wonjoon.choi\}@samsung.com)} 
	
	\markboth
	{J. Mo \headeretal: Beam Codebook Design for 5G mmWave Terminals}
	{J. Mo \headeretal: Beam Codebook Design for 5G mmWave Terminals}
	
	\corresp{Corresponding author: Jianhua Mo (e-mail: jianhua.m@samsung.com).}

	
	\begin{abstract}
        A beam codebook of 5G millimeter wave (mmWave) for data communication consists of multiple high-peak-gain beams to compensate the high pathloss at the mmWave bands. These beams also have to point to different angular directions, such that by performing beam searching over the codebook, a good mmWave signal coverage over the full sphere around the terminal (spherical coverage) can be achieved.
        A model-based beam codebook design that assumes ideal omni-directional antenna pattern, and neglects the impact of terminal housing around the antenna, does not work well because the radiation pattern of a practical mmWave antenna combined with the impact of terminal housing is highly irregular.
        In this paper, we propose a novel and efficient data-driven method to generate a beam codebook to boost the spherical coverage of mmWave terminals. The method takes as inputs the measured or simulated electric field response data of each antenna and provides the codebook according to the requirements on the codebook size, spherical coverage, etc. The method can be applied in a straightforward manner to different antenna type, antenna array configuration, placement and terminal housing design. Our simulation results show that the proposed method generates a codebook better than the benchmark and 802.15.3c codebooks in terms of the spherical coverage.
	\end{abstract}
	
	
	\titlepgskip=-15pt
	
	\maketitle
	
	\section{Introduction}
	In 5G cellular networks, beamforming is necessary for overcoming large channel pathloss when a user equipment (UE) tries to establish a connection with a base station (BS) in millimeter wave (mmWave) bands such as the 28 GHz, 39 GHz, or 60 GHz bands \cite{Andrews_JSAC14,Alkhateeb_COMM14, Raghavan_COMM18}. To compensate for the smaller angular coverage due to the narrow analog beamwidth in mmWave, beam sweeping can be employed to enable wider angular signal reception or transmission coverage for the UE \cite{Hong_Wonbin_COMM14, Kutty_CST16, Hong_Wei_TAP17}. A beam codebook comprises a set of beams or codewords, where a codeword is a set of analog phase shift values, or a set of magnitude plus phase shift values, applied to the antenna elements, in order to form an analog beam. The 3rd Generation Partnership Project (3GPP) is the 5G standardization body that specifies the minimum peak  \ac{EIRP} and the spherical coverage requirements of UE defined as a certain percentile of the \ac{CDF} over the full sphere around the UE. There are a total of four UE power classes defined for various use cases or deployment scenarios; and the minimum peak EIRP and the spherical coverage requirements are different for different UE power classes. For example, it has been specified for the first generation (Release 15) of 5G mmWave handheld UE (power class 3) that the minimum peak EIRP is 22.4 dBm (20.6 dBm) and the minimum EIRP at the 50th percentile CDF over the full sphere around the UE is 11.5 dBm (8 dBm) for 28 GHz bands (39 GHz band) \cite[Table 6.2.1.3-3]{3gpp.38.101-2}. This paper describes a novel codebook generation procedure and algorithms to obtain a beam codebook given a set of requirements and performance criteria.
	
    \subsection{Related Work}
    \revision{Beam codebook design has been extensively considered in both academia and industry \cite{Kutty_CST16, Hong_Wei_TAP17}. 
    A beam codebook design was provided in 802.15.3c \cite[Chapter 13]{802.15.3c-2009}, assuming 1-D and 2-D arrays with uniform spacing of half-wavelength. 
    The beam searching or training process is divided into three stages, namely, link-level device discovery, sector-level alignment, and beam-level refinement. The omni or quasi-omni radiation pattern, wide beam, and narrow beam are designed respectively to fulfill the requirements of these three phases \cite{Wang_Junyi_JSAC09, Chen_Li_ICST11}.
    The same 3-stage training process was adopted in \cite{Feng_Wei_WCNC13} where codebooks for a 2-ring circular array were proposed. The inner small ring generates quasi-omni and sector radiation patterns while the outer larger ring generates the last-stage directional beam patterns.}
    
    \revision{The idea of 3-stage codebook design was extended to general hierarchical codebook design where the number of stages or layer are not limited to be three. The analog codebook design was considered in \cite{Hur_TCOM13} where the sub-array method used to generate a ``flatted'' wide beam. The paper \cite{Raghavan_JSTSP16} proposed a heuristic method where the uniform linear array is divided into 2, 3, or 4 sub-arrays and the steering direction and length of the sub-arrays are numerically optimized to maximize the minimum beamforming gain in the required coverage region. The work \cite{He_Tong_MobileNetwAppl15} proposed a deactivation approach where the antenna elements are adaptively deactivated to create beam with various beamwidth. Combined with the deactivation method,
    the sub-array based hierarchical codebook generation was optimized in \cite{Xiao_Zhenyu_TWC16} where either all or a half of the antenna elements are activated. The method of \cite{Xiao_Zhenyu_TWC16} was further enhanced in \cite{Xiao_Zhenyu_TVT18} where the deactivation method is dropped, and all the antenna are always activated to increase the maximal total transmission power. The sub-array method was adopted in \cite{Zhu_Lipeng_WCL19} to design a 3-D wide beam for uniform planar array.  To design beams with a small ripple in both the main and side lobes, a beam pattern optimization problem was formulated in \cite{Zhang_Jianjun_TCOM17}. However, the optimization problem considered the total power constraints rather than the individual antenna power constraint, thereby the resultant beam has a large peak-to-average-power ratio, which implies low power efficiency.}
    
    \revision{Besides analog precoding, hierarchical codebook design for hybrid analog-digital precoding was considered in \cite{Alkhateeb_JSTSP14, Xiao_Zhenyu_TWC17,Noh_TWC17}. Given the required angular region to cover, the analog and digital precoder design was formulated in \cite{Alkhateeb_JSTSP14} as a sparse approximation problem, and solved by a variant of of orthogonal matching pursuit algorithms. The authors of \cite{Noh_TWC17} proposed a DFT-based multilevel codebook design where the adjacent phase-shifted DFT beams are summed up to construct wide beams. Last, the sub-array method was altered in \cite{Xiao_Zhenyu_TWC17} to support the hybrid precoding.}
    
    \revision{Considering the high cost, power consumption and form factor of radio frequency (RF) chain, the mmWave terminals are not likely to adopt hybrid or fully digital beamforming where more than one RF chains are needed for a single array. Therefore, analog beamforming for each antenna array is assumed throughout this paper.
    }
    
    
    \revision{In this paper, we focus on the beam codebook for data transmission, i.e, the third stage codebook in 802.15.3c for beam searching \cite{802.15.3c-2009}, or the bottom layer fine codebook in a hierarchical codebook design \cite{He_Tong_MobileNetwAppl15, Xiao_Zhenyu_TWC16, Noh_TWC17}. In 802.15.3c, the codebooks are generated with 2-bit phase shifters without amplitude adjustments for the consideration of the hardware complexity. 
    To reduce the gain loss at the intersections of two beams, the number of beams should be twice the number of array elements \cite{Wang_Junyi_VTC09}. In the 802.11ad document \cite[Section 6.6]{11ad_channel_models}, the beam codebook design is formulated as a geometric problem to cover the sphere sector with circles by assuming that the main lobe of the beam has a circular shape. The assumption of circular shape, however, does not hold when a beam is beamforming towards directions away from the broadside direction and therefore results in coverage gaps between beams.}
    \revision{In \cite{Alkhateeb_JSTSP14, He_Tong_MobileNetwAppl15, Xiao_Zhenyu_TWC16, Noh_TWC17}, the last layer beams are pointing to directions uniformly distributed in the angular domain or spatial frequency. In such cases, the beamforming vector is just the steering vector for a given beamforming direction (or an approximation of it if there is a phase shifters resolution constraint). For example, for a simple linear array with spacing $d$, the beamforming weights would have the progressive phases as  $\frac{2 \pi i}{\lambda} d \cos \theta$ where $\lambda$ is the wavelength, $i$ is the antenna index, and $\theta$ is the beamforming direction with respect to the axis of the array. The codeword for 2-D planar array is then the Kronecker product of two codewords for 1-D linear arrays, as done in \cite{Mao_WCL18,Zhu_Lipeng_WCL19}.} 
    
    \revision{All these work \cite{802.15.3c-2009,Wang_Junyi_JSAC09, Wang_Junyi_VTC09, Chen_Li_ICST11,11ad_channel_models,Feng_Wei_WCNC13,Hur_TCOM13, Alkhateeb_JSTSP14, Noh_TWC17, He_Tong_MobileNetwAppl15, Xiao_Zhenyu_TWC16, Xiao_Zhenyu_TWC17, Xiao_Zhenyu_TVT18, Zhang_Jianjun_TCOM17, Raghavan_JSTSP16, Mao_WCL18,Zhu_Lipeng_WCL19} assumed an ideal isotropic radiation pattern and considered rather regular antenna setup, i.e., uniform linear, uniform planar or uniform circular array. We call these designs, which are based on simple theoretical assumptions, as \textbf{model-based} approach hereafter.
    Such designs, however, ignore many practical issues as described next.}
    
	Antenna for mmWave bands is intrinsically directional. For example, the patch antenna usually has a high front-to-back ratio and consequently can cover at most half-sphere \cite{Balanis_Book12}. The directional element radiation pattern will also result in the drift of the peak gain direction from the intended one if the beamforming vector is merely designed based on the steering vector. In addition, when placed inside mobile handsets, the radiation gain of the mmWave antenna is less than the free-space case due to blockage loss and the radiation pattern shape is also changed \cite{Hong_Wonbin_TAP17}.
    
    Antenna placement and antenna spacing may not be regular. For example, the planar array may not have the half-wavelength spacing between adjacent elements due to form-factor constraints. \revision{Another reason is related to the multi-frequency bands that the mmWave terminal has to support. The mmWave bands for 5G deployment in US will include 24 GHz, 28 GHz and 39 GHz, etc\footnote{Federal Communications Commission's Facilitate America's Superiority in 5G Technology (the 5G FAST Plan).}. The same antenna arrays, however, are likely to be used at all these carrier frequency bands. Therefore, a half-wavelength spacing at a frequency band will result in less than (or more than) half-wavelength spacing at other lower (or higher) frequency bands.}
    
    A 5G mmWave capable UE is typically equipped with multiple antenna arrays. For example, in a design given in \cite{Raghavan_TCOM19}, there are at most four mmWave modules mounted on the top, bottom, left and right edges of the phone, respectively. Multiple mmWave antenna arrays are necessary to enable a good spherical coverage over the whole sphere and to circumvent human body blocking. In a benchmark codebook design, the beam codewords are designed independently for each array as assumed in \cite{Raghavan_TCOM19}, which is a suboptimal solution since the interaction and coordination between the arrays are ignored. 
	
    \subsection{Contributions}
    
    A practical beam codebook design should at least take into account the following factors,
    \begin{enumerate}
        \item Antenna element type and gain (e.g. isotropic, dipole, microstrip patch);
        \item Array layout (e.g. linear, rectangular, circular, cylinder) and placement if there are multiple arrays;
        \item Requirements of codebook (e.g. codebook size, required coverage regions, phase shifter resolution);
        \item Consideration about UE housing (e.g., display screen, battery);
        \item \revision{The coordination among different arrays mounted on the same terminal.}
    \end{enumerate}

    \revision{Although it might be able to model the antenna element type by approximation models \cite{Balanis_Book12}, it is difficult, if not impossible, to analytically model the other factors, including the housing effects caused by a plurality of components inside the mmWave terminal with various size, shape and electromagnetic properties.} Faced with aforementioned challenges, it is generally difficult to find an analytical method to generate the codebook. It is also impossible to find the optimal codebook by an exhaustive search because of its exponential complexity as $\mathcal{O}\left(2^{bLK}\right)$ where $b$ is the phase shifter resolutions, $L$ is the antenna array size, and $K$ is the codebook size. For instance, for a small array where $b=2$, $L=4$ and $K=4$, there are $2^{32}$ possible codebooks assuming that the \revision{analog beamforming} codewords are ordered. 
    
    In this paper, we present a \textbf{data-driven} codebook design method. An important advantage of our method is that it can be applied agnostically with any antenna type, array layout and placement. The antenna information required for our method is simply the electrical field (E-field) response of each antenna element in a given layout, which can be obtained through electromagnetic simulation software (for example, \ac{HFSS} by Ansys) or through measurements.
    
    Our first algorithm is a \textit{Greedy} algorithm, which sequentially selects the beam codewords to augment the spherical coverage. The performance of this algorithm relies on the quality of the candidate codewords pool as well as the codeword selection criterion. 
    Our second algorithm is based on an unsupervised machine learning algorithm, namely, \textit{K-Means}. In this algorithm, the angular directions are clustered based on their E-field response and then the beam codewords are optimized to improve the average gain of the clustered points. This clustering and optimization procedure is repeated until convergence.
    
    \revision{The contributions of this paper are summarized as follows.
    \begin{enumerate}
        \item We formulate a beam codebook design problem from the perspective of maximizing the spherical coverage. The optimization problem takes into account the amplitude and phase resolution constraint, as well as the codebook size. In addition, compared with the previous work (e.g., \cite{Raghavan_TCOM19}), the two polarization components are both considered in the design.
        \item We propose a data-driven approach for codebook design. The proposed approach, which takes as inputs the E-field response data from simulations or measurements, automatically generate the codebook without request of modeling the antenna element pattern, the housing effects, etc. 
        \item An upper bound of the composite radiation pattern is derived. The upper bound provides a reference for evaluating the performance of the designed codebook.
        \item Comprehensive numerical simulations are provided to confirm the effectiveness and superiority of the proposed codebook design.
    \end{enumerate}}
	
	The paper is organized as follows. In \secref{sec:beamcodebookdesign}, we formulate the problem of beam codebook design. In \secref{sec:single_beam}, we present the design of a single beam with power and phase constraints, which lays the foundation for discussions of our algorithms.
	The upper bound of the composite radiation pattern is discussed in \secref{sec:UB}.
    Our two heuristic algorithms are provided in \secref{sec:Greedy} and \secref{sec:Kmean}. The simulation results are shown in \secref{sec:simulations}. Further discussions on the additional advantages of our algorithms are provided in \secref{sec:other_advantages}. A comparison with other model-based method based on simplified E-field response data is given in \secref{sec:comparison}. The paper is concluded in \secref{sec:conclusion}. 
	
	\emph{Notation}: Bold uppercase letter $\mathbf{A}$ and bold lowercase letter $\mathbf{a}$ represents a matrix and a column vector, respectively. $\bA\geq 0$ implies that $\bA$ is a positive semi-definite matrix.
    $\left(\cdot \right)^T, \left(\cdot\right)^*, \left(\cdot\right)^H$ denotes the transpose, conjugate and Hermitian of a vector or matrix, respectively. $\|\ba\|$ is the norm of the vector $\ba$. $\arg(\cdot) \in [0, 2\pi)$ denotes the phase of a complex-valued input. $\mathrm{mod}(a,n)$ stands for the remainder of $a$ divided by $n$.
	
	\section{Beam Codebook Design Problem} \label{sec:beamcodebookdesign}
	
    \begin{figure}[t]
        \centering
        \includegraphics[width=1.0\linewidth]{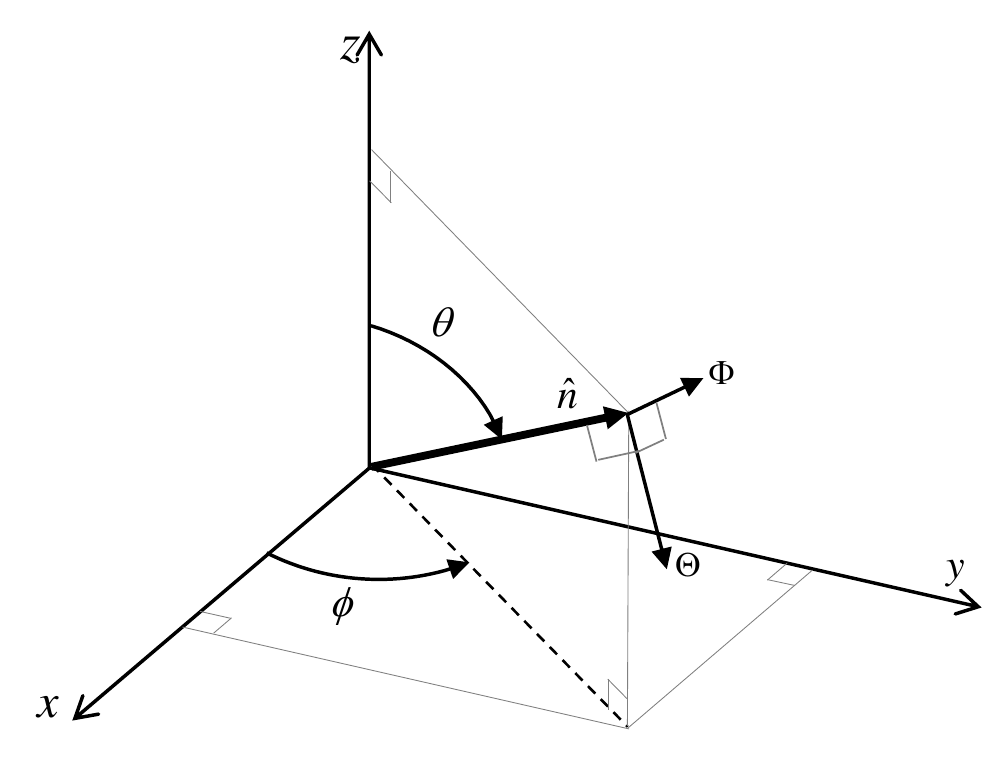}
        \caption{The coordinate system for E-field analysis.}
        \label{fig:Coordinate}
    \end{figure}

%
	The coordinate system used throughout this paper is shown in \figref{fig:Coordinate}. The UE is placed around the origin. $\theta$ ($\phi$) is defined as the zenith (azimuth) angle. Since the electrical field is a vector field, it is represented by three orthogonal components, denoted as $(E_R, E_\Theta, E_\Phi)$, at each observation point on the surface of a sphere. We consider the E-field response in the far-field (Fraunhofer) region where the electromagnetic wave appears locally as a plane wave in any specified direction. As a result, the radial component, i.e., $E_R$, is zero or vanishingly small compared to the other two components, i.e., $E_\Theta$ and $E_\Phi$ \cite{Balanis_Book12}. Therefore, for a given direction $\widehat{\mathbf{n}}$, we only consider the E-field $\Theta$ component and $\Phi$ component, which are perpendicular to $\widehat{\mathbf{n}}$ as shown in \figref{fig:Coordinate}. 
    
    Assume there are $L$ antenna elements in an array. Let $e_{\ell}^{\Theta}(\theta, \phi)$ and $e_{\ell}^{\Phi}(\theta, \phi)$, $\ell \in \left\{1,2,\cdots,L\right\}$, denote the complex-valued E-field response of the $\ell$-th antenna element for the $\Theta$ component and the $\Phi$ component, respectively, at the direction $(\theta, \phi)$\footnote{The E-field response in this paper denotes a product, i.e., $r \cdot E_X$, where $E_X \left(X=\Theta, \Phi \right)$ is the E-field strength measured at a distance $r$ to the origin when the incident power to the antenna element is 1 Watt. Since $E_X \propto \frac{1}{r}$ in the far-field region \cite{Balanis_Book12}, the E-field response is independent of the distance $r$. Note that the E-field response in this paper corresponds to the term `rE' in HFSS.}.
	Denote the E-field data in vectors as,
	\begin{align}
		\mathbf{e}_{\Theta}(\theta, \phi) \triangleq [e_1^{\Theta}(\theta, \phi), e_2^{\Theta}(\theta, \phi), \cdots, e_{L}^{\Theta}(\theta, \phi)]^T, \\
		\mathbf{e}_{\Phi}(\theta, \phi) \triangleq [e_1^{\Phi}(\theta, \phi), e_2^{\Phi}(\theta, \phi), \cdots, e_{L}^{\Phi}(\theta, \phi)]^T.
	\end{align}
	As mentioned in the introduction, the E-field data can be obtained through electromagnetic simulation or measurement, which is usually sampled on a mesh grid, for example, $[\theta, \phi] = [0^\circ: q_{\theta} : 180^\circ] \times [0^\circ: q_{\phi} :360^\circ)$, where $q_{\theta}$, $q_{\phi}$ are the simulation or measurement step sizes.
	
	Let $\mathbf{w} \triangleq [w_1,w_2,\cdots,w_L ]^T$ denote the complex-valued weights applied on the antenna elements. Without loss of generality, we will assume the beamforming codeword $\bw$ always has unit-norm throughout this paper, i.e., $\|\bw\|^2=1$.
    According to the superposition principle, the E-fields for the $\Theta$ and $\Phi$ components after applying the beamforming weights are given by
	\revision{
    \begin{align} 
		\mathcal{E}_{\Theta, \bw} \left(\theta, \phi \right) = \sum_{\ell=1}^{L} w_{\ell}^* e_{\ell}^{\Theta}(\theta, \phi) = \mathbf{w}^H \mathbf{e}_{\Theta}(\theta, \phi) , \label{eq:E_field_Theta_sum}\\
		\mathcal{E}_{\Phi, \bw} \left(\theta, \phi \right) = \sum_{\ell=1}^{L} w_{\ell}^* e_{\ell}^{\Phi}(\theta, \phi) = \mathbf{w}^H \mathbf{e}_{\Phi}(\theta, \phi). \label{eq:E_field_Phi_sum}
	\end{align}}
    
	The realized beamforming gain is the sum of the realized gains of the $\Theta$ and $\Phi$ components \cite{Balanis_Book12},
	\revision{
        \begin{align}
		&G_{\bw} (\theta, \phi) \nonumber\\
		=& \frac{4 \pi}{\| \bw\|^2} \cdot \frac{1}{2 \eta_0}\left(\left| \mathcal{E}_{\Theta, \bw} \left(\theta, \phi \right) \right|^2 +\left|\mathcal{E}_{\Phi, \bw} \left(\theta, \phi \right) \right|^2 \right)\\
		=& \frac{2 \pi}{\eta_0} \left(\mathbf{w}^H \left( \mathbf{e}_{\Theta} \left(\theta,\phi \right) \mathbf{e}^H_{\Theta} \left(\theta,\phi \right) + \mathbf{e}_{\Phi} \left(\theta,\phi \right) \mathbf{e}^H_{\Phi} \left(\theta,\phi \right) \right) \mathbf{w} \right) \label{eq:gain_sum}\\
		=& \frac{2 \pi}{\eta_0} \mathbf{w}^H \mathbf{M}(\theta, \phi) \mathbf{w} . 
	\end{align}}
	where $\eta_0 \approx 377 $ $\Omega$ is the impedance of the free space, \revision{$\mathbf{M}(\theta, \phi)\triangleq \mathbf{e}_{\Theta}(\theta,\phi) \mathbf{e}^H_{\Theta}(\theta,\phi) + \mathbf{e}_{\Phi}(\theta,\phi) \mathbf{e}^H_{\Phi}(\theta,\phi)$}, and \eqref{eq:gain_sum} is obtained by plugging in \eqref{eq:E_field_Theta_sum} and \eqref{eq:E_field_Phi_sum} and noticing the unit-norm assumption of $\bw$.

	We assume that the phase shifters are constrained to $b$ bits, and the codebook $\mathcal{W}_c$ has a size limitation $K$, i.e., $\mathcal{W}_c \triangleq \l\{\bw_1, \bw_2, \cdots, \bw_K \r\}$. A codebook of small size will help reduce the beam sweeping time, power consumption as well as the memory space in the modem. There is also a requirement on the composite radiation gain pattern, which is the maximum over all the gain patterns of the codewords and is denoted as $S \left(\mathcal{W}_c, \theta, \phi \right)$. The composite radiation pattern indicates the wellness of the spherical coverage of the codebook. Specifically, it can be used to identify coverage holes. 
    The beam codebook design problem is formulated as below.
	\begin{subequations}
	\begin{align}
		\left(\mathbf{P1}\right) \, \max_{\mathcal{W}_c} \quad  & U\big( S \left({\mathcal{W}_c}, \theta, \phi \right) \big) \\
		s.t. \quad & S \left( \mathcal{W}_c, \theta, \phi \right) = \frac{2 \pi}{\eta_0} \max_{\bw_k \in \mathcal{W}_c} \bw_k^H \bM(\theta, \phi) \bw_k, \\
		&  \l(\sqrt{L} w_{k \ell}\r)^{2^b}=1, \enspace \forall k, \ell, \label{eq:elm_cons}
	\end{align}
	\end{subequations}
	where the last equation \eqref{eq:elm_cons} encapsulates the magnitude constraint $\left|w_{k\ell} \right| = \frac{1}{\sqrt{L}}$ as well as the phase constraint $\arg \left(w_{k\ell} \right)  \in \l\{0, \frac{2 \pi}{2^b}, \cdots, \l(2^b-1\r) \frac{2 \pi}{2^b} \r\}$.

	When there are multiple arrays, we assume that only one of the antenna arrays is activated at a given time, which is a typical implementation assumption. As a result, the problem formulation is similar to $\left(\mathbf{P1}\right)$ with the exception that the composite radiation pattern is the maximum over all the codewords of all the arrays.
	
	The utility function $U(\cdot)$ can be defined as the average gain across the whole sphere, or the $x$th-percentile of the gain over the unit-sphere (e.g., $x=20, 50$). As mentioned in the introduction, the 3GPP specifies the spherical converage requirement for handheld UE (power class 3) in terms of the 50th percentile \ac{EIRP} \cite[Table 6.2.1.3-3]{3gpp.38.101-2}\footnote{Note that \ac{EIRP} is equal to the sum of the realized beamforming gain and the incident power towards the antennas in the log scale. We normalize the incident power as one throughout this paper and thus the optimization of \ac{EIRP} is equivalent to the optimization of beamforming gain.}. Throughout this paper, the utility function is defined over a uniform sampling over the sphere or a specified angular region. 
	In particular, the \ac{CDF} of the gain over the sphere is defined as,
	\begin{align}
		& F_S(s) \nonumber \\
		= & \frac{1}{4 \pi} \int_{\theta=0}^{\pi} \int_{\phi=0}^{2 \pi} \mathbbm{1} \l\{ S \l( \mathcal{W}_c, \theta, \phi \r) \leq s \r\} \sin \theta d \theta d \phi \\
		\approx & \frac{1}{N_p} \sum_{i=1}^{N_p}\mathbbm{1} \l\{ S \l( \mathcal{W}_c, \theta_i, \phi_i \r) \leq s \r\}.
	\end{align}
	\revision{Maximizing a particular percentile value (e.g., 50\% requested by 3GPP) of the distribution is not an easy task when considering the mathematical tractability. A more tractable utility function is the average gain over the sphere, which is defined as}
    \begin{align}
    & \mathbb{E}_{\theta, \phi} \left[ S \left( \mathcal{W}_c, \theta, \phi \right) \right] \nonumber \\
    = & \frac{1}{4 \pi} \int_{\theta=0}^{\pi} \int_{\phi=0}^{2 \pi} S \l( \mathcal{W}_c, \theta, \phi \r) \sin \theta d \theta d \phi \\
    \approx & \frac{1}{N_p} \sum_{i=1}^{N_p} S(\mathcal{W}_c, \theta_i, \phi_i), \label{eq:s_approx}
    \end{align}
    where the approximation in \eqref{eq:s_approx} comes from a set of $N_p$ uniformly distributed sampling points on the sphere, which can be obtained through a Fibonacci grid \cite{Fibonacci_Swinbank06}. 
	
	The problem $\mathbf{P1}$ is non-convex and NP-hard due to the constraint \eqref{eq:elm_cons}. In this paper, we provide two heuristic algorithms. As verified through our simulation based on practical phone design, the proposed heuristic algorithms have low complexities and provide satisfactory performance.
    
    \revision{In the problem formulation $\mathbf{P1}$, the flatness of each beam is ignored for several reasons. First, we are designing high-peak-gain narrow beams for data transmission instead of quasi-omni or wide beam for initial device discovery and sector-level searching, therefore there is no flatness issue in our designed narrow beams. Second, to establish a successful mmWave connection, the spherical coverage of the composite radiation pattern is a more relevant and effective metric than the flatness of the individual beam. Last, but not least, if the radiation pattern is severely irregular due to element pattern or strong housing effects, then it is impossible to design a wide beam with flat gain.}

	\section{\revision{Data-Driven Design of a Single Beam}} \label{sec:single_beam}
	Before presenting our beam codebook design, we first present the approach on the design of a single beam, which will be used in the Greedy and K-Means algorithms.
	
	
	\revision{It is noteworthy that the beam design in the data-driven codebook is quite different from the conventional model-based method in two-fold. First, the beam codeword is designed based on the E-field data, instead of being a steering vector \footnote{A steering vector of an array at a direction $(\theta, \phi)$ has the form $ w_{\ell} = \frac{1}{\sqrt{L}}\exp \left( \j \frac{2\pi}{\lambda} \widehat{\bn}^T \bx_\ell \right)$ where $\widehat{\mathbf{n}} = (\sin \theta \cos \phi, \sin \theta \sin \phi, \cos \theta)^T$ (see the definition of $\theta$ and $\phi$ in \figref{fig:Coordinate}), $\lambda$ is the wavelength, and $\bx_\ell$ is the 3-D coordinate of the $\ell$-th element.}, or a weighted sum of a few steering vectors \cite{Noh_TWC17}, or a concatenation of steering vectors on sub-arrays \cite{Hur_TCOM13, Xiao_Zhenyu_TWC16, Raghavan_JSTSP16, Zhu_Lipeng_WCL19}. Due to the directional element pattern and housing effects, the beam radiation pattern may point away from the intended direction.
    Second, the beam is carefully designed to take into account two polarization components, which is not considered in the prior work, e.g., \cite{Raghavan_TCOM19}.}
	
	Given the E-field response matrix $\bM \left(\theta, \phi \right)$ at a given direction $(\theta, \phi)$, or the sum E-field response over a set of directions, i.e., $\bM=\sum_{(\theta, \phi) \in \mathcal{A}}\bM(\theta, \phi)$, we want to design a beamforming vector to maximize the beamforming gain $\bw^* \bM \bw$ under different constraints.
	
	First, consider a simple case with sum power constraint. The optimization problem is as follows.
	\begin{subequations} \label{eq:B1}
	\begin{align} 
	{B}_1 \left(\bM \right) \triangleq &\max_{||\bw||\leq 1} \bw^H \bM \bw \\
	=& \lambda_{\max} \l( \bM \r),
	\end{align}
	\end{subequations}
	where $\lambda_{\max}$ represents the maximal eigenvalue and the optimal $\bw$ is the eigenvector corresponding to the largest eigenvalue. The solution value is denoted as $B_1$ and $\frac{2 \pi}{\eta_0} B_1$ is the maximum achievable beamforming gain.
	\subsection{Continuous-Phase Unimodular Beam Design}
	In our beam codebook design, the beam codeword is subject to per-element constant power constraints. The problem with such constraint can be formulated as,
	\begin{align} \label{eq:B2}
	B_2 \left( \bM \right) \triangleq \max_{\bw: |w_i| \leq \frac{1}{\sqrt{L}}, \forall i} \bw^H \bM \bw.
	\end{align}
	
	First, it is not hard to see that the optimal $\bw$ should fully utilize the power, i.e., $|w_i| = \frac{1}{\sqrt{L}}$. A proof can be found in \cite[Corollary 2]{Pi_Zhouyue_ICC12}. 
	Second, if $\mathrm{rank}(\bM)=1$, i.e., $\bM = \bm \bm^H$, then the optimal solution is the co-phasing beamforming, i.e., $w_i^{\star} = \frac{m_i} {|m_i|\sqrt{L}}$ and $B_2  = \frac{1}{L} \l(\sum_i |m_i|\r)^2$ \cite{Raghavan_TCOM19}. However, since $\mathbf{e}_{\Theta}(\theta, \phi)$ is not a scaled vector of $\mathbf{e}_{\Phi}(\theta, \phi)$ almost surely, $\mathrm{rank}(\bM)$ is larger than one almost surely and thus there is no closed-form solution. 
	
    In fact, since both the objective function and the constraints ($w_i^* w_i = \frac{1}{L}, \forall i$) are quadratic functions  and $\bM$ is positive semi-definite, \eqref{eq:B2} is a non-convex \ac{QCQP}, which is in general an NP-hard problem proved by reducing an NP-complete matrix partitioning problem \cite{Zhang_Shuzhong_SIAM06}. An approximate solution can be found by using the prevailing \ac{SDR} method \cite{Goemans_ACM95} as follows.
    
    Denote $\bD_i$ as an $L \times L$ all-zero matrix except that the $i$-th diagonal element is 1. We relax \eqref{eq:B2} as a \ac{SDP} as follows,
	\begin{subequations} \label{eq:SDP}
		\begin{align} 
	 	\overline{B}_2(\bM) \triangleq \max_{\bW} \quad & \mathrm{tr} \left(\bM \bW \right) \\
		s.t. \quad & \mathrm{tr} \left( \bD_i \bW \right) = \frac{1}{L}, \quad 1\leq i \leq L, \\
		& \bW \geq \b0.
		\end{align}
	\end{subequations}
	
	A standard interior point method \cite{Helmberg_SIAM96} or a more efficient row-by-row block coordinate descend method \cite{Wen_Zaiwen_09,Wai_ICASSP11} can be applied to solve this convex \ac{SDP} problem. \revision{The worst-case complexity to solve a SDP is $\mathcal{O}(L^{4.5})$, while the customized row-by-row method has a complexity of $\mathcal{O}(L^3)$ \cite{Luo_Zhiquan_SPM10}.} If the obtained optimal solution $\bW_0$ is of rank one, then we can write $\bW_0 = \bw_0 \bw_0^{H}$, and $\bw_0$ is a feasible optimal solution. On the other hand, if the rank of $\bW_0$ is larger than one, then a random approximation procedure \cite{Zhang_Shuzhong_SIAM06, So_MP07, Luo_Zhiquan_SPM10} can be used to find an approximate optimal solution. The details of the procedure is shown in Algorithm \ref{alg:GRP}, where $N_G$ realizations of ${\bw} \sim \mathcal{CN} \left( \b0, \bW_0 \right)$ are generated and the best one is selected and denoted as $\widetilde{\bw}_0$. The \revision{theoretical} approximation accuracy is $\frac{\pi}{4}$, i.e., the expectation of ${\widetilde{\bw}_0}^H \bM \widetilde{\bw}_0$ is no less than $\frac{\pi}{4}$ of the global optimum \cite{Zhang_Shuzhong_SIAM06, So_MP07}. \revision{In our simulation setup where $L=4$ and $\mathrm{rank}(\bM)=2$, a rank-one solution is obtained in more than $99\%$ of the cases. That is to say, SDR provides the optimal unimodular beam in more than $99\%$ of simulation cases.}
	\begin{algorithm} 
		\caption{Gaussian Randomization Procedure (GRP) \cite{Luo_Zhiquan_SPM10}} 
		\label{alg:GRP} 
		Inputs: \ac{SDP} solution ${\bW}_0$ with $\mathrm{rank}(\bW_0) > 1$, and the number of randomizations $N_G$.
		\begin{enumerate} 
			\item Compute the eigenvalue decomposition of $\bW_0$, i.e., $\bW_0 = \bU \mb{\Lambda} \bU^H$.
			\item For $1\leq n \leq N_G$, generate $\bw^{(n)} = \bU \mb{\Lambda}^{\frac{1}{2}} \mathbf{\xi}^{(n)}$, where $\mathbf{\xi}^{(n)} \sim \mathcal{CN} \left( \b0, \bI \right)$ are complexed-valued Gaussian random vectors.
			\item Construct $N_G$ feasible solutions, 
			\begin{align} \label{eq:feasible_power}
			\widetilde{\bw}^{(n)} = \frac{1}{\sqrt{L}} \exp \left( \j \arg \left( \bw^{(n)} \right) \right).
			\end{align}
			\item Determine $\widetilde{\bw}_0 = {\arg \max_ {\widetilde{\bw}^{(n)}}} \left(\widetilde{\bw}^{(n)} \right)^H \bM \widetilde{\bw}^{(n)}$.
		\end{enumerate}
	\end{algorithm}
	
	Besides SDP-GRP, another sub-optimal but more time-efficient algorithm is to sequentially optimize the phase of each element \cite[Table 1]{Pi_Zhouyue_ICC12} \cite[Algorithm 2]{Cui_Guolong_TSP17}. The details of the iterative algorithm is given in Algorithm \ref{alg:iterative} for completeness. The solution of Algorithm 2 is guaranteed to converge to a stationary local optimal solution satisfying the \ac{KKT} condition \cite{Pi_Zhouyue_ICC12}, but may not be the optimal solution. The performance of the iterative algorithm depends on the choice of the initialization in Step 1. As an option, we can choose the initial $\bw$ as the eigenvector associated with the largest eigenvalue of $\bM$ to increase the likelihood of convergence to the global optimum. Another option is to set the solution from SDP-GRP as the initial $\bw$, i.e., concatenate SDP-GRP and Algorithm \ref{alg:iterative}. By doing so, the sequential optimization method is used to further improve the quality of the SDP-GRP solution and therefore the chance of finding the global optimum. 
	\revision{The complexity of the iterative algorithm is $\mathcal{O}(L^2)$ \cite{Cui_Guolong_TSP17}. Another alternative of Algorithm \ref{alg:iterative} is the power method-like approach given in \cite[Section III]{Soltanalian_TSP14} which, however, has a higher complexity as $\mathcal{O}(L^3)$.}
    
     
	\begin{algorithm} 
		\caption{Iterative coordinate descent algorithm for beamforming design with per-element power constraint \cite{Pi_Zhouyue_ICC12, Cui_Guolong_TSP17}} 
		\label{alg:iterative} 
		\begin{enumerate}
			\item Initialize $\bw$ and $i \leftarrow 1$.
			\item Update $w_i$ as
			\begin{align}
			w_i \leftarrow \frac{1}{\sqrt{L}} \exp \left(\j \arg \left( \sum_{k\neq i} M_{ik} w_k \right) \right).
			\end{align}
			\item Check convergence of the beamforming gain. If yes, stop; if not, let $i \leftarrow \mathrm{mod}(i, L) + 1$ and go back to Step 2.
		\end{enumerate}
	\end{algorithm}
	
	\subsection{Discrete-Phase Unimodular Beam Design}
	In practice, there is also a resolution constraint on the phase shifters. By taking into account both the per-element power and phase constraints, the beam design problem is,
	\begin{align} \label{eq:B3}
	{B}_3(\bM) \triangleq \max_{\bw: \ \l( \sqrt{L} w_i \r)^{2^b}= 1, \forall i} \bw^H \bM \bw.
	\end{align}
    
	If $\bM$ is not rank-deficient, it is proven that \eqref{eq:B3} is also a NP-hard problem \cite{Zhang_Shuzhong_SIAM06}, since it includes the max-cut problem and max-3-cut problem which are known to be NP-hard. An approximate solution can be obtained by applying the SDP-GRP technique shown above but with minor modifications. 
    First, in this discrete-phase case, we have to run the Gaussian randomization procedure even if $\bW_0$ is a rank-one matrix because of the phase constraints. Second, when constructing feasible solutions inside the Gaussian randomization procedure, the phases need be quantized. This can be done by replacing \eqref{eq:feasible_power} with the following equation,
	\begin{align} \label{eq:feasible_power_phase}
	\widetilde{\bw}^{(n)} = \frac{1}{\sqrt{L}}\exp \left( \j \mathcal{Q}_b \left( \arg \left( \bw^{(n)} \right) \right) \right),
	\end{align}
	where the function $\mathcal{Q}_b \left(\cdot \right)$ quantizes the phase from $[0, 2 \pi)$ to $\left\{0, \frac{2\pi}{2^b}, \cdots, \l(2^b-1\r) \frac{2 \pi}{2^b} \right\}$. The approximation accuracy of the \ac{SDP}-GRP solution is $\frac{ \left( 2^b \sin \left( \frac{\pi}{2^b} \right) \right)^2}{4 \pi}$ \cite{So_MP07}, which is same as the case of continuous phase, i.e., $\frac{\pi}{4}$, when $b \rightarrow \infty$.
	
	Similar to the case of continuous phase, the sequential optimization method can be used to improve the quality of the SDP-GRP solution. The iterative process is done in Step 2-3 in Algorithm \ref{alg:iterative_discrete}. 
    It is not hard to see that Algorithm \ref{alg:iterative_discrete} will definitely converge since in each iteration, the phase of $w_i$ is assigned as the optimal value from the discrete set $\left\{0, \frac{2\pi}{2^b}, \cdots, \l(2^b-1\r) \frac{2 \pi}{2^b} \right\}$ to maximize $\bw^H \bM \bw$. \revision{The overall complexity of Algorithm \ref{alg:iterative_discrete} is $\mathcal{O}(L^3)$.}
	
	\begin{algorithm} 
	\caption{Algorithm for beamforming design with per-element power and phase constraints} 
	\label{alg:iterative_discrete} 
	\begin{enumerate} 
		\item Solve the SDP given in \eqref{eq:SDP} and perform the Gaussian randomization procedure shown in Algorithm \ref{alg:GRP} where \eqref{eq:feasible_power} is replaced with \eqref{eq:feasible_power_phase} to obtain a discrete-phase solution $\bw$.
		\item Update $w_i$ as
		\begin{align}
		w_i \leftarrow \frac{1}{\sqrt{L}} \exp \left(\j  \mathcal{Q}_b \left( \arg \left( \sum_{k\neq i} M_{ik} w_k \right) \right) \right).
		\end{align}
		\item Check convergence of the beamforming gain. If yes, stop; if not, let $i \leftarrow \mathrm{mod}(i, L)+1$ and go back to Step 2.
	\end{enumerate}
\end{algorithm}

	On the other hand, when $\bM$ is rank-deficient, the problem \eqref{eq:B3} can be solved with polynomial complexity of $L$, i.e., $\mathcal{O} \left( \left( \frac{ 2^b L}{2} \right)^{2 \, \mr{rank}(\bM)} \right)$ \cite{Kyrillidis_ICASSP11}. Unfortunately, such algorithms for discrete phases have exponential complexity with respect to $b$ and cannot be extended to the continuous phase case in \eqref{eq:B2} where $2^b$ is approaching infinity. In addition, the runtime of the algorithm when $b$ is larger, e.g., $b=5$, is much longer than solving a SDP problem. Therefore, we do not adopt this method in this paper.

	Last, it is not hard to see that $B_1(\bM) \geq B_2(\bM) \ge B_3(\bM)$ since the set of feasible solutions shrinks from \eqref{eq:B1}, to \eqref{eq:B2} and \eqref{eq:B3}.

    \section{Upper Bound of the Composite Gain Pattern} \label{sec:UB}
    \revision{In the prior work, the upper bound has a uniform gain across the whole sphere by assuming all the elements are omni-directional. However, this is not the case at mmWave band where the antenna element has an inherent directional radiation pattern.}
    
    In this section, we provide an upper bound for the composite radiation pattern. The upper bound is directly derived from antenna element E-field response data (i.e., $\mathbf{e}_{\Theta}$ and $\mathbf{e}_{\Phi}$) and independent of codebook size $K$. It provides a good reference for evaluation of codebooks. For example, the number of beams required for the composite pattern to approach the upper bound can be evaluated.
    
    
    Mathematically, the upper bound is obtained by solving the following problem over the whole sphere $(\theta, \phi) \in [0^{\circ}, 180^\circ] \times [0, 360^\circ)$.
    \begin{subequations}
        \begin{align}
        \overline{S}(\theta, \phi) \triangleq & \frac{2 \pi}{\eta_0} \max_{||\bw||\leq 1}   \bw^H \bM(\theta, \phi) \bw \\
        =& \frac{2 \pi}{\eta_0} \lambda_{\max} \l( \bM(\theta, \phi) \r).
        \end{align}
    \end{subequations}
    
    The upper bound can only be achieved by a beam codebook consisting of the maximal eigenvector of $\bM(\theta, \phi)$ for every direction. In other words, we have to remove codebook size limitation, per-element power constraint, and the discrete-phase constraint to construct a codebook being able to attain the upper bound.
    
    
    Last, the upper bound for a multi-array setup is simply taken as the maximum over the upper bounds of each individual array.
    	
	\section{Greedy Algorithm} \label{sec:Greedy}
	In this section, we present a Greedy algorithm for the beam codebook design. The proposed algorithm greedily selects codewords from a candidate set as shown in Algorithm \ref{alg:Greedy}. 
	
	
	\begin{algorithm} 
	\caption{Greedy algorithm for beam codebook design} 
	\label{alg:Greedy} 
	\begin{enumerate} 
		\item Generate the candidate beam set, denoted as $\mathcal{W}_d$. Initialize the beam codebook as an empty set, i.e., $\mathcal{W}_c = \varnothing$.
		\item 
		Find a beam codeword maximizing the utility function of the spherical coverage, i.e.,
		\begin{align}
		\bw^{\star} = \argmax_{\bw \in \mathcal{W}_d \setminus \mathcal{W}_c} \ U \left(S \left(\mathcal{W}_c \cup \bw, \theta, \phi \right) \right).
		\end{align}
		Insert the selected beam codeword into the beam codebook,
		\begin{align}
			\mathcal{W}_c \leftarrow \mathcal{W}_c \ \cup \bw^{\star}.
		\end{align} 
	
		\item Stop if a certain stopping criterion is met; go back to Step 2 otherwise.
	\end{enumerate}
	\end{algorithm}

	In Step 1 of the Greedy algorithm, candidate beam codewords are generated. We provide two possible methods to generate the candidate beam codewords in \secref{sec:candidate}.
	In Step 2, given certain performance criteria, a beam from the candidate set is selected.
	In Step 3, check if the stopping condition is met. If the answer is yes, the algorithm is terminated and the selected beam codewords constitute the final codebook. Otherwise, Step 2 is repeated.
    
    \revision{Denote the size of the candidate beam set as $N_d$ and the codebook size as $K$.
    The complexity of the Greedy algorithm is mainly from Step 2, whose runtime is proportional to $\sum_{i=0}^{K-1} (N_d-i)$. 
    Along with the complexity of $\mathcal{O}(N_d L^3)$ to generate $N_d$ candidate beams, the total complexity is $\mathcal{O}(N_d (L^3 + K))$.
    }

\begin{figure*}[t]
    \centering
    \subfigure[Pattern of the first beam]{
        \includegraphics[width= 0.32\linewidth]{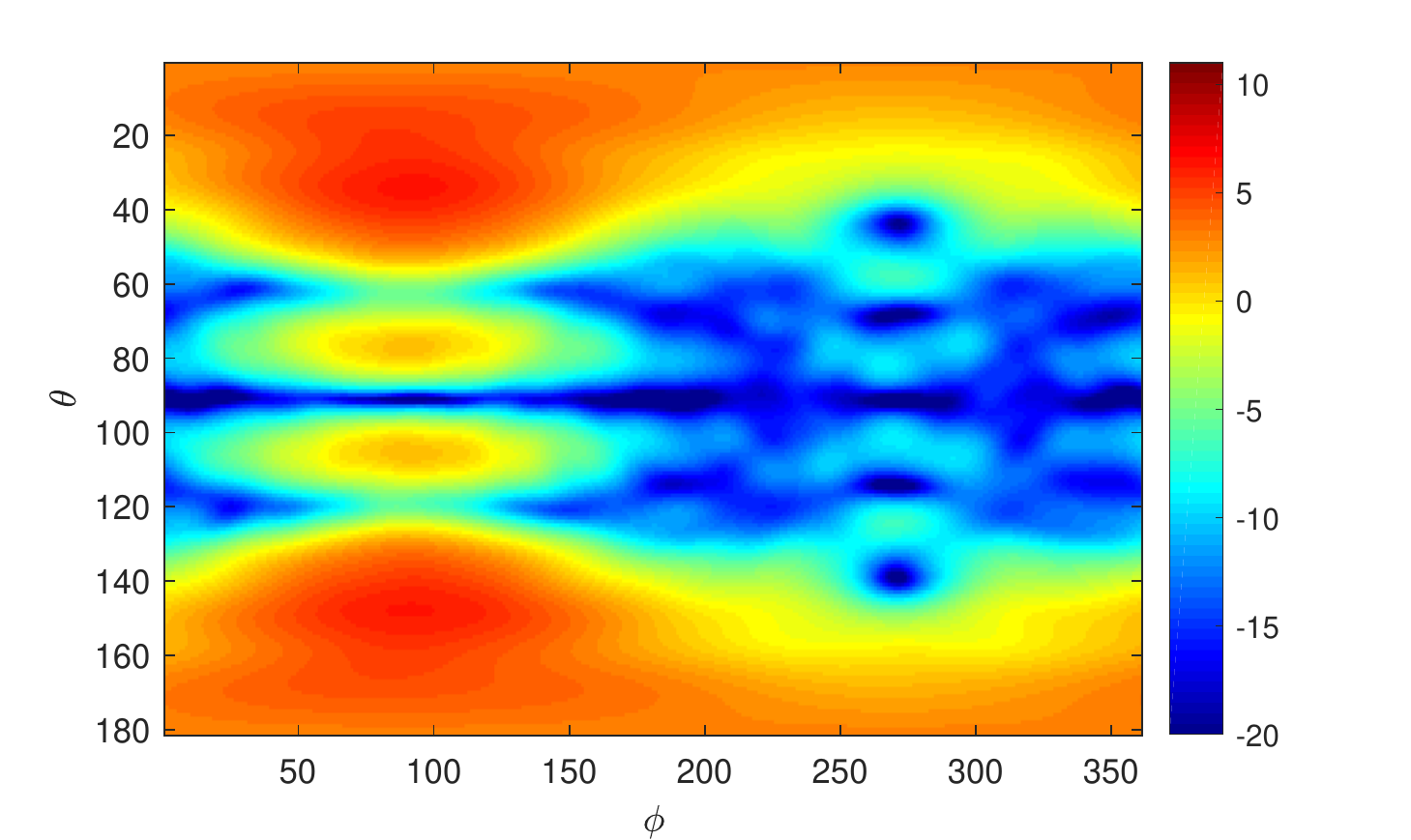}
        \label{fig:Greedy_1Beam}}
    \subfigure[Composite pattern of the first 2 beams]{
        \includegraphics[width= 0.32\linewidth]{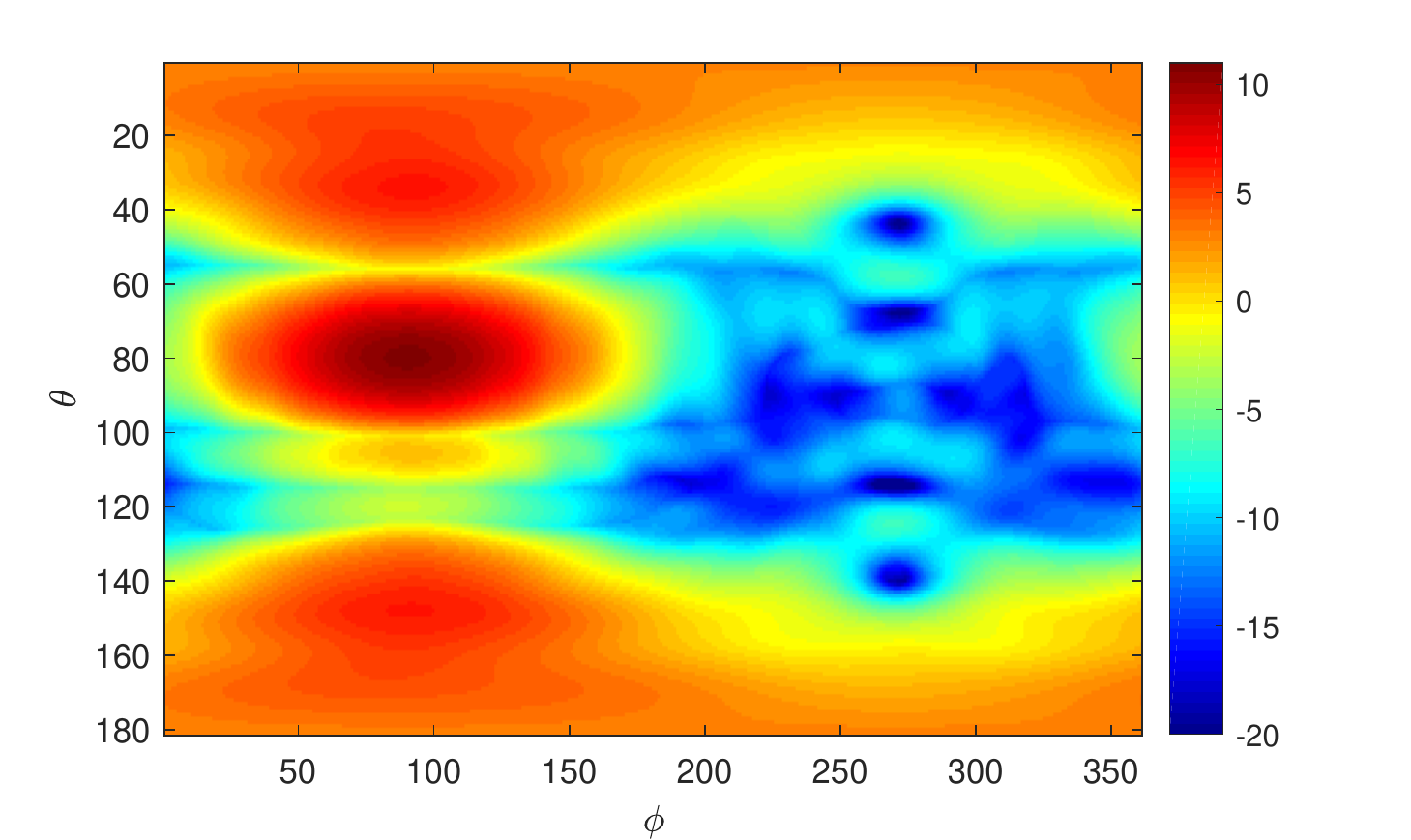}
        \label{fig:Greedy_2Beams}}
    \subfigure[Composite pattern of the first 3 beams]{
        \includegraphics[width= 0.32\linewidth]{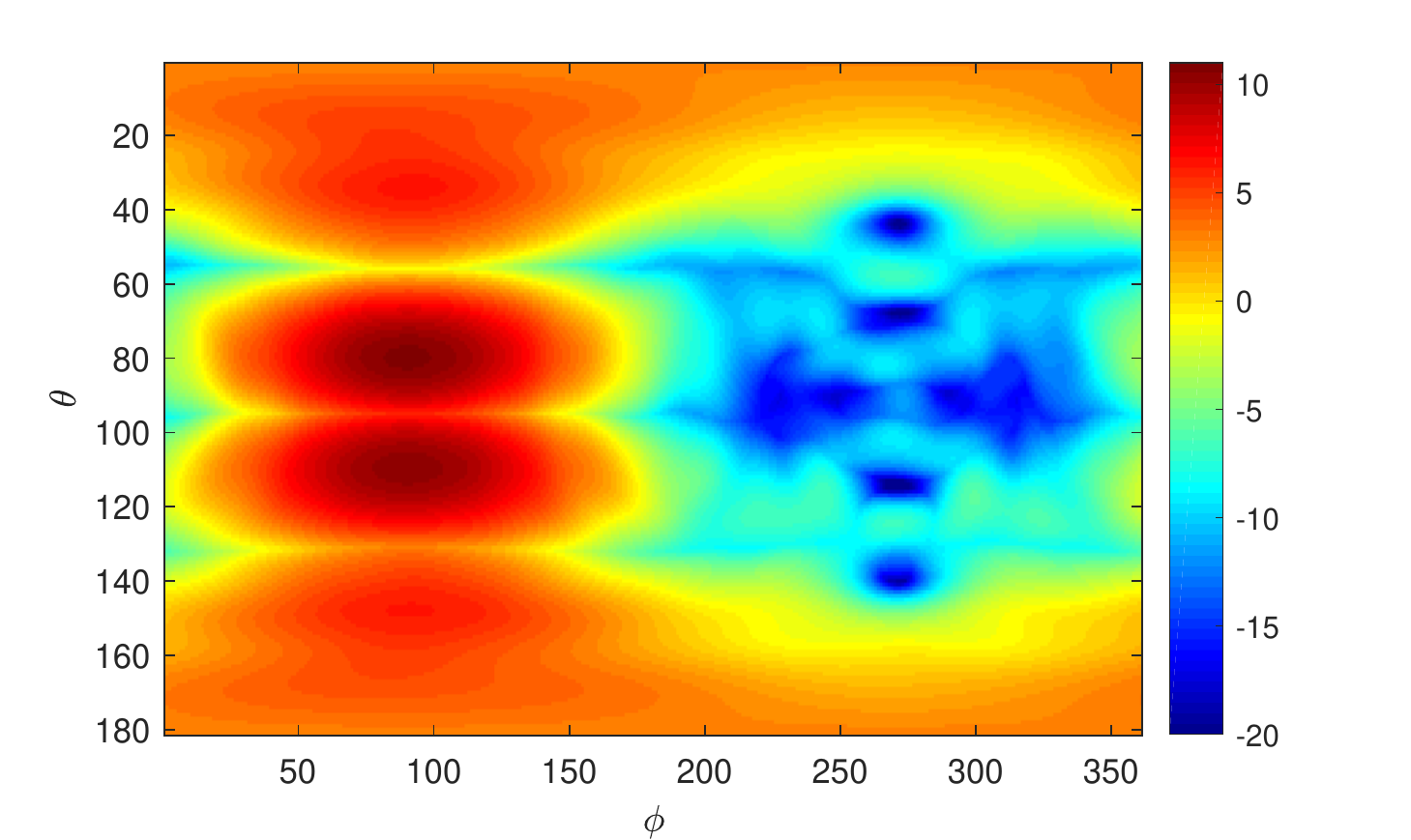}
        \label{fig:Greedy_3Beams}}
    \subfigure[Composite pattern of the first 4 beams]{
        \includegraphics[width= 0.32\linewidth]{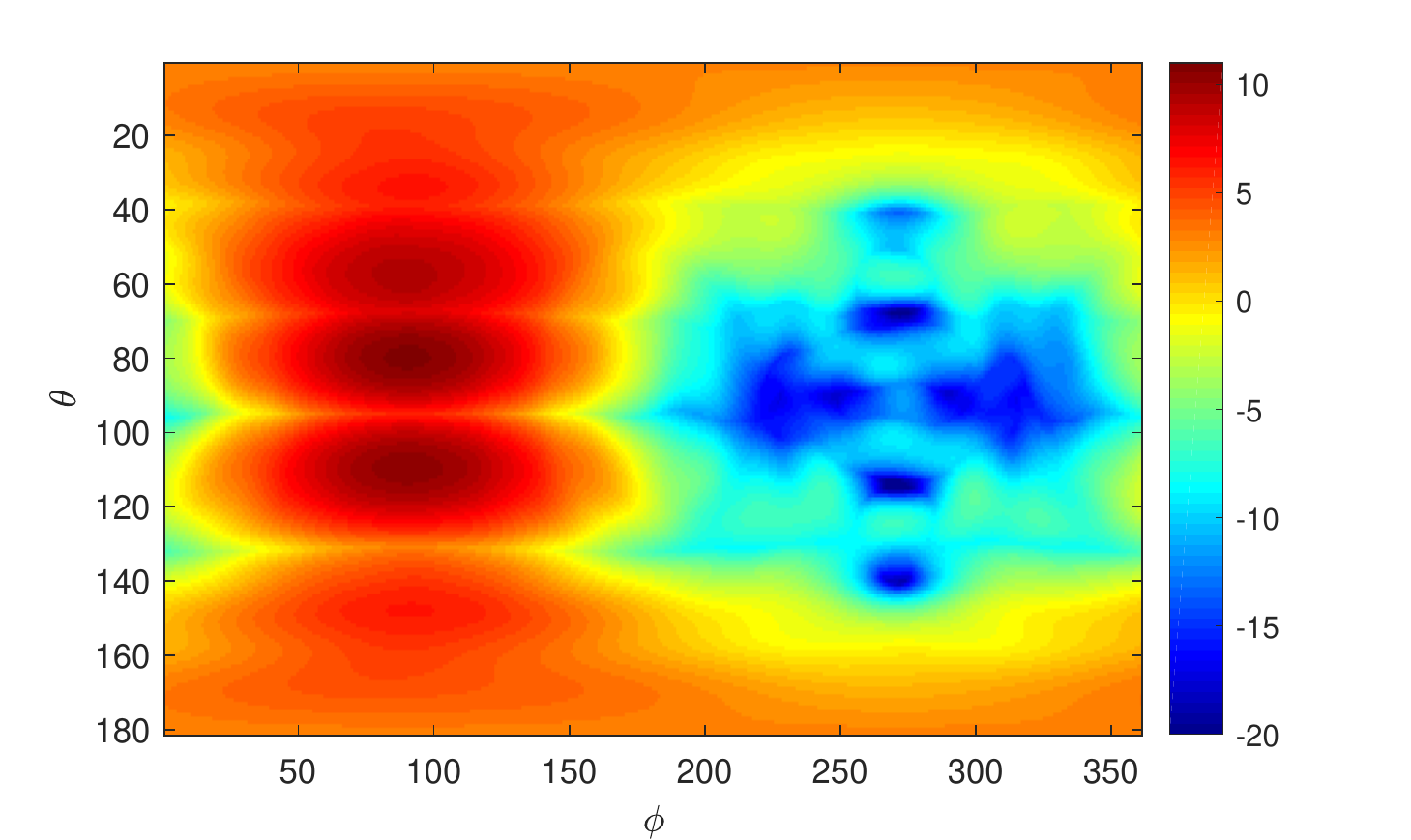}
        \label{fig:Greedy_4Beams}}
    \subfigure[Composite pattern of the first 5 beams]{
        \includegraphics[width= 0.32\linewidth]{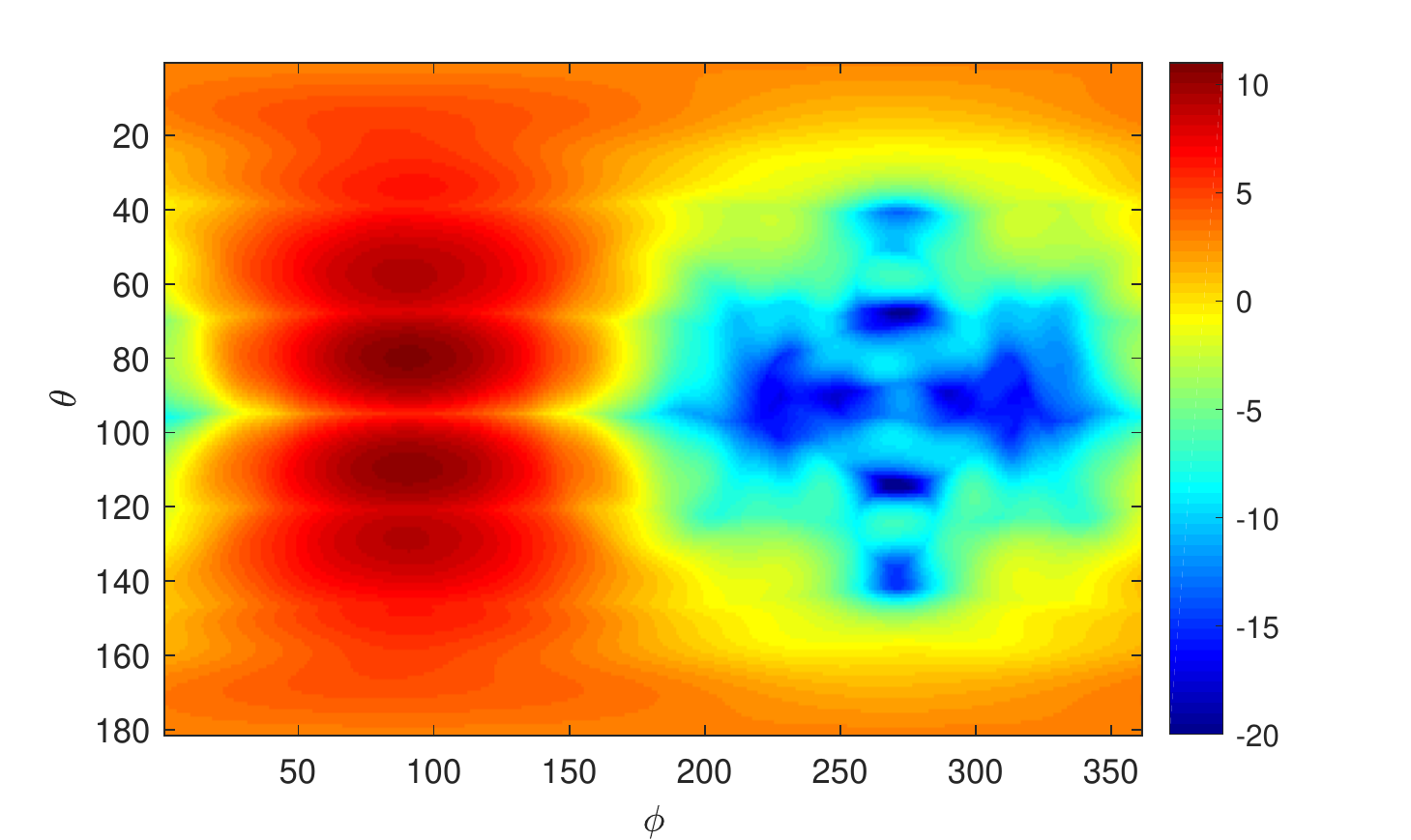}
        \label{fig:Greedy_5Beams}}
    \subfigure[3-dB contour of the beams]{
        \includegraphics[width= 0.32\linewidth]{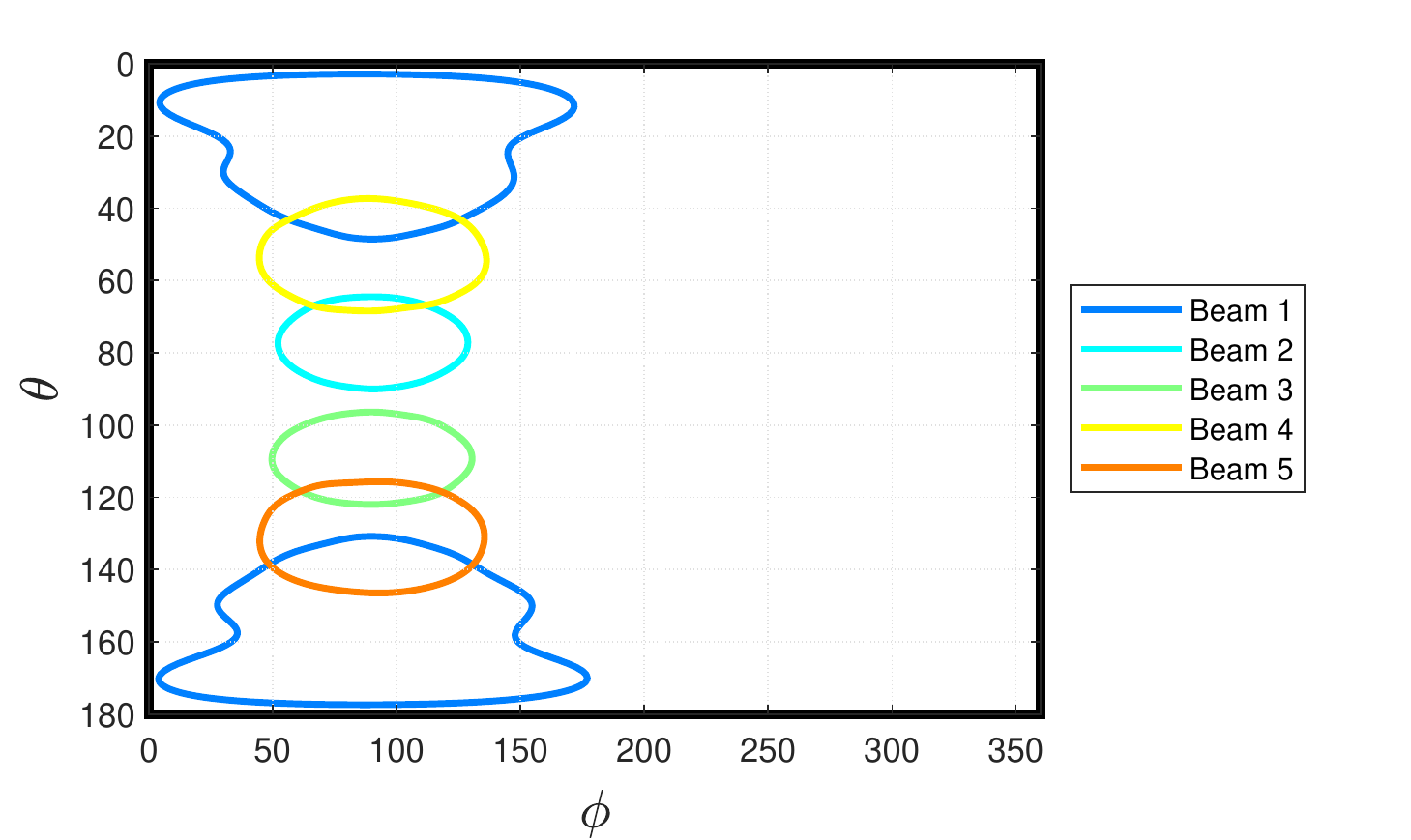}
        \label{fig:Greedy_3dB_contour}}
    \subfigure[CDF]{
        \includegraphics[width= 0.32\linewidth]{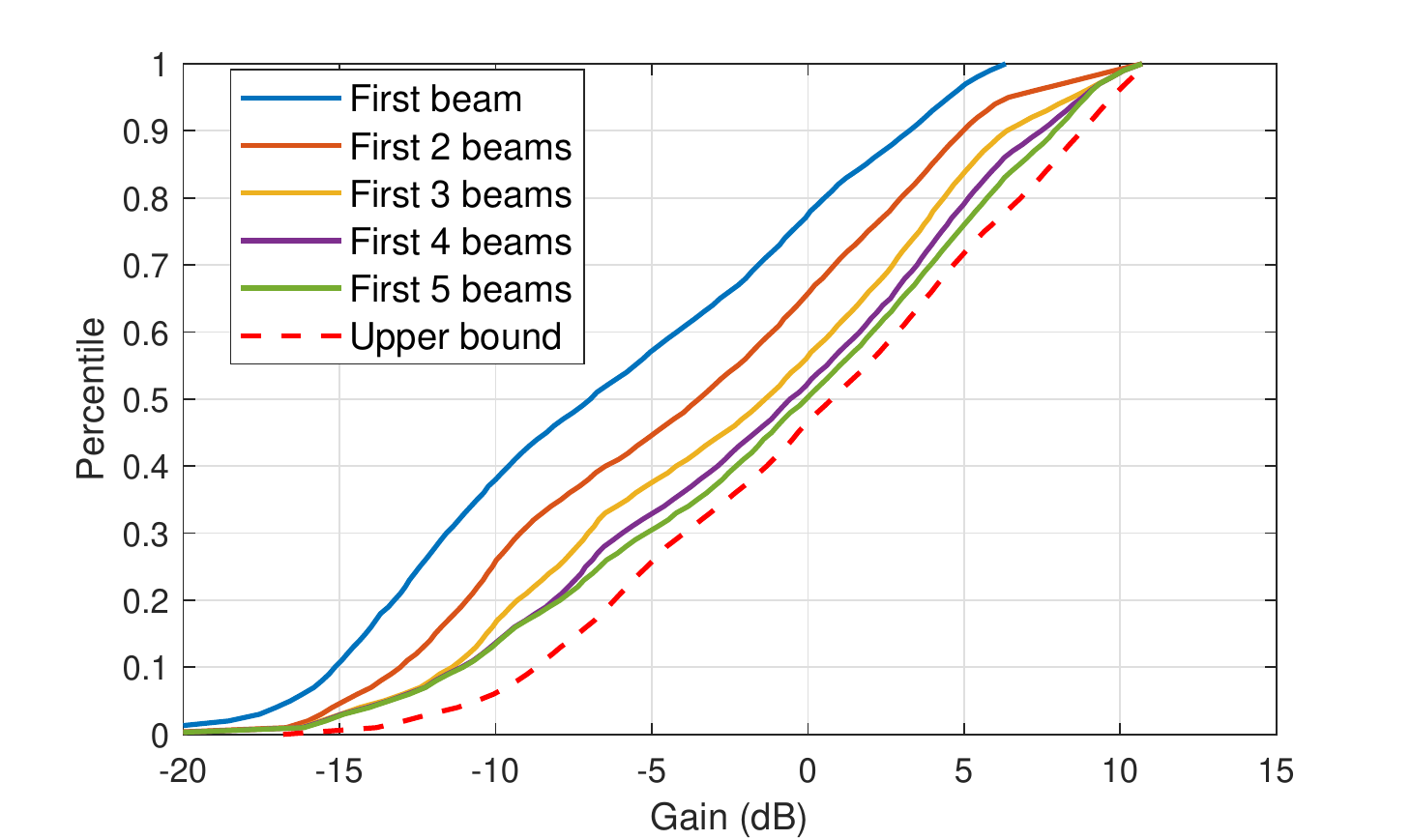}
        \label{fig:Greedy_eg_CDF}}
    \subfigure[Upper bound]{
        \includegraphics[width= 0.32\linewidth]{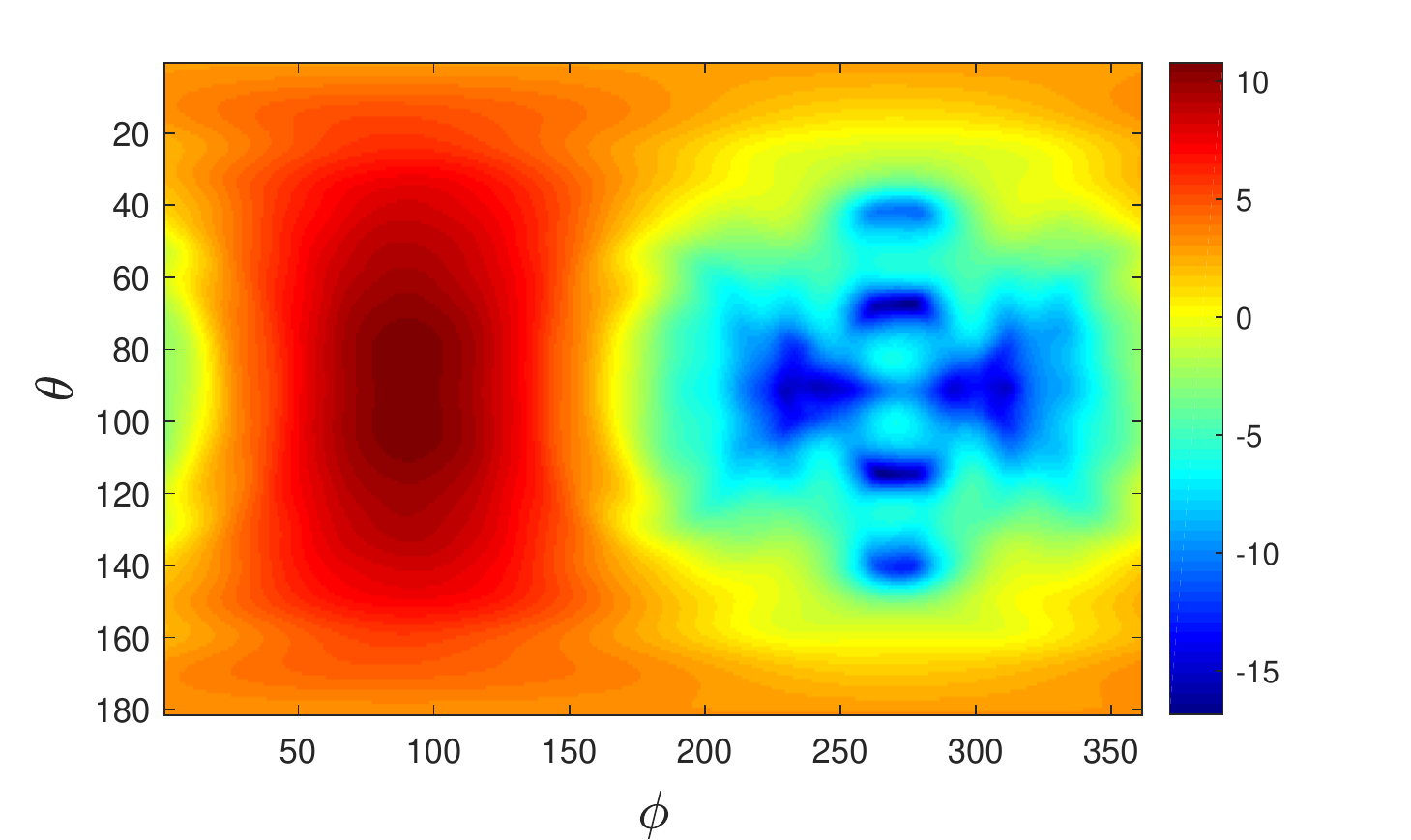}
        \label{fig:Upper_bound_right_2D}}
    \subfigure[Gap to the upper bound]{
        \includegraphics[width= 0.32\linewidth]{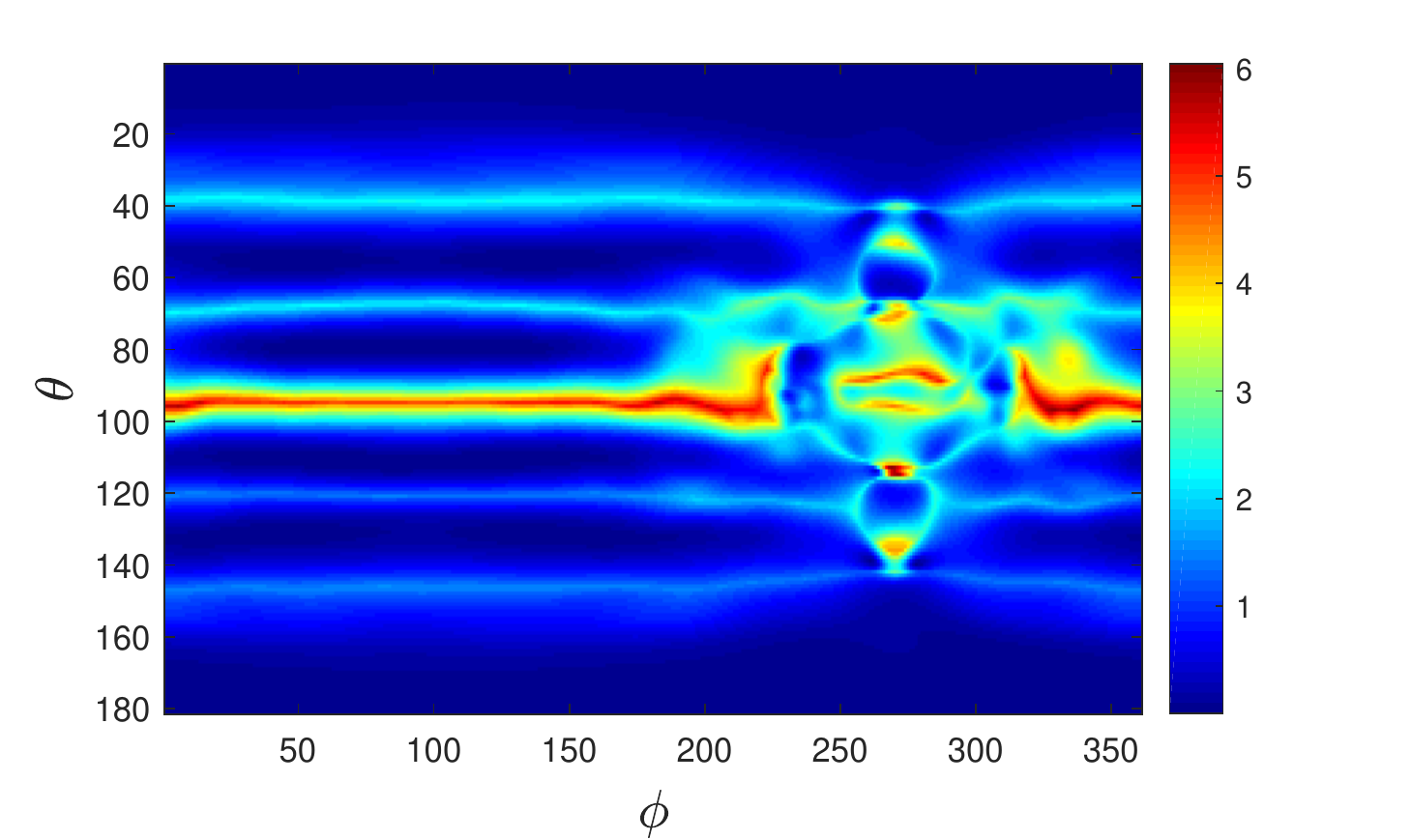}
        \label{fig:Greedy_eg_Gap}}
    \caption{Greedy algorithm operation on a linear $1 \times 4$ patch antenna array with 5-bit phase shifters. In this example, the candidate codewords are phase-quantized and magnitude-normalized eigenvectors corresponding to the maximal eigenvalues, and the selection criterion is the mean gain. Note that the beamforming gain is plotted in the dB scale.}
    \label{fig:Greedy_example}
    \end{figure*}

    \figref{fig:Greedy_example} shows an example of the Greedy algorithm operation. A linear $1 \times 4$ patch antenna array along the z-axis with broadside direction being $\left(\theta=90^\circ, \phi = 90^\circ \right)$, according to the coordinate system in \figref{fig:Coordinate}, is simulated by HFSS. 
    The 5 codewords are selected one at a time from a candidate set of 363 codewords to boost the composite radiation pattern (see \figref{fig:Greedy_1Beam}-\figref{fig:Greedy_5Beams}).  It is important to note that the main lobe of the selected beams shown in \figref{fig:Greedy_3dB_contour} are naturally pointing to different directions without explicit or manual enforcement. In each step, the selected codeword naturally targets the region with the poorest coverage thus far. It is observed in \figref{fig:Greedy_eg_CDF} that the spherical coverage improves with increasing number of beams.
    The composite radiation pattern is compared with the upper bound in \figref{fig:Greedy_eg_Gap}. The gap is less than 2 dB excluding a coverage hole located around $\theta = 95^\circ$. The codewords selected in the next iterations are expected to cover this hole.
    
    Note that for this simulated patch array, the antenna elements do not assume omni-directional radiation patterns. As clearly seen from the upper bound shown in \figref{fig:Upper_bound_right_2D}, the array can cover only a half sphere ($0^\circ \leq \phi \leq 180^\circ$) because of the high front-to-back ratio of the patch antenna. \revision{As seen in the CDF curve shown in \figref{fig:Greedy_eg_CDF},
    the dynamic range of the upper bound is from -15 dB to 10 dB, namely, the front-to-back ratio is around 25 dB.}
    In addition, the 3-dB contour of Beam 1, which is pointing to the directions away from the broadside, does not have a regular circular or ellipsoid shape.
    
    
	In the following subsections, we discuss more details on the candidate codewords generation in Step 1, selection criteria in Step 2 and the stopping condition in Step 3.

	\subsection{Candidate codewords Generation} \label{sec:candidate}
    We provide two methods to generate the candidate codewords based on the E-field data. We first pick a set of sampling points on the sphere. This can be done by generating a Fibonacci grid $(\widetilde{\theta}, \widetilde{\phi})$ \cite{Fibonacci_Swinbank06} and rounding the points to the nearest simulated (or measured) ones $(\theta, \phi)$. Then we generate the beamforming vectors pointing to these points. One option is to find the optimal or near-optimal codewords according to Algorithm \ref{alg:iterative_discrete}. The second option is to first find the eigenvector of $\bM(\theta, \phi)$ corresponding to the maximal eigenvalue and then obtain the beamforming vector by scaling the magnitude and quantizing the phase of the eigenvector to meet the per-element power and phase constraint. We will compare the performances of these two candidate codewords generation methods in Section \ref{sec:simulations}.

	\subsection{Codeword Selection Criteria}
	Codeword selection is performed based on the performance optimization criterion which defines the utility function the algorithm tries to maximize.
	
	A possible design goal is to maximize the mean gain, i.e. the average gain of the composite radiation pattern over a given spatial coverage region. In each step, the codeword that maximizes the improvement of the mean composite gain is selected, i.e.,
	\begin{align} \label{eq:greedy_criterion_mean_A}
	\bw^{\star} = \argmax_{\bw \in \mathcal{W}_d \setminus \mathcal{W}_c} \  \mathbb{E}_{\left( \theta, \phi \right) \in \mathcal{A}} \left[ S\left(\mathcal{W}_c \cup \bw, \theta, \phi \right) \right].
	\end{align}
	where $\mathcal{A}$ is the coverage region of interest on the sphere.
	
	Another option can be the maximization of gain value at one or more percentile points. For optimization with multiple percentile points, weighted average of percentile points of interest can be considered, i.e.,
	\begin{align}
		\sum_i \beta_i F_S^{-1} \left(X_i \%\right),
	\end{align}
	where $\beta_i$ is the weight.
	Single percentile point optimization is a special case with all but one percentile point set to be zero weight. 
	
	In the simulations shown in \figref{fig:Greedy_example}, the mean gain over the whole sphere is assumed as the selection criterion. Generally, the performance optimization criterion is a design choice which depends on the spherical coverage CDF requirements and link budget analysis, etc.
		
	\subsection{Algorithm Stopping Condition}
	The algorithm stopping condition can be taken from the codebook requirements.
	If there is a limitation on the codebook size, then the algorithm stops picking new codeword once enough codewords are selected. 
	
	The Greedy algorithm can also generate codebooks with variable size. In this case, stopping conditions are based on the spherical coverage performance, i.e., the algorithm is terminated once a required spherical coverage has been reached. For example, the requirement could be the average gain over an angular region $\mathcal{A}$,
	\begin{align}
	\mathbb{E}_{\left(\theta, \phi\right) \in \mathcal{A}} \left[S \left( \mathcal{W}_c, \theta, \phi \right) \right]> Y,
	\end{align}
	where $Y$ is a threshold that can be assigned. Another example of stopping condition is related to the spherical coverage CDF requirement, where the selection stops when the gain value at one (or more) $X\%$-tile is larger than a threshold  $Y'$,
	\begin{align}
	F_S^{-1} \left( X \% \right) > Y'.
	\end{align}
    
	\section{K-Means Algorithm} \label{sec:Kmean}
    
	For the Greedy algorithm, care is needed to ensure that the candidate codewords sufficiently cover the whole sphere (or the angular region of interest). If the set of candidate beam codewords offered for selection does not cover certain directions well, the resulting codebook performance can be poor, for example, coverage holes at certain directions. In this section, we propose another algorithm called \textit{K-Means}, which does not require careful constructions of the candidate codewords. As the name suggests, the core idea of this algorithm is based on the K-Means clustering, which is an unsupervised machine learning algorithm \cite{Macqueen_67}.	

	Given an initial set of $K$ beams $\{\bw_1, \bw_2, \cdots, \bw_K\}$ and a set of $N_p$  directions of interest
    \begin{align} 
    \mathcal{D}=\left\{\left(\theta_1, \phi_1\right), \left(\theta_2, \phi_2\right), \cdots, \left(\theta_{N_p}, \phi_{N_p}\right)\right\},
    \end{align}
    the algorithm proceeds by alternating between two steps:
	\begin{enumerate}
		\item \textit{Assignment step}: Assign each direction to the beam, which provides the largest gain. Mathematically, this means partitioning the set of directions $\mathcal{D}$ into $K$ subsets, denoted as $\mathcal{D}_1, \mathcal{D}_2, \cdots, \mathcal{D}_K$. 
        \revision{The set of directions $\mathcal{D}_k$ is served by the beam $\mathbf{w}_k$ and is defined as follows,
        \begin{align}
        \mathcal{D}_k = \left\{(\theta, \phi) \bigg| k=\argmax_{1\leq i\leq K} \bw_i^H \bM(\theta, \phi)\bw_i \right\}.
        \end{align}}
		\item \textit{Update step}: Optimize the beams to serve the directions in their associated subsets. This is done by solving the following optimization problem for $1\leq k \leq K$, 
		\begin{align}
		\bw_k = \argmax_{\bw:\left(\sqrt{L} w_\ell \right)^{2^b} = 1, \forall \ell} & \quad \bw^H \left( \sum_{(\theta, \phi) \in \mathcal{D}_k} \bM(\theta, \phi) \right) \bw. \label{eq:LM_Update}
		\end{align}
		Problem \eqref{eq:LM_Update} is similar to the optimization problem \eqref{eq:B3} discussed in Section \ref{sec:single_beam}. 
		Hence, Algorithm \ref{alg:iterative_discrete} is adopted to solve this problem. 
		
	\end{enumerate}
	
	The algorithm is terminated when the average gain of the composite pattern no longer improves or assignments no longer changes.

    \revision{The complexity of the K-Means algorithm concentrates on the \textit{Update step} where the Algorithm \ref{alg:iterative_discrete} is run for $K$ times in each iteration. Overall, the K-Means has a complexity of $\mathcal{O}(L^3 K N_p N_I)$  where $N_I$ is the number of iterations needed until convergence \cite{Hartigan_79}. As found in our simulations, the K-Means algorithm converges very quickly and $N_I$ is usually very small, i.e., less than 20.}

    \begin{figure*}[t]
        \centering
        \subfigure[Initial coverage]{
            \includegraphics[width= 0.32\linewidth]{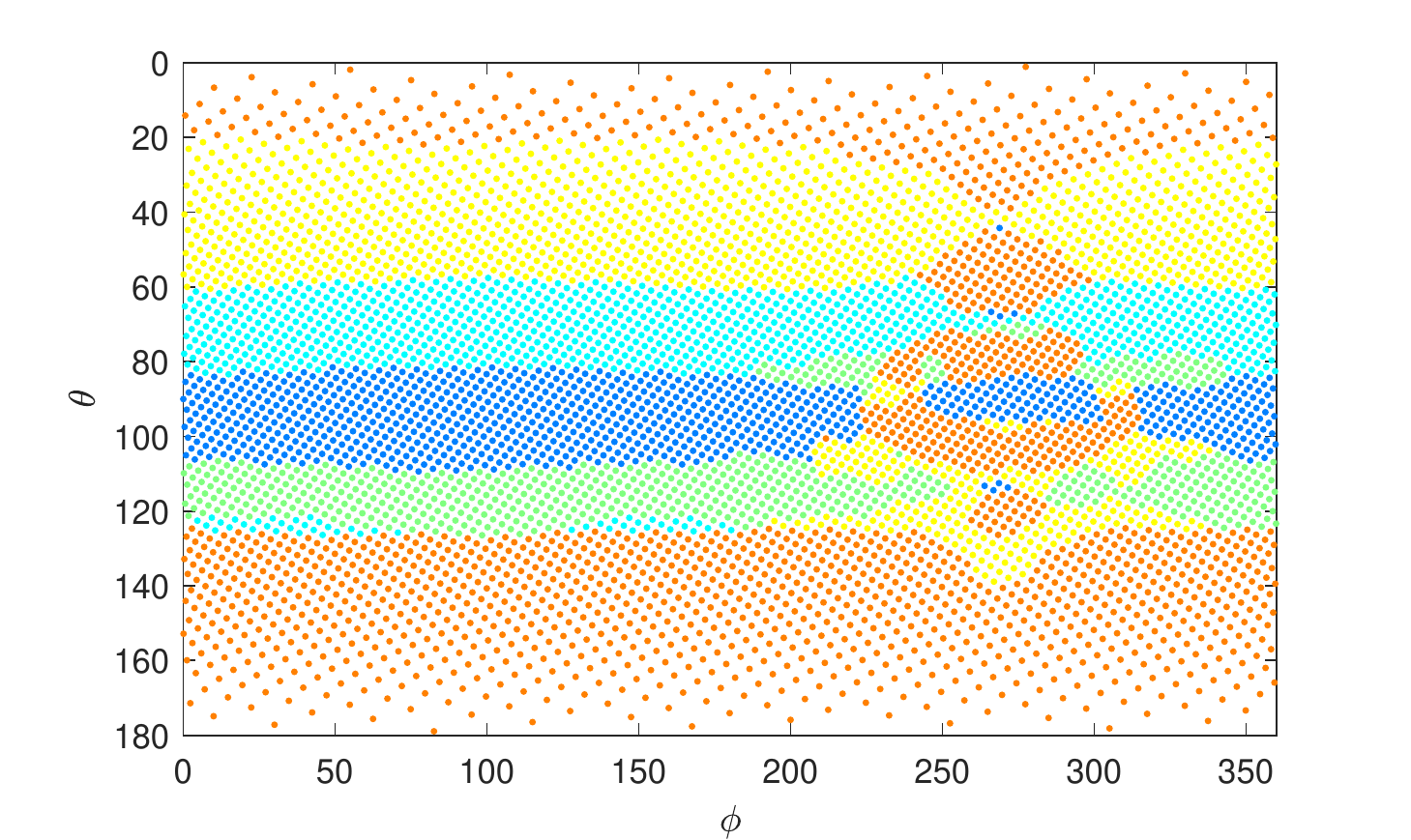}
            \label{fig:KMeans_Iter0}}
        \subfigure[Coverage at the 4th iteration]{
            \includegraphics[width= 0.32\linewidth]{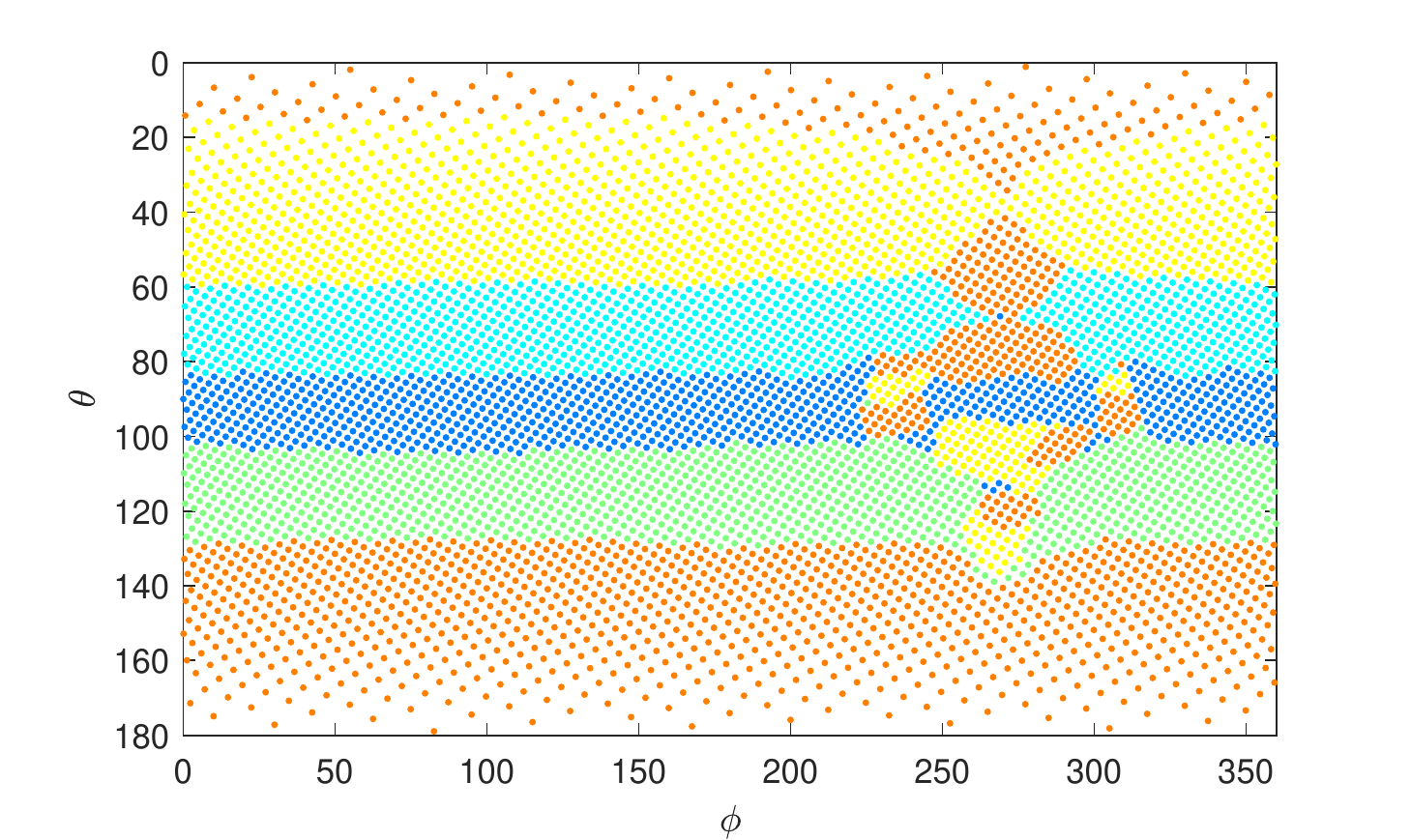}
            \label{fig:KMeans_Iter4}}
        \subfigure[3-dB contour of the beams]{
            \includegraphics[width= 0.32\linewidth]{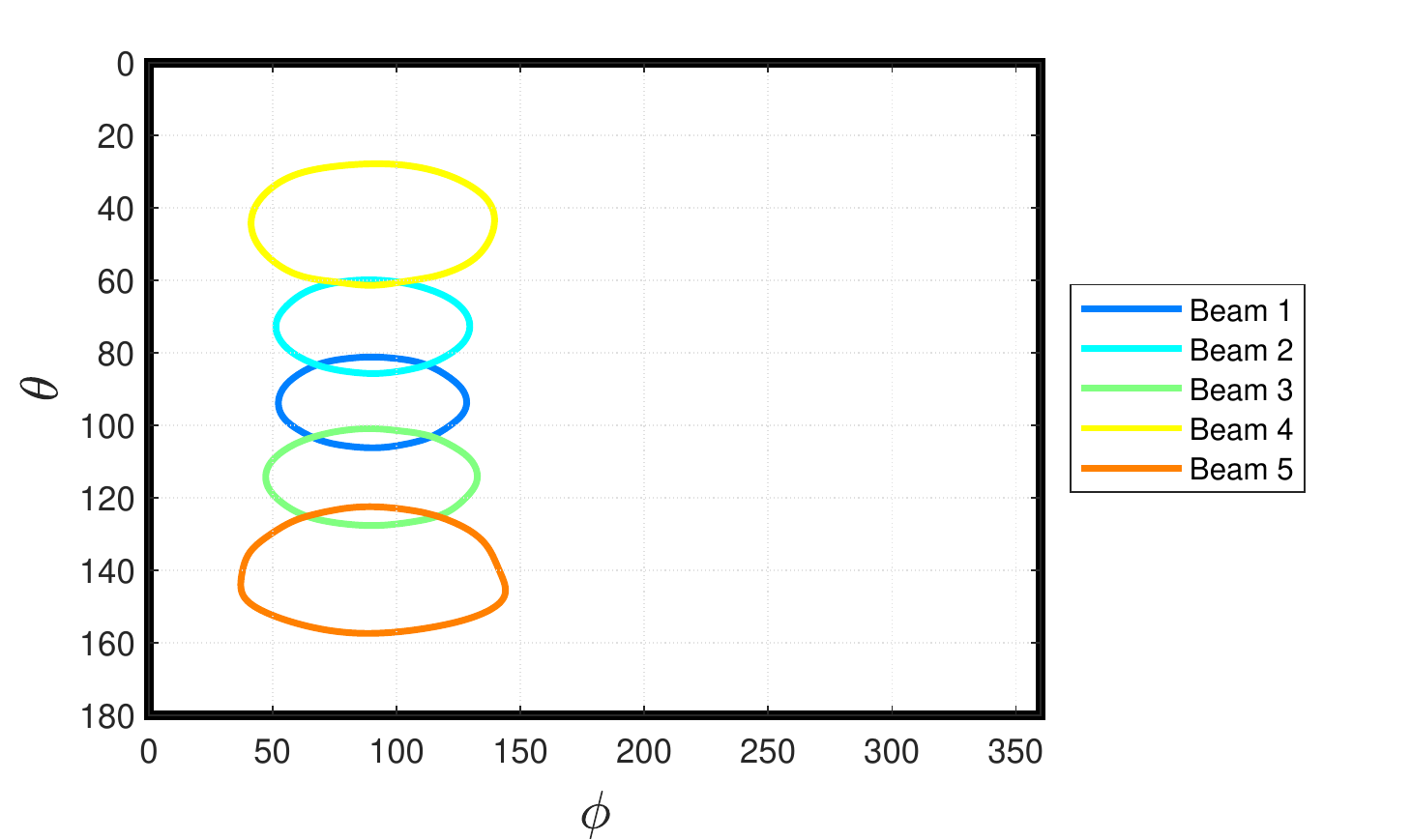}
            \label{fig:KMeans_3dB_contour}}
        \subfigure[The increase of the mean gain]{
            \includegraphics[width= 0.32\linewidth]{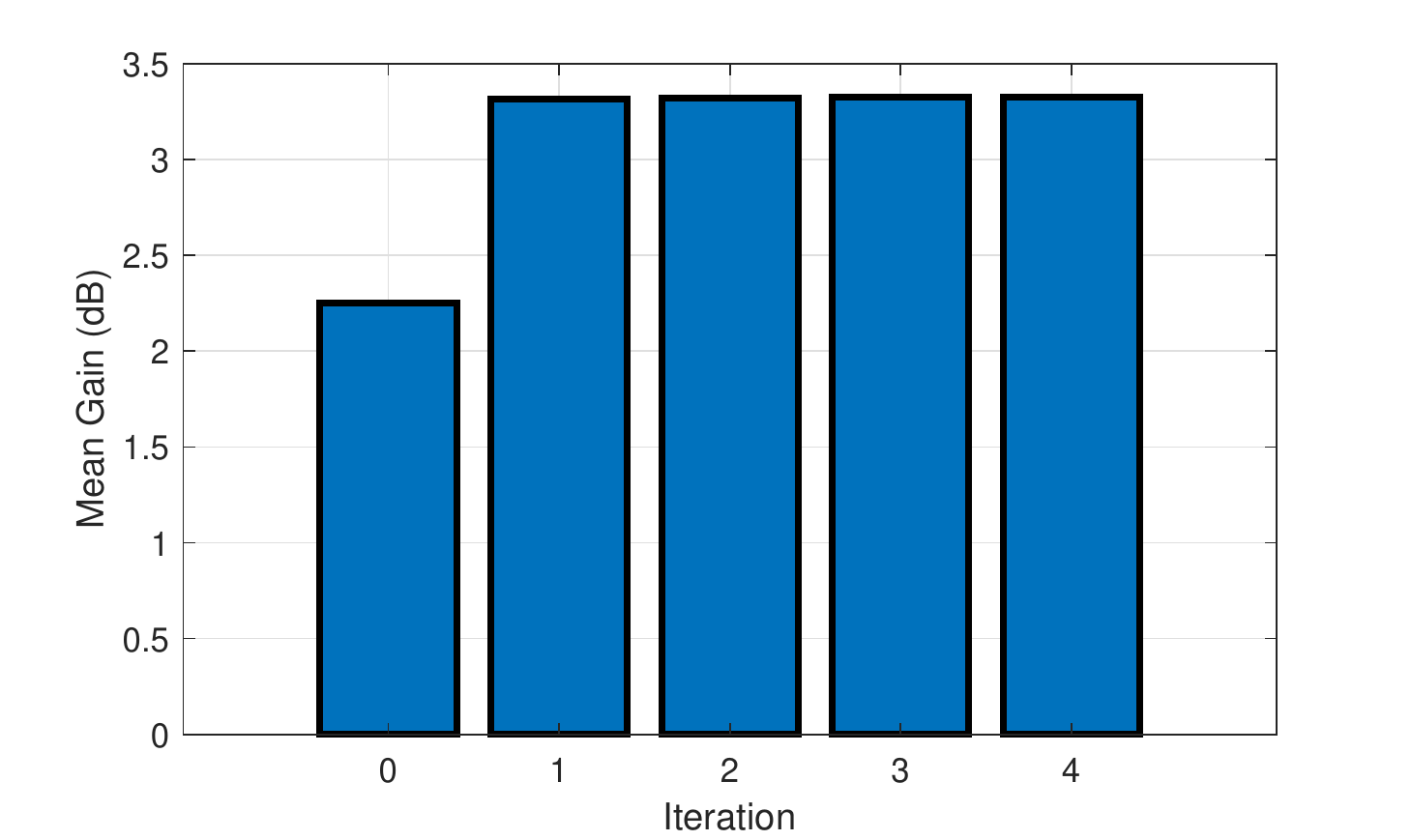}
            \label{fig:KMeans_Mean_Gain}}
        \subfigure[Composite pattern]{
            \includegraphics[width= 0.32\linewidth]{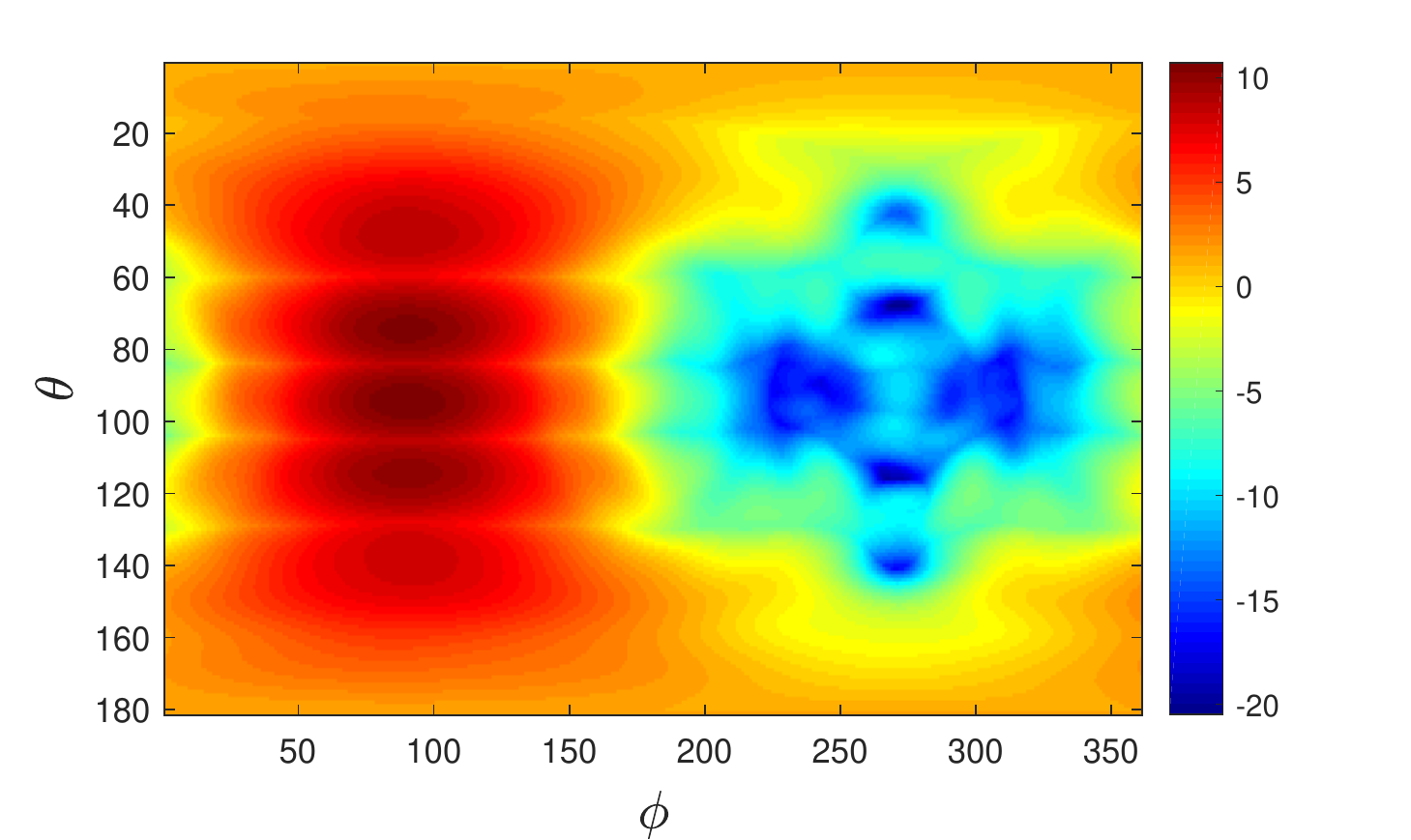}
            \label{fig:KMeans_Composition_Pattern}}
        \subfigure[Gap to the upper bound]{
            \includegraphics[width= 0.32\linewidth]{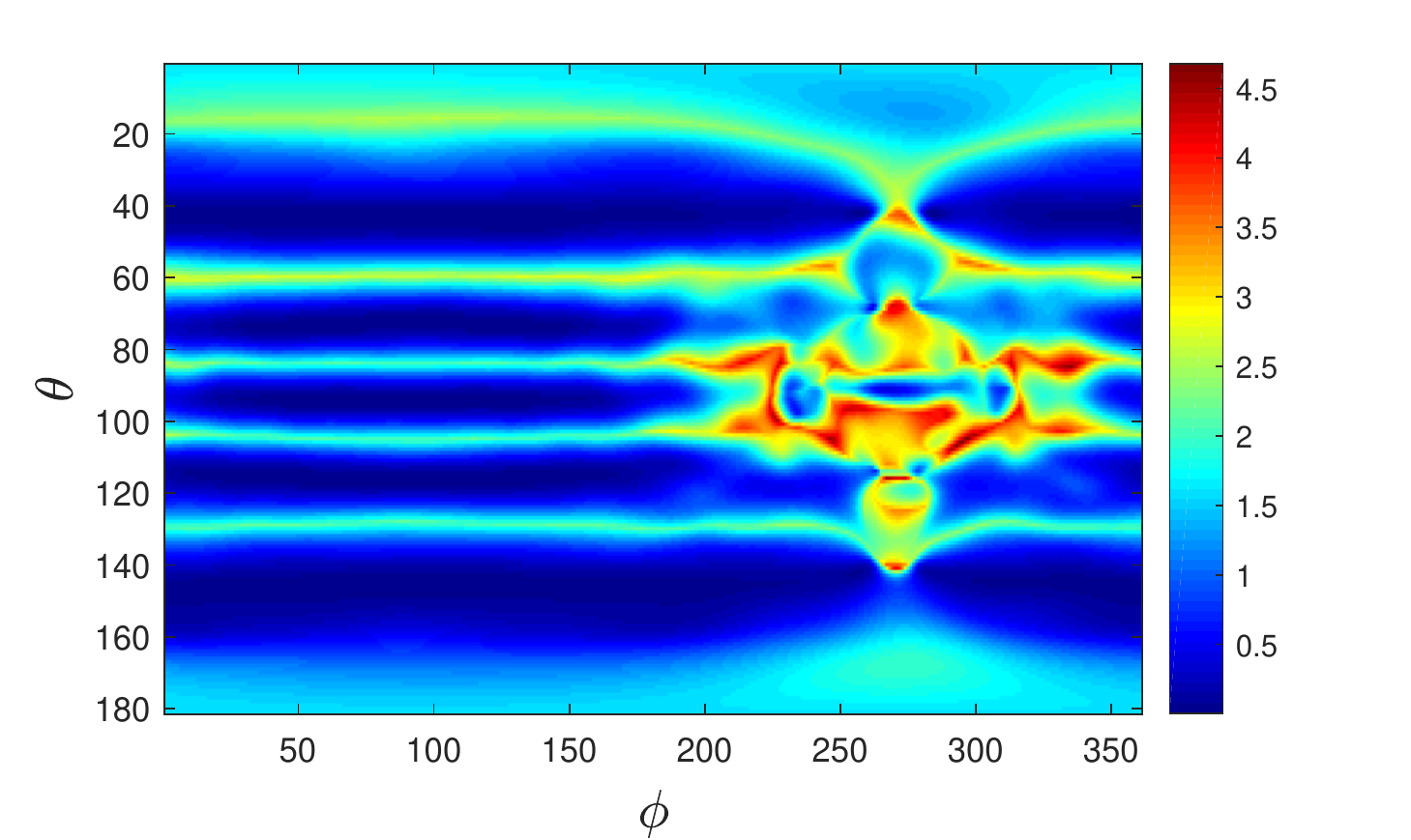}
            \label{fig:KMeans_eg_Gap}}
        \caption{K-Means algorithm operation on a linear $1 \times 4$ patch antenna array with 5-bit phase shifters. The `Uniform' initialization is used in this example.}
        \label{fig:KMeans_example}
    \end{figure*}

    \fig{fig:KMeans_example} shows an example of the K-Means algorithm. The same linear $1 \times 4$ patch array considered in \fig{fig:Greedy_example} is assumed. Each colored point in \fig{fig:KMeans_Iter0}-\fig{fig:KMeans_Iter4} represents one direction to cover. Note that for a uniform distribution of points on the sphere, there are less points around the polar regions than the equator region. The five different colors correspond to the five codewords. As seen in \figref{fig:KMeans_Iter0}-\figref{fig:KMeans_Iter4}, the coverage regions change as the K-Means algorithm updates the codebook iteratively. Compared to the Greedy algorithm example in \figref{fig:Greedy_eg_Gap}, there is no deep coverage hole in \figref{fig:KMeans_eg_Gap}. There are some directions with gap as large as 4.5 dB. However, the directions fall within the back-of-the-panel regions, which has less gain as well as interest.
    
	\subsection{Convergence of the K-Means Algorithm}
	The proposed K-Means algorithm is guaranteed to converge. This can be seen as follows. In the first step, the algorithm finds the best beam for each direction. In other words, the best assignment for a given beam codebook is obtained. Hence, the average gain increases (or keeps same) in this step. In the second step, for the directions served by the same beam, the algorithm finds out an optimal (or local optimal) beam to maximize the average gain over these directions. The average gain increases (or keeps same) in the second step as well.
	
	Since the mean gain is monotonically nondecreasing in each iteration, and there is an upper bound on the mean gain, we can conclude that the K-Means algorithm always converges. 
    
    In the above example, \fig{fig:KMeans_Mean_Gain} shows the convergence of the mean gain of the codebook. It is seen that the algorithm converges quickly within 4 iterations.

	\subsection{Initialization of the K-Means Algorithm}
	For the initialization of the codebook $\{\bw_1, \bw_2, \cdots, \bw_K\}$, two options are considered.
	\begin{enumerate}
	
	\item `Greedy': The initial codebook is generated from the Greedy algorithm shown in Section \ref{sec:Greedy}. In other words, we concatenate the two algorithms. We first run the Greedy algorithm and then take the output of the Greedy algorithm as the initialization of the K-Means algorithm.
	
	\item `Uniform': First generate $K$ uniformly distributed points on the sphere or the required coverage region. Then compute the codewords by normalizing the magnitude and quantizing the phase of the maximal eigenvector of the $\bM$ matrix at these directions. This procedure is similar to one of the methods of generating candidate codewords given in \secref{sec:candidate}, but a small number of codewords are generated.
    The idea underlying this option is to ensure that the initial codewords are pointing to different and well-separated directions.

	\end{enumerate}
	The `Uniform' initialization is employed in the example shown in \fig{fig:KMeans_example}. Although the mean gain of this initial codebook is not good (see the mean gain at the $0$-th iteration in \figref{fig:KMeans_Mean_Gain}), the mean gain increases substantially with a single iteration comprising an assignment step and an update step. We will compare these two initializations in Section \ref{sec:simulations}.

	\section{Simulation Results} \label{sec:simulations}    
	\subsection{Simulation Setup and Data Generation}
\begin{figure}[t]
    \centering
    \includegraphics[width=0.7\linewidth]{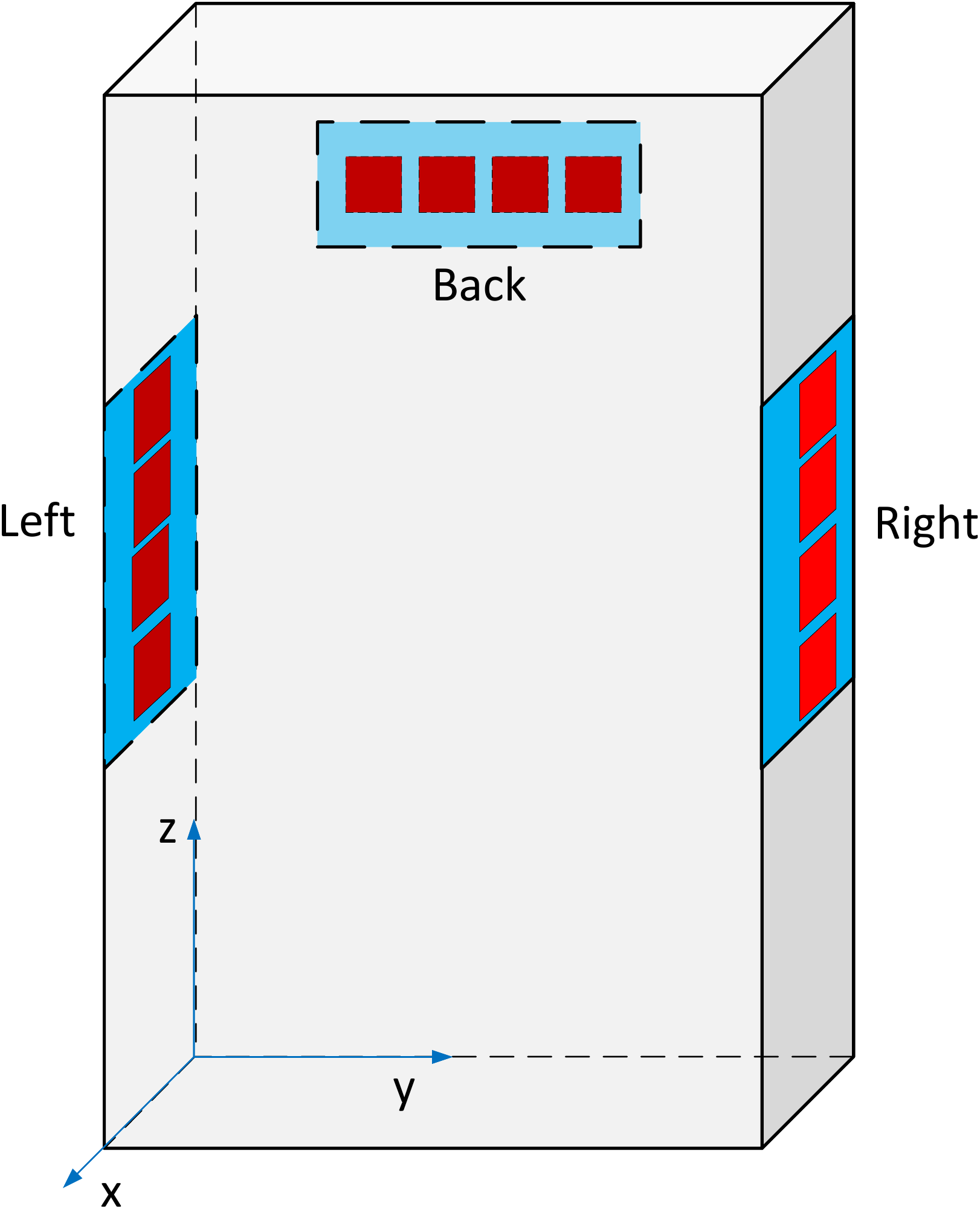}
    \caption{A terminal with three patch arrays on the left edge, right edge and the back, respectively.}
    \label{fig:Phone_left_right_back}   
    \end{figure}

    In the simulation, we consider a terminal operating at 28 GHz with three antenna arrays, where the first is placed on the left edge, the second  is placed at the right edge and the third is placed at the back of the terminal as shown in \figref{fig:Phone_left_right_back}. All the three arrays are $1 \times 4 $ linear patch antenna arrays with half-wavelength spacing. Assume that the terminal is placed vertically in the y-z plane with the front facing $+x$ direction. The three arrays are pointing to the $-y$, $+y$, $-x$ directions, respectively.

    \begin{figure}[t]
        \centering
        \subfigure[Upper bound plotted in 3-D]{
            \includegraphics[width= \linewidth]{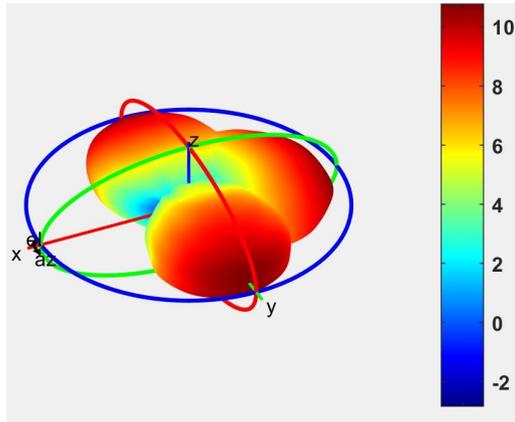}
            \label{fig:Upperbound_left_right_back_3D}}
        \subfigure[Upper bound plotted in 2-D]{
            \includegraphics[width= \linewidth]{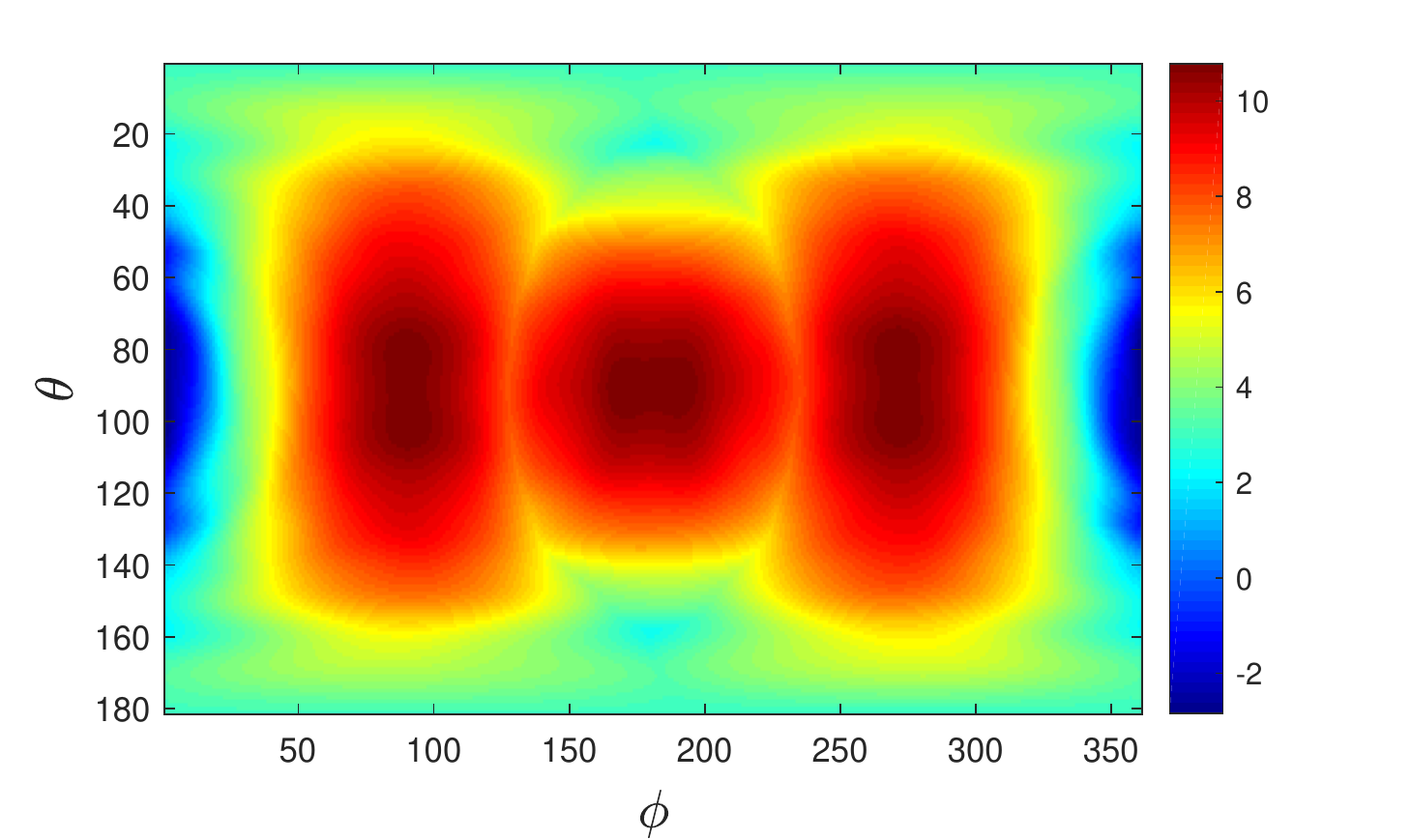}
            \label{fig:Upperbound_left_right_back_2D}}
        \caption{Upper bound of the radiation pattern. The three panels are pointing to $-y$, $+y$, $-x$ directions, respectively. }
        \label{fig:Upperbound_left_right_back}
    \end{figure}

    We assume that  each antenna element is supplied with the same power, i.e. the per-element power constraint holds. The resolution of the phase shifters is assumed to be 5 bits. In addition, we assume that only one of the antenna arrays can be activated at a given time, which is a common practice.

    The E-field data used in simulations is generated using finite-element electromagnetic simulator HFSS by Ansys. We assume the E-field data of each antenna element are available, i.e., $\mathbf{e}_{\Theta}(\theta, \phi)$, $\mathbf{e}_{\Phi}(\theta, \phi)$, in the form of discrete samples on a mesh grid, e.g. $(\theta, \phi) = [0^\circ: q_{\theta} : 180^\circ] \times [0^\circ: q_{\phi} :360^\circ)$. We assume $q_{\theta}=q_{\phi}=1^\circ$ for results illustration in this paper; however it should be noted that this is not a necessary assumption for the algorithms, and other values of $q$ can also be assumed, such as $5^\circ$ or $15^\circ$.
    
    \figref{fig:Upperbound_left_right_back_3D} and \figref{fig:Upperbound_left_right_back_2D} show the upper bound of the composite pattern in 3-D and 2-D, respectively. The three arrays cover the angular regions around $(\theta=90^\circ, \phi=270^\circ)$, $(\theta=90^\circ, \phi=90^\circ)$ and $(\theta=90^\circ, \phi=180^\circ)$, respectively.
    The mean and median gain of the upper bound is $7.56$ dB and $7.41$ dB, respectively.

    \subsection{Performance of the Proposed Algorithms}
    In this subsection, we compare the algorithms proposed in this paper. Depending on different initializations, we consider five implementations of the proposed algorithms as follows.
    \begin{enumerate}
    \item \textbf{Greedy(Eigen)}: Greedy algorithm where the candidate codewords are phase-quantized and magnitude-normalized eigenvectors corresponding to the maximal eigenvalues.
    
    \item \textbf{Greedy(Iterative)}: Greedy algorithm where the candidate codewords are generated using Algorithm \ref{alg:iterative_discrete}.
    
    
    \item \textbf{K-Means(Greedy(Eigen))}: K-Means algorithm initialized by the `Greedy(Eigen)' algorithm.
    
    \item \textbf{K-Means(Greedy(Iterative))}: K-Means algorithm initialized by the `Greedy(Iterative)' algorithm.
    
    
    \item \textbf{K-Means(Uniform)}: K-Means algorithm where the initial codewords are beamforming to $K$ uniformly distributed directions.
    \end{enumerate}

    For the two implementations of the Greedy algorithm listed above, the selection criterion is assumed to be the mean gain over the whole sphere, which is aligned with the optimization metric of the K-Means implementations for the sake of a fair comparison. 363 candidate codewords pointing to quasi-uniformly distributed 363 directions are generated \revision{by either `Eigen' or `Iterative' approach.} The angle separation of adjacent directions is around $10^\circ$. The Greedy algorithms stop selecting new codewords when the codebook size limitation is reached. \revision{For the K-Means algorithm, the beams are updated by Algorithm 3 where the number of randomization in the first step is chosen as $N_G=1000$.}
    
\begin{figure}[t]
    \centering
    \subfigure[Mean gain]{
        \includegraphics[width= 1.0\linewidth]{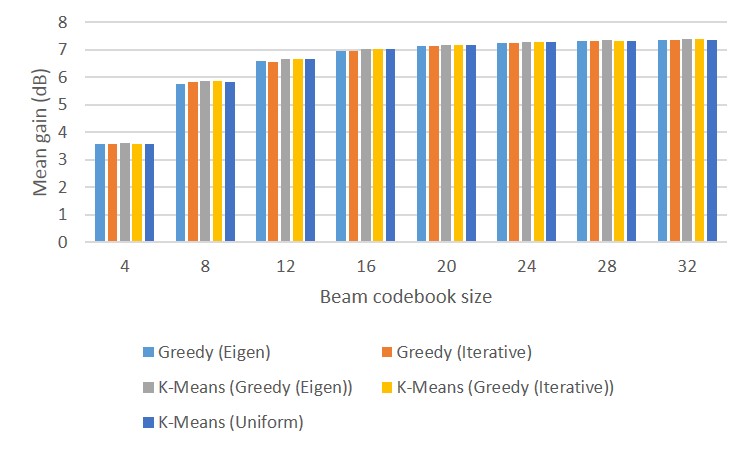}
        \label{fig:Algorithms_Comparison_Mean}}
    \subfigure[Median gain]{
        \includegraphics[width= 1.0\linewidth]{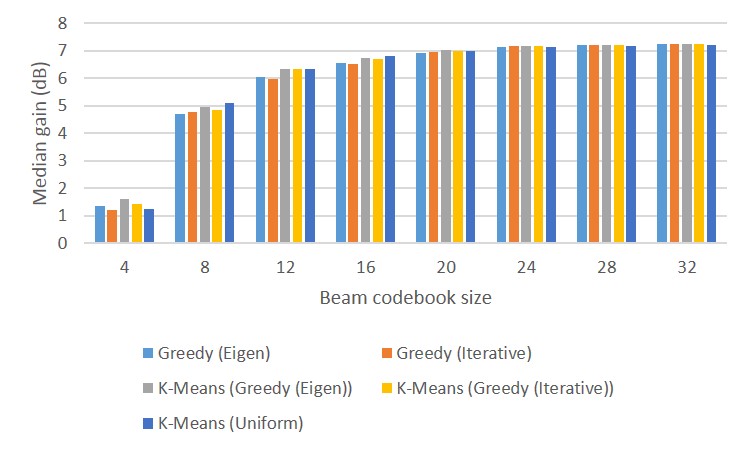}
        \label{fig:Algorithms_Comparison_50th}}
    \caption{The mean gain and median gain of the codebooks generated by different implementations.
    } 
    \label{fig:Algorithms_Comparison}
    \end{figure}

    In \figref{fig:Algorithms_Comparison_Mean}, we compare the mean beamforming gain over the sphere. Our first observation is that the choice of the candidate codewords, i.e.,`Iterative' and `Eigen', does not result in a significant performance difference. This can be explained by noting that, although an `Iterative' codeword may be slightly better than the `Eigen' codeword in a given direction, it may be worse than the `Eigen' codeword when considering the average gain of the surrounding region of this direction. Our second observation is that there is nearly no performance difference across these implementations. Nevertheless, there may be meaningful performance difference for other antenna and terminal designs. Therefore, it is expected that the choice of the algorithm needs to be considered on a case-by-case basis. Finally, the mean gain increases with the codebook size and saturates when the codebook size is larger than 24. Actually, when the codebook size is 32, the mean gain is around $7.39$ dB, which is very close to the mean gain of the upper bound, i.e., $7.56$ dB.

    In \figref{fig:Algorithms_Comparison_50th}, we show the median gains produced by the algorithms since the 3GPP has defined the requirement of spherical coverage for handheld UE in terms of the median gain. Unlike the case with mean gains, there are more variations among different algorithms. This is reasonable since we set the mean gain as the common optimization metric and distributions with the same mean value could have very different median values. Furthermore, the difference in the median gains is very small when the codebook size is larger than 20,  implying that all algorithms are converging to similar spherical coverage as the codebook size increases.
    
    In our simulations, we find that the K-Means algorithm generally provides slightly better performance than the Greedy algorithm, but not in all cases. In particular, it is less likely to find coverage holes in the codebooks generated by K-Means algorithm than those by Greedy algorithm. However, the Greedy algorithm is much more flexible than K-Means algorithm. First, there are many possible choices about the utility function and stopping condition in the Greedy algorithm, while the K-Means has limited options. In this paper, we only consider the metric of mean gain for the K-Means algorithm. Second, the Greedy algorithm can generate variable-sized codebooks, whereas the codebook size has to be determined before running the K-Means algorithm. To sum up, the choice of the algorithm depends on the codebook requirements as well as the E-field data.
    
\section{Advantages of the proposed method}  \label{sec:other_advantages}
    Our proposed method can be easily configured to deal with different antenna setups and to meet a variety of requirements on the beam codebook. In this section, we show several such cases.

    In this subsection, we compare the proposed method with two other designs. The K-Means(Greedy(Iterative)) algorithm is assumed here for the performance comparison purpose. 
    
    The benchmark beams are designed to point to certain directions. In this paper, we assume that a benchmark codebook for a single linear array consist of $K'$ beams pointing to directions $\arccos \left( -1+
    \frac{2k-1}{K'} \right)$, where $k=1, 2, \dots, K'$ \cite{Xiao_Zhenyu_TWC16, Xiao_Zhenyu_TWC17, Xiao_Zhenyu_TVT18}.
    For example, a beam codebook of size 4 consists of codewords pointing to $138.6^\circ$, $104.5^\circ$, $75.5^\circ$, $41.4^\circ$ with respect to the array axis, respectively. The beam codewords are computed as,
    \revision{
    \begin{align}
    w(\ell, k) = \frac{1}{\sqrt{L}}\exp \left( \j \mathcal{Q}_b \left( \frac{2\pi d \ell}{\lambda} \left(-1+ \frac{2k-1}{K'} \right) \right) \right),
    \end{align}
    where the function $\mathcal{Q}_b \left(\cdot \right)$ quantizes the phase from $[0, 2 \pi)$ to $\left\{0, \frac{2\pi}{2^b}, \cdots, \l(2^b-1\r) \frac{2 \pi}{2^b} \right\}$.
    }

    In additional to the benchmark method, the 802.15.3c codebook is also included here for comparison.
    The original 802.15.3c codebooks are designed for 2-bit phase-shifters. To have a fair comparison, we adopt its generalization to $2^b$-phase codebook shown as follows \cite{Zou_Weixia_ChinaCom14}, 
    \begin{align}
    &w(\ell, k) \nn \\
    = &\frac{1}{\sqrt{L}}\exp\left({\j \frac{2 \pi}{2^b} \lfloor\frac{(\ell-1) \times \mathrm{mod}\left(k-1+ \frac{K'}{2}, K' \right)}{K'/2^b} \rfloor}\right),
    \end{align}
    for $1\leq \ell \leq L, 1\leq k \leq K'$, where $\lfloor x \rfloor$ rounds $x$ to the nearest integer less than or equal to $x$. 
    The 802.15.3c codebook basically consists of $K'$ codewords having (approximately) progressive phases.

    \subsection{Joint Design of Multi-array Codebook}
    
    \begin{table}[t]
        \centering
        \caption{\revision{Comparison of the different codebooks in terms of mean gain and median gain. The three arrays are mounted at the left edge, right edge and back of the terminal.}}
        \label{tab:Benchmark_3c_KMeans_12_24Beams_left_right_back}
        \revision{
        \begin{tabular}{|c|c|c|c|c|c|}
            \hline 
            {\multirow{2}{*}{Beamforming gain (dB)}} & \multicolumn{5}{c|}{Codebook size $K$} \\ 
            \cline{2-6} 
            & 12 & 15 & 18 & 21 & 24\\ 
            \hline 
            Mean Gain of Benchmark.  & 6.556 & 6.854  &  7.045 & 7.171 & 7.248\\ 
            \hline
            Mean Gain of 802.15.3c.  & 6.423 & 6.816  &  7.030 & 7.156 & 7.246\\ 
            \hline
            Mean Gain of Proposed. & \textbf{6.665} & \textbf{6.945} & \textbf{7.102} & \textbf{7.216} & \textbf{7.279}\\
            \hhline{|=|=|=|=|=|=|} 
            Median gain of Benchmark. & 6.305 & \textbf{6.744} & \textbf{6.934} & 7.047 & 7.120\\
            \hline
            Median gain of 802.15.3c. & 5.804 & 6.362 & 6.745 & 6.964 & 7.128\\
            \hline
            Median gain of Proposed. & \textbf{6.322} & 6.659 & 6.870 & \textbf{7.118} & \textbf{7.171}\\
            \hline
        \end{tabular}
        }
    \end{table}

    

    \begin{figure*}[t]
        \centering
        \subfigure[Benchmark codebook]{
            \includegraphics[width=0.32 \linewidth]{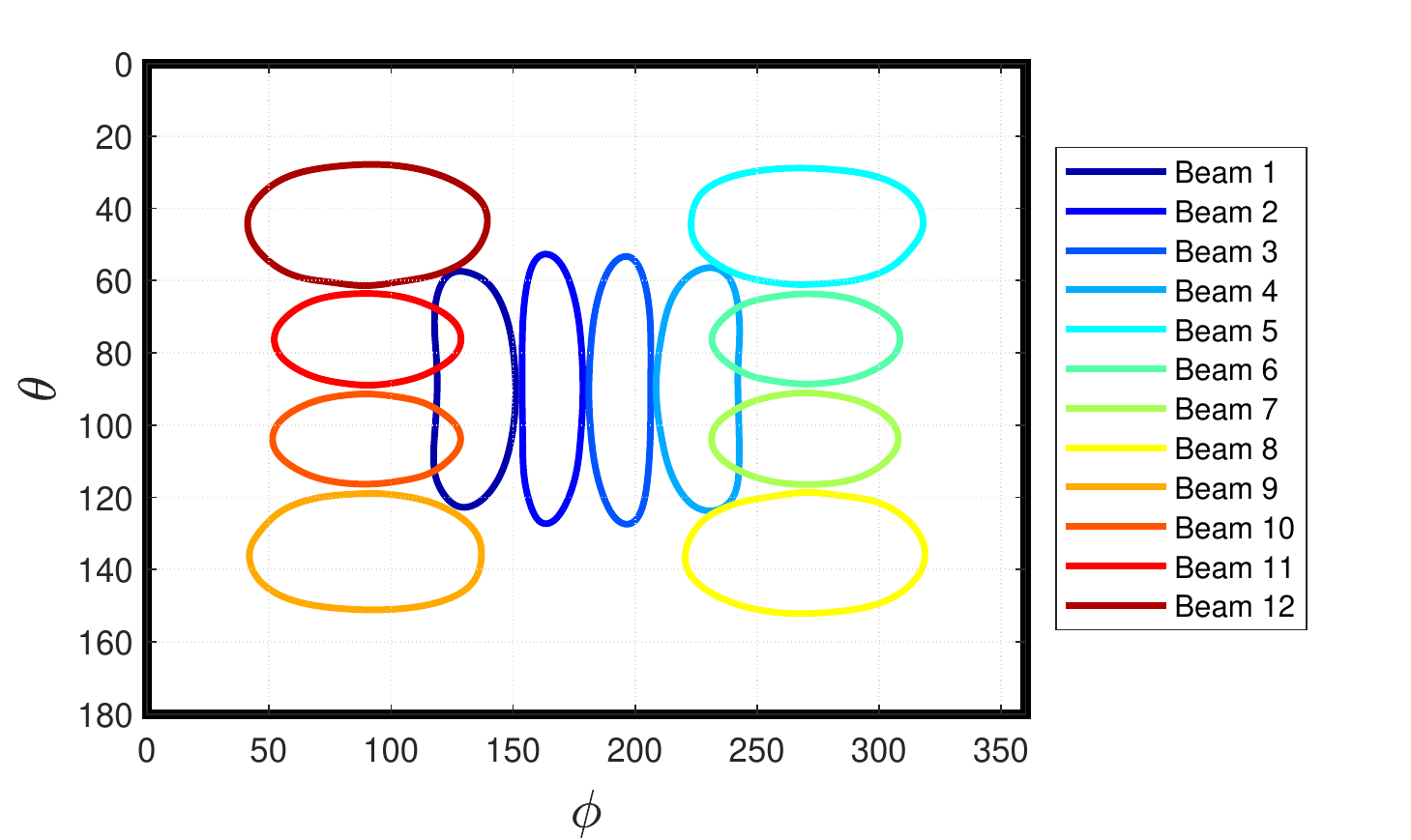}
            \label{fig:Benchmark_Left_Right_Back_3dB_contour}}
        \subfigure[802.15.3c codebook]{
            \includegraphics[width=0.32 \linewidth]{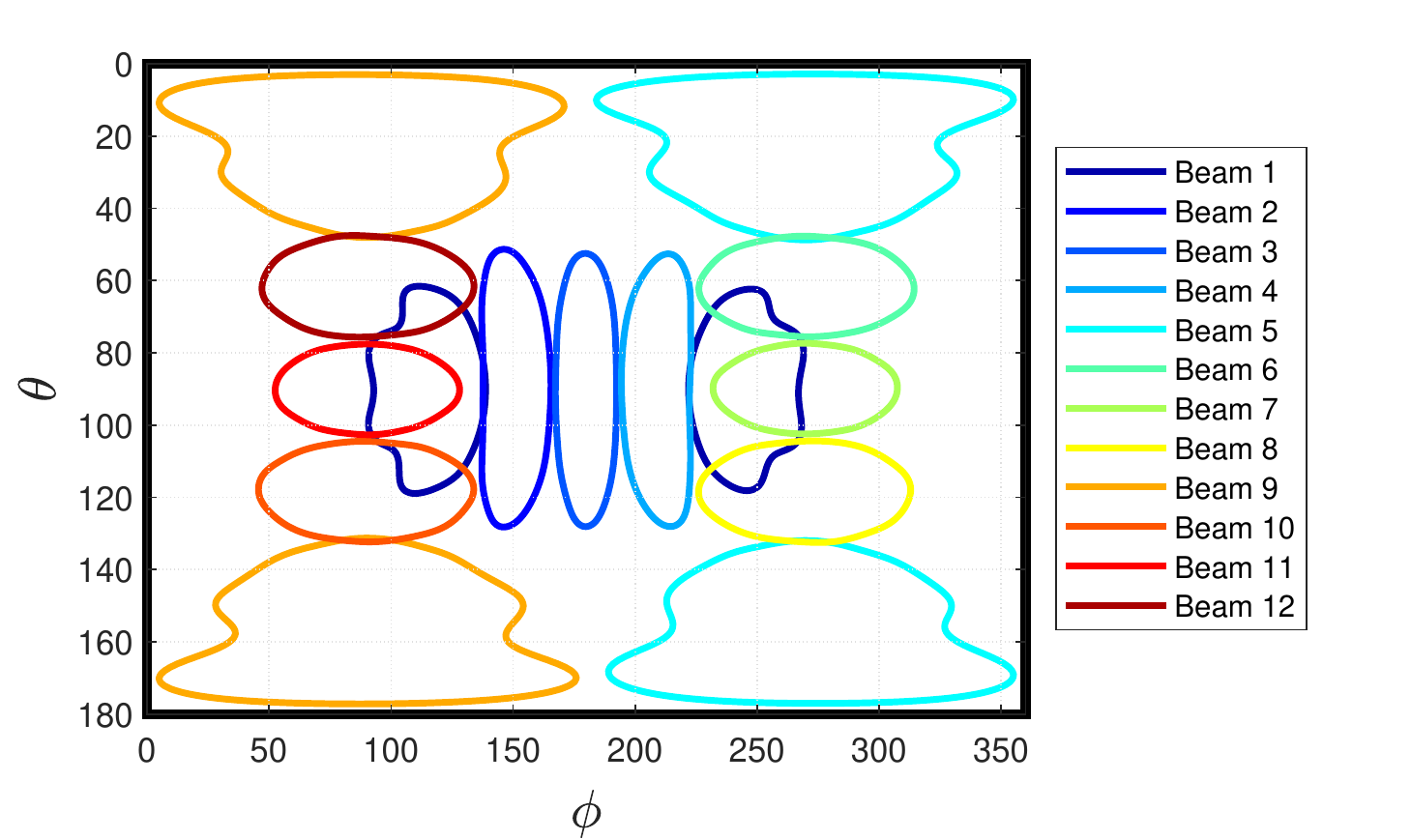}
            \label{fig:3c_Left_Right_Back_3dB_contour}}
        \subfigure[Proposed codebook]{
            \includegraphics[width=0.32 \linewidth]{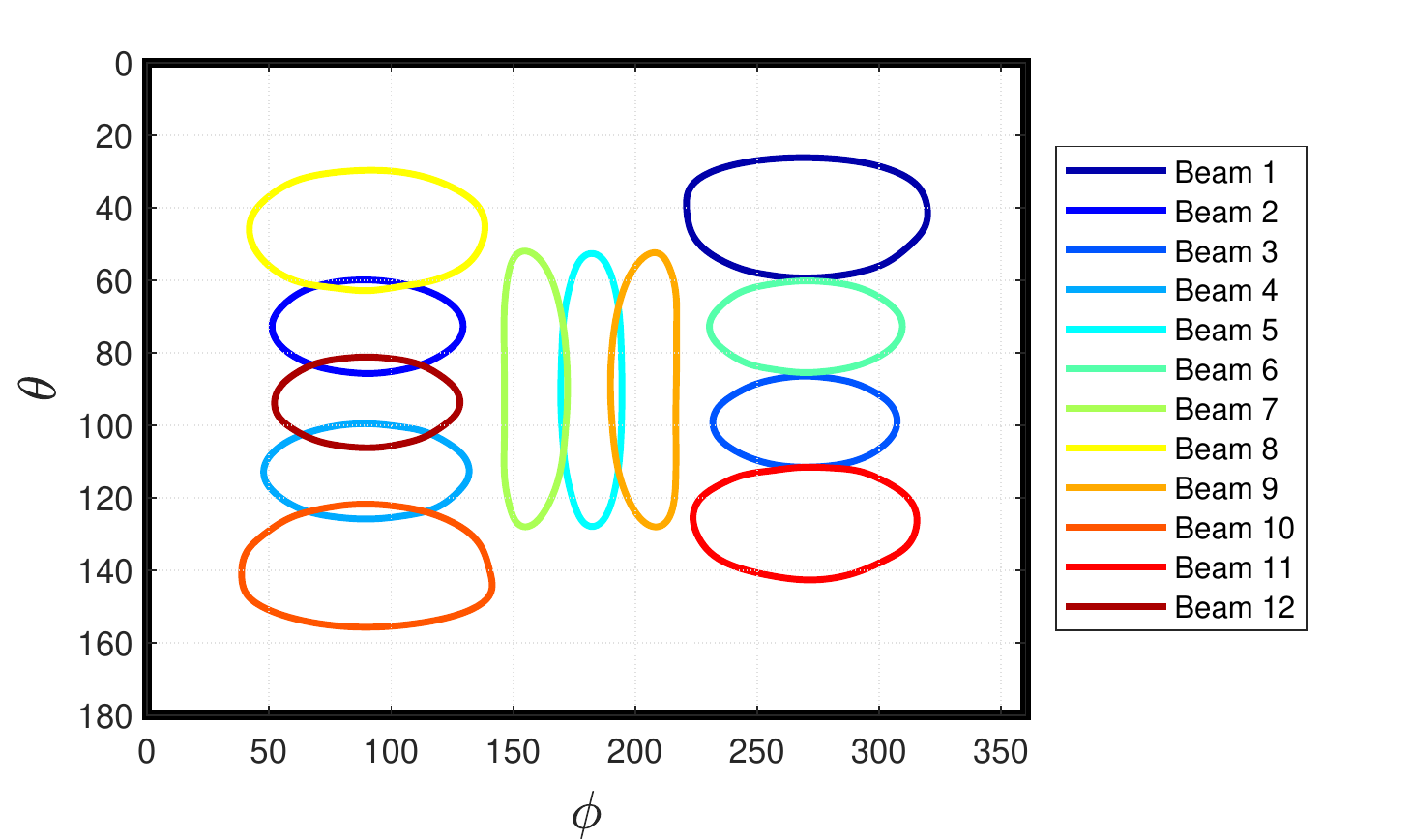}
            \label{fig:KMeans_Greedy_Iter_Left_Right_Back_3dB_contour}}
        \caption{The 3-dB beam contours of the benchmark and proposed codebooks when $K=12$. The three arrays are mounted at the left edge, right edge and back of the terminal.}
        \label{fig:Comparison_Left_Right_Back_3dB_contour}
    \end{figure*}
    
    When there are multiple arrays mounted on the terminal, such as the three patch arrays as shown in \figref{fig:Phone_left_right_back}, a conventional design may assume the same set of codewords for each array \cite{Raghavan_TCOM19}. This assumption restricts the codebook size to be an integer multiples of the number of arrays. In contrast, there is no limitation on the choice of codebook size in our proposed algorithm, as seen in \figref{fig:Algorithms_Comparison}. More importantly, the conventional design does not take into account the possible overlapping of the coverage regions of different arrays and therefore the generated codebook may include codewords pointing to similar directions.

    As shown in \tabref{tab:Benchmark_3c_KMeans_12_24Beams_left_right_back}, the proposed codebook is better than benchmark and 802.15.3c codebooks in terms of the mean and median gains, in most of the cases. The advantage of the proposed algorithm is large especially when the codebook size is small, i.e, $K=12$. When the codebook size increases, the performance of all the algorithms approach the upper bound and thus are similar to each other.
    The advantage of the proposed algorithm is clearly seen in \figref{fig:Comparison_Left_Right_Back_3dB_contour} where the 3-dB beam contours are shown. In the benchmark codebook, beam 1 and beam 4 associated with the back array is pointing to the regions which are also partially covered by the beams from the other two arrays, i.e. beam 10-11 and beam 6-7. Similarly for the 802.15.3c codebook, beam 1 associated with the back array is pointing to the regions which are also covered by the beams from the other two arrays, i.e. beam 6-8 and beam 10-12.
    By contrast, the proposed codebook displays a much better coordination among different arrays by automatically allocating different number of beams to the arrays and avoiding the beam overlapping.

	\subsection{Flexible Adaptation to Required Coverage Region}
    \begin{figure*}[t]
        \centering
        \subfigure[3-dB contour of beams]{
            \includegraphics[width= 0.32\linewidth]{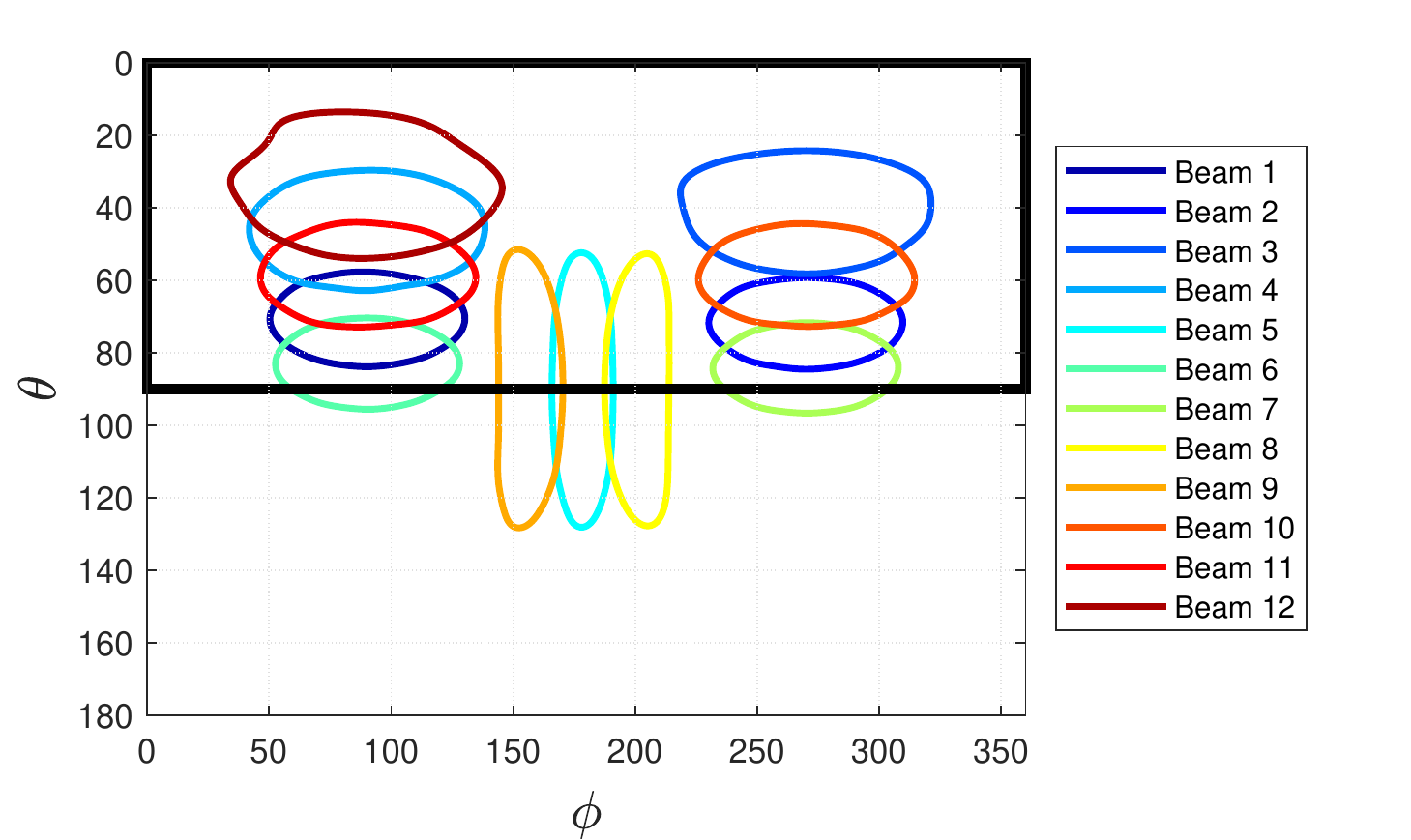}
            \label{fig:Coverage_eg1_3dB_contour}}
        \subfigure[Composite pattern]{
            \includegraphics[width= 0.32\linewidth]{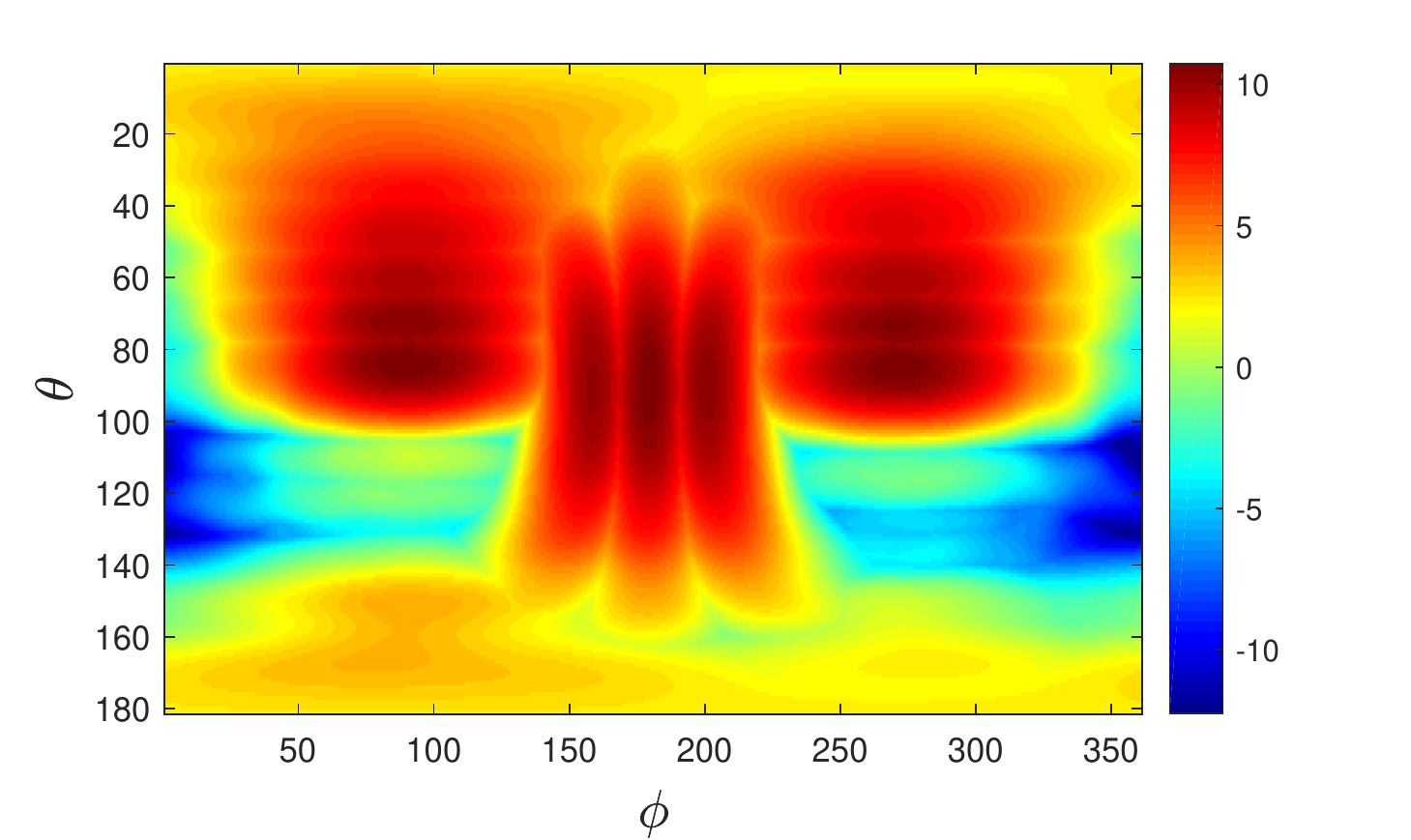}
            \label{fig:Coverage_eg1_pattern}}
        \subfigure[Gap to the upper bound]{
            \includegraphics[width= 0.32\linewidth]{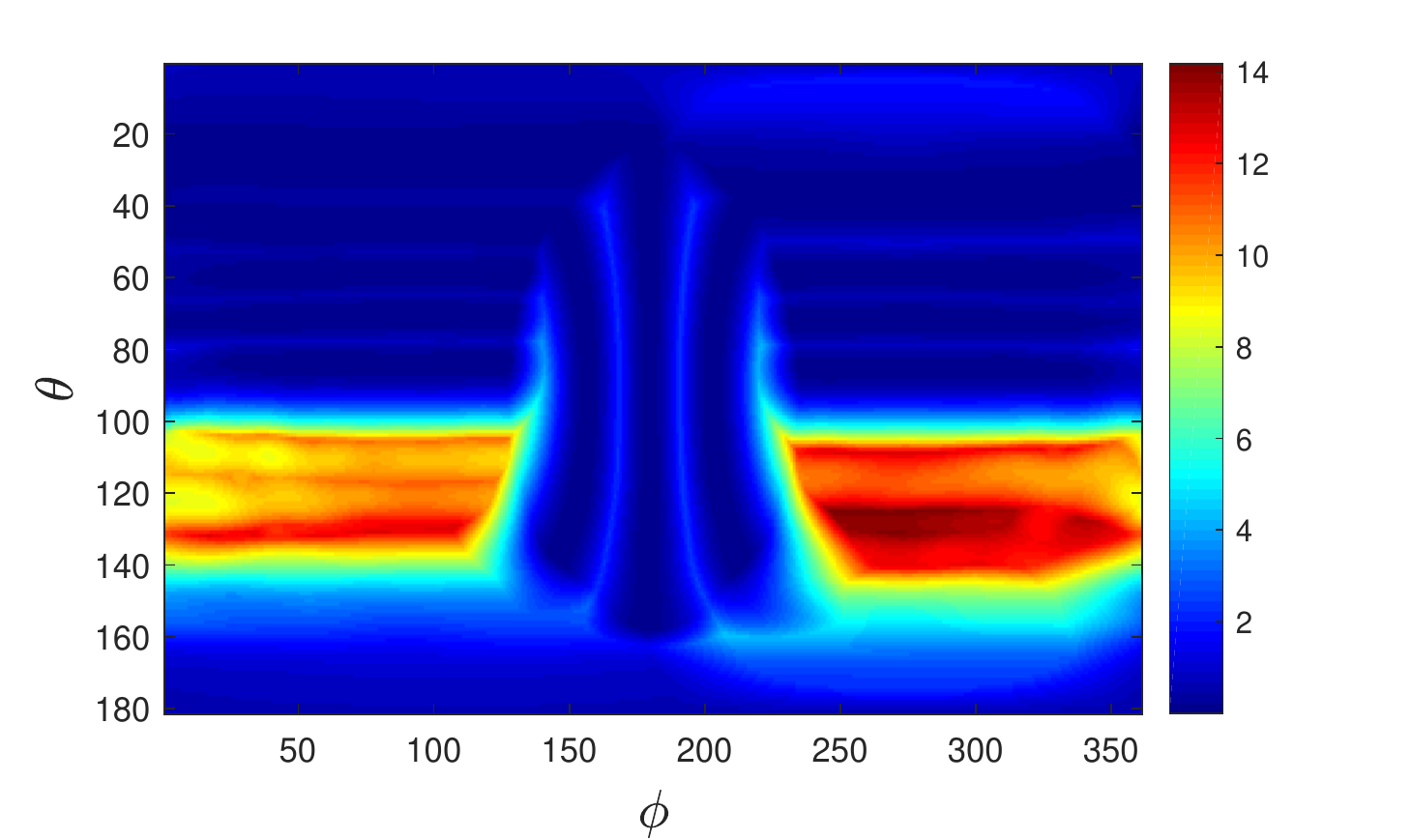}
            \label{fig:Coverage_eg1_gap}}
        \subfigure[3-dB contour of beams]{
            \includegraphics[width= 0.32\linewidth]{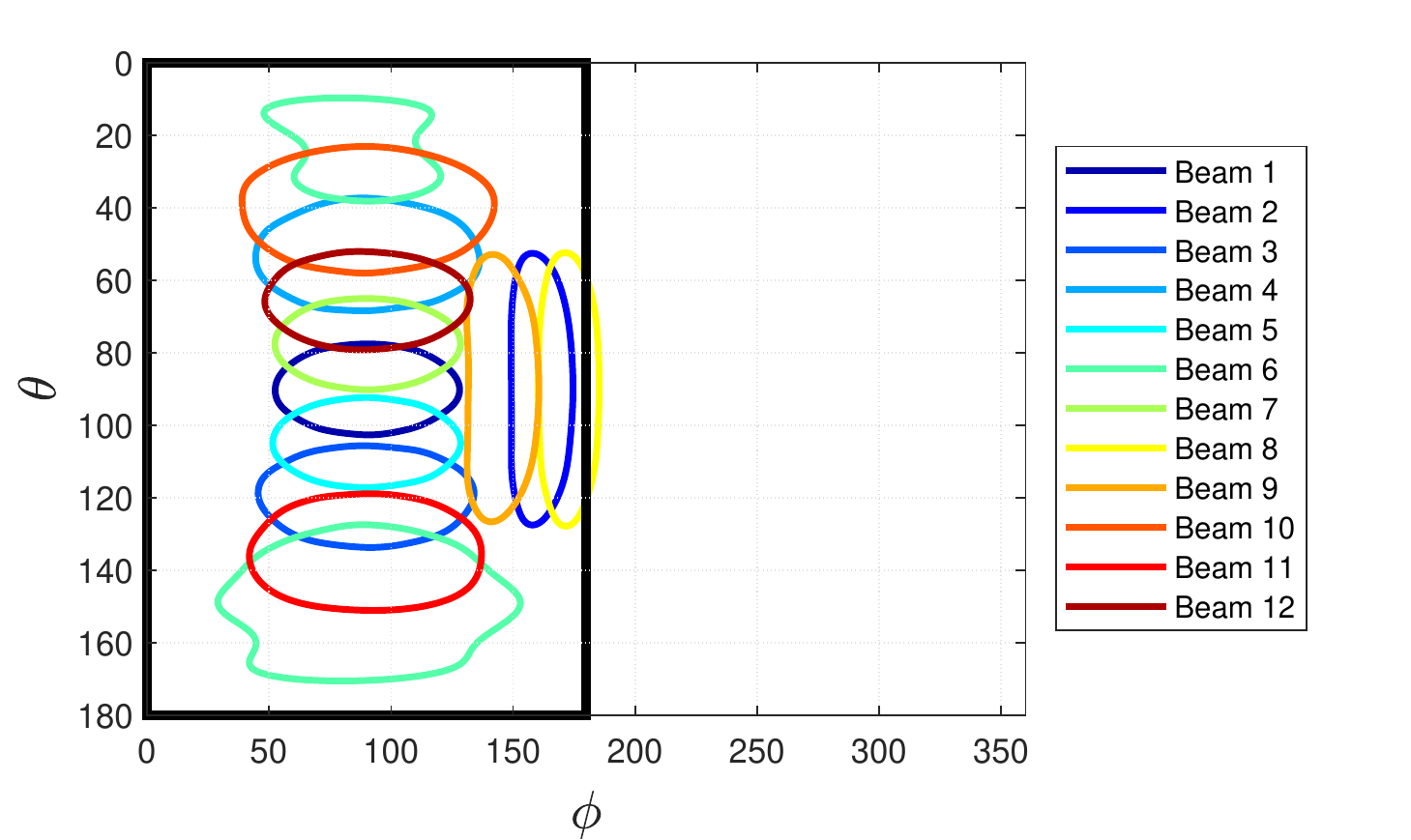}
            \label{fig:Coverage_eg2_3dB_contour}}
        \subfigure[Composite pattern]{
            \includegraphics[width= 0.32\linewidth]{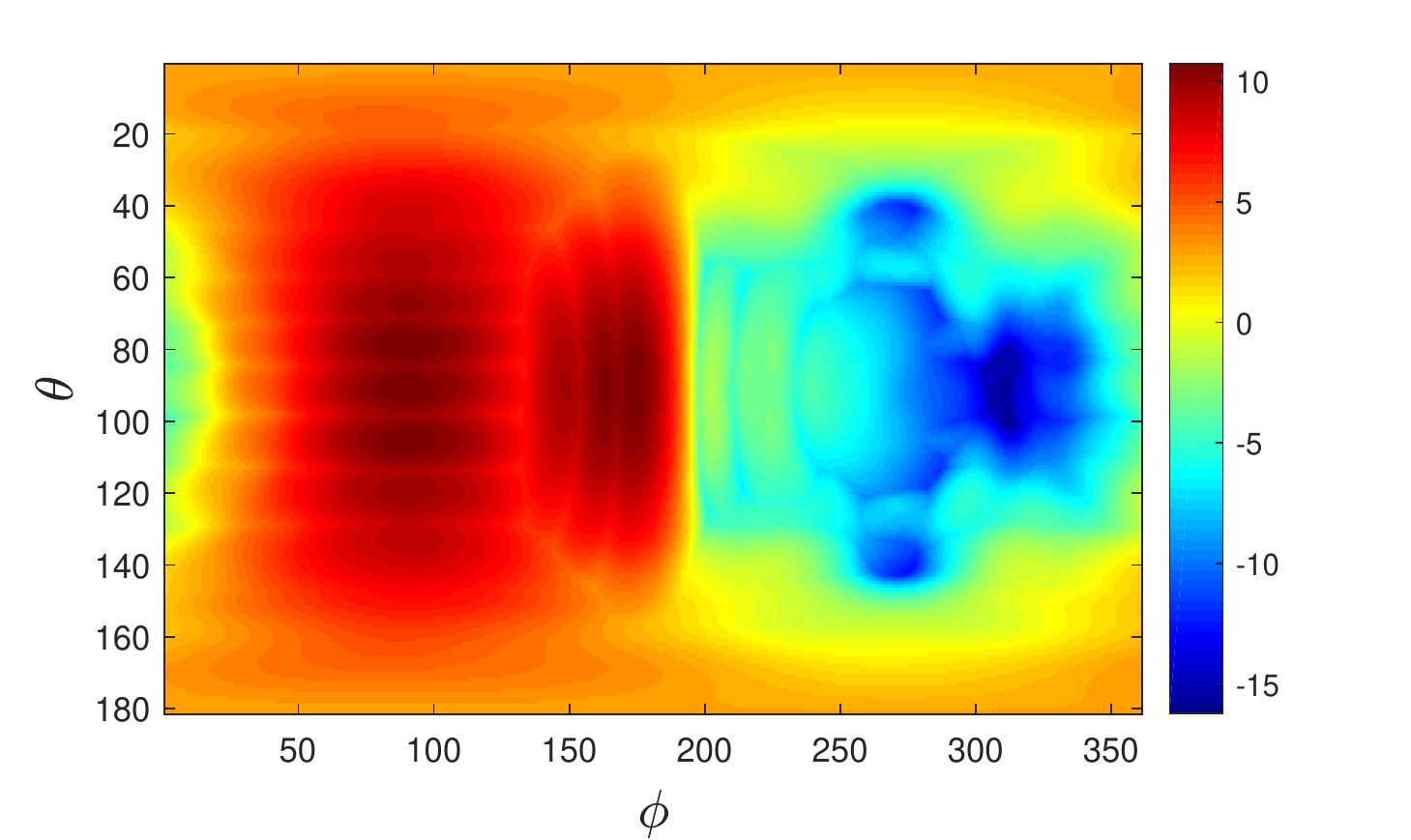}
            \label{fig:Coverage_eg2_pattern}}
        \subfigure[Gap to the upper bound]{
            \includegraphics[width= 0.32\linewidth]{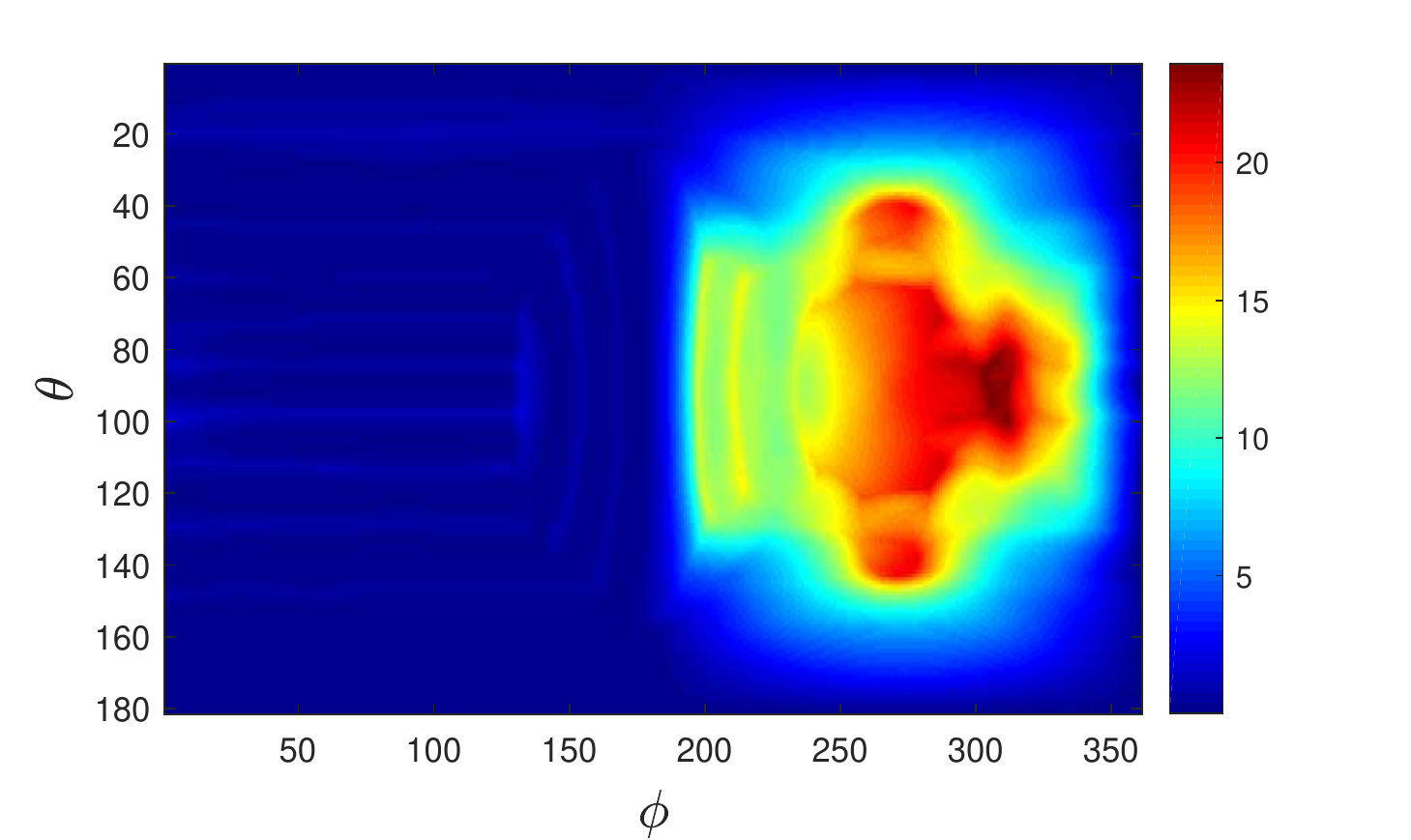}
            \label{fig:Coverage_eg2_gap}}
        \caption{The beam codebook adaptation to the coverage region requirements. In \figref{fig:Coverage_eg1_3dB_contour}-\figref{fig:Coverage_eg1_gap}, the required coverage region is $(\theta, \phi) \in [0^\circ, 90^\circ] \times [0^\circ, 360^\circ)$ whereas in \figref{fig:Coverage_eg2_3dB_contour}-\figref{fig:Coverage_eg2_gap}, the required coverage region is $(\theta, \phi) \in [0^\circ, 180^\circ] \times [0^\circ, 180^\circ]$.}
        \label{fig:Coverage_eg}
    \end{figure*}
    
    In certain scenarios, the beam codebook is required to cover a part of the sphere rather than the whole sphere. For instance, when the user is holding the phone next to the head to make a call, the phone should not beam towards user's head, because of the high blockage loss of human body to mmWave signals and the radio frequency exposure compliance \cite{FCC_47_2_1093, FCC_47_1_1310, ICNIRP_98}.
    
    Our proposed method is capable of adapting to varying coverage region requirement. In the Greedy algorithm, we can define the utility function $U$ over the region of interest instead of the whole sphere. For example, the mean gain over a region could be optimized as shown in \eqref{eq:greedy_criterion_mean_A}. For the K-Means algorithm, we can filter the set of directions $\mathcal{D}$ to keep only the directions within the required coverage region.
     
    \figref{fig:Coverage_eg} illustrates two cases with required coverage region being $(\theta, \phi) \in [0^\circ, 90^\circ] \times [0^\circ 360^\circ)$ and $(\theta, \phi) \in [0^\circ, 180^\circ] \times [0^\circ 180^\circ]$, respectively. The regions are highlighted by black boxes drawn in \figref{fig:Coverage_eg1_3dB_contour} and \figref{fig:Coverage_eg2_3dB_contour}.
    As seen in the figures, the resulting beam codewords are naturally concentrated in the required region. As a result, the composite patterns in these two cases have a less than 2 dB gap to the upper bound in the required region in contrast to a more than 10 dB gap out of the required region. When the coverage region is the half-sphere $0^\circ \leq \phi \leq 180^\circ$, we find that the array on the left edge of the phone is turned off automatically as shown \figref{fig:Coverage_eg2_3dB_contour}, since the required coverage region is at the back of it.

    \subsection{Straightforward Extension to Different Module Placement}
    
    \begin{table}[t]
    \centering
    \caption{\revision{Comparison of the different codebooks in terms of mean gain and median gain. The three arrays are mounted at the left edge, right edge and top edge of the terminal.}}
    \label{tab:Benchmark_3c_KMeans_12_24Beams_left_right_top}
    \revision{
    \begin{tabular}{|c|c|c|c|c|c|}
        \hline 
        {\multirow{2}{*}{Beamforming gain (dB)}} & \multicolumn{5}{c|}{Codebook size $K$} \\ 
        \cline{2-6} 
        & 12 & 15 & 18 & 21 & 24\\ 
        \hline 
        Mean Gain of Benchmark.  & 6.320 & 6.646  &  6.847 & 6.983 & 7.062\\ 
        \hline
        Mean Gain of 802.15.3c.  & 6.277 & 6.649  &  6.842 & 6.963 & 7.052\\ 
        \hline
        Mean Gain of Proposed. & \textbf{6.635} & \textbf{6.887} & \textbf{7.016} & \textbf{7.088} & \textbf{7.150}\\
        \hhline{|=|=|=|=|=|=|} 
        Median gain of Benchmark. & 5.921 & 6.334 & 6.584 & 6.733 & 6.768\\
        \hline
        Median gain of 802.15.3c. & 5.735 & 6.133 & 6.376 & 6.592 & 6.748\\
        \hline
        Median gain of Proposed. & \textbf{6.131} & \textbf{6.551} & \textbf{6.768} & \textbf{6.839} & \textbf{6.885}\\
        \hline
    \end{tabular}}
    \end{table}

    

\begin{figure*}[t]
    \centering
    \subfigure[Benchmark codebook]{
        \includegraphics[width=0.32 \linewidth]{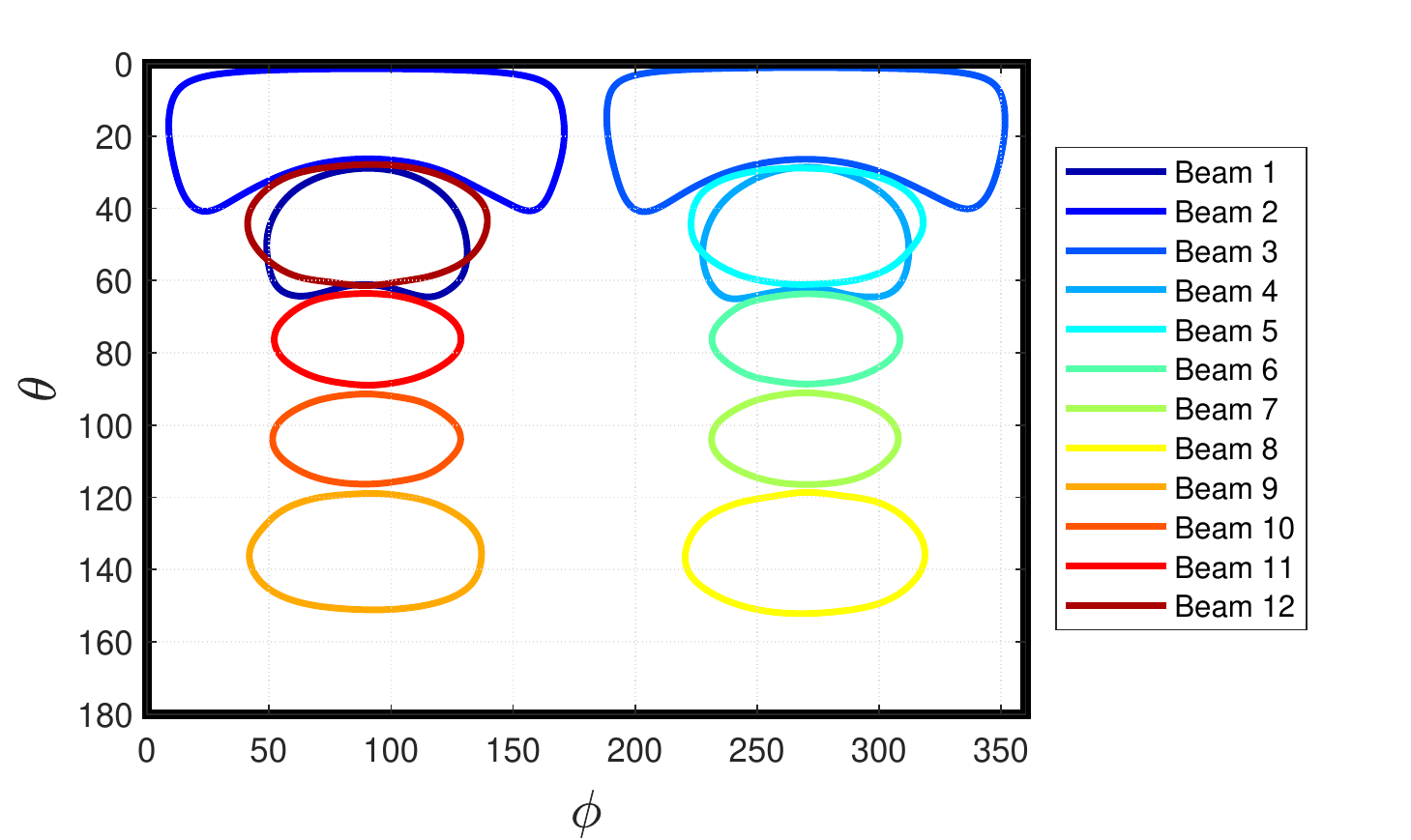}
        \label{fig:Benchmark_Left_Right_Top_3dB_contour}}
    \subfigure[802.15.3c codebook]{
        \includegraphics[width=0.32 \linewidth]{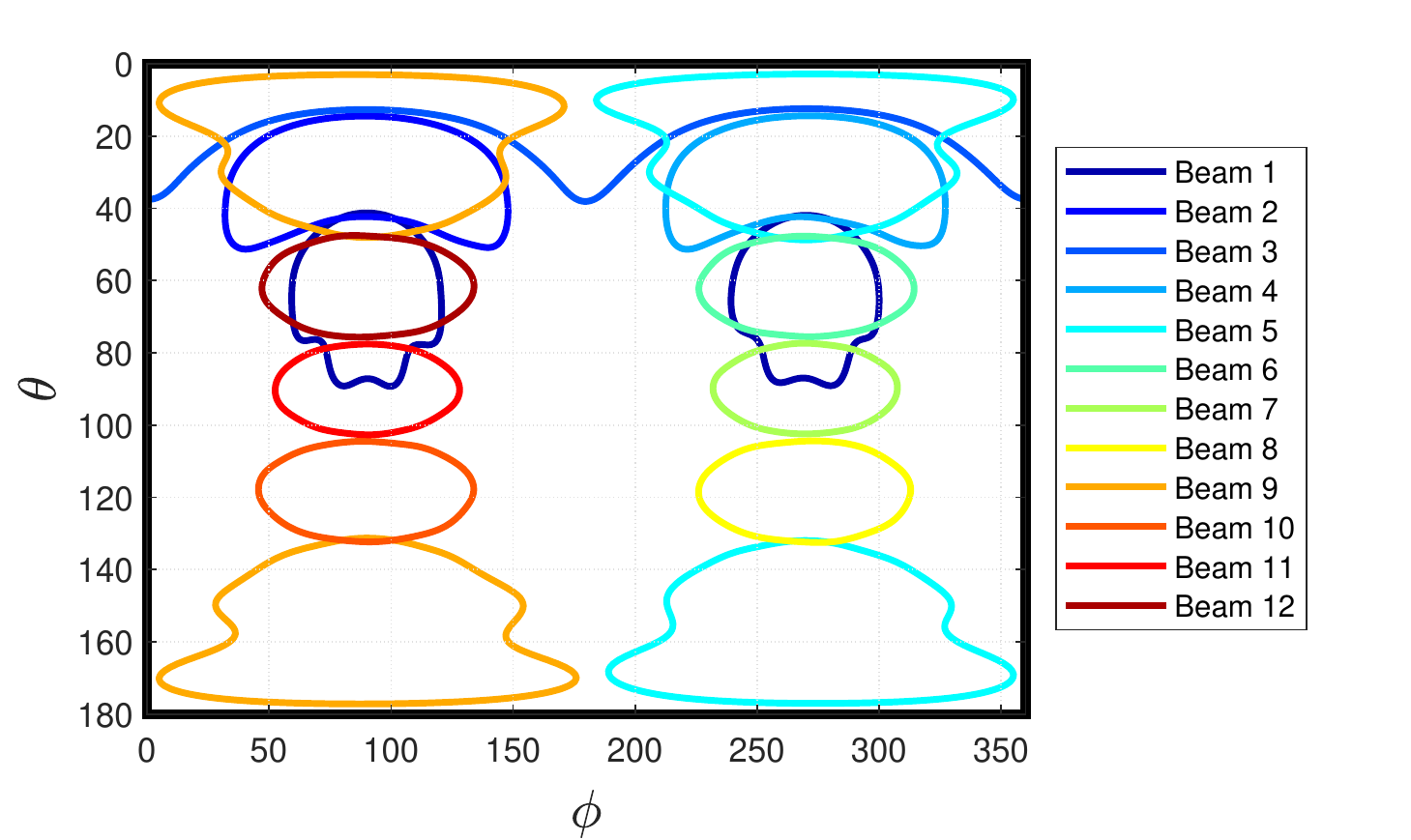}
        \label{fig:3c_Left_Right_Top_3dB_contour}}
    \subfigure[Proposed codebook]{
        \includegraphics[width=0.32 \linewidth]{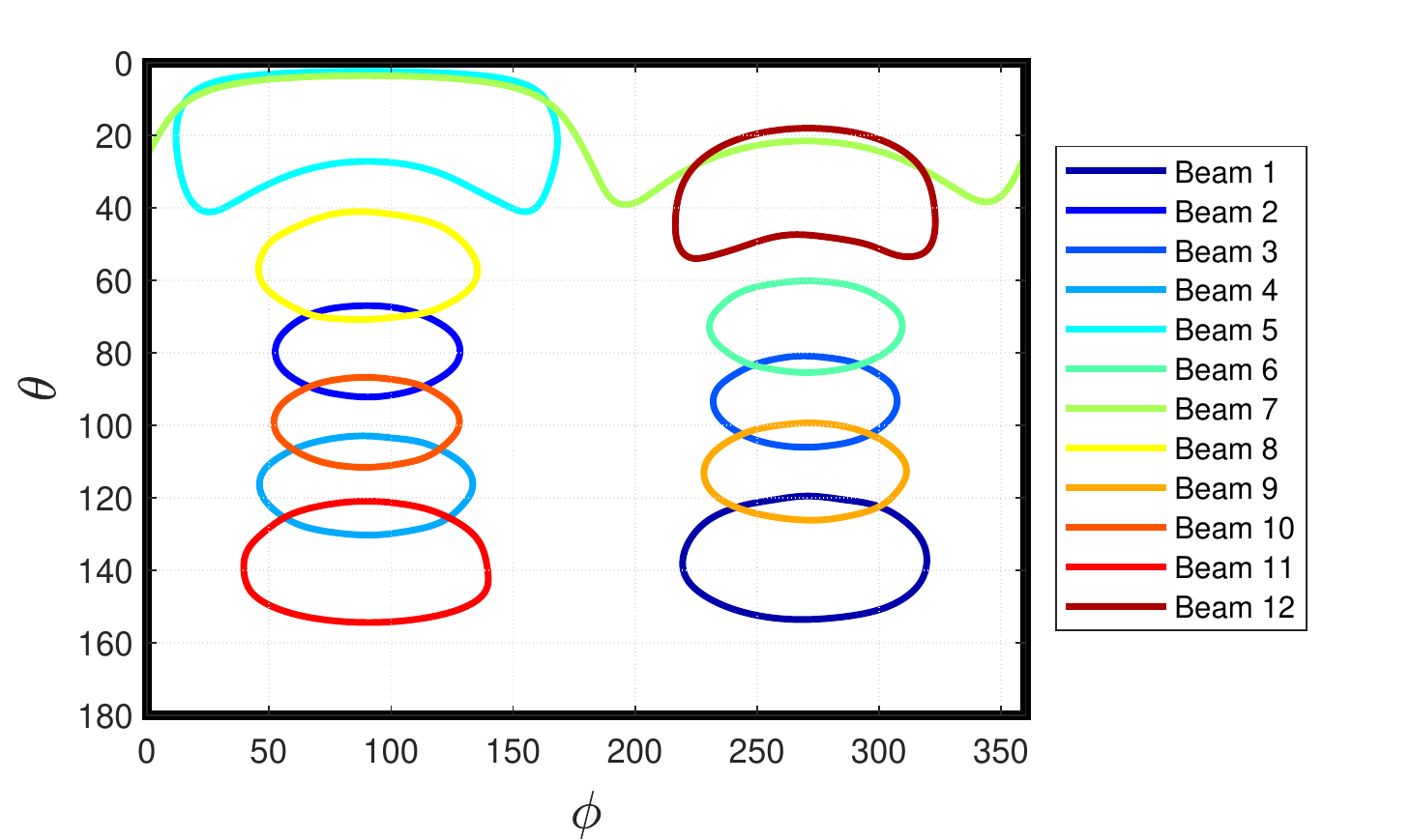}
        \label{fig:KMeans_Greedy_Iter_Left_Right_Top_3dB_contour}}
    \caption{The 3-dB beam contours of the benchmark and proposed codebooks when $K=12$. \revision{The three arrays are mounted at the left, right and top edge of the terminal.}}
    \label{fig:Comparison_Left_Right_Top_3dB_contour}
    \end{figure*}

    Our proposed method can deal with any kind of antenna type and placement on the terminal. Here, we show an example with the same antenna arrays as \figref{fig:Phone_left_right_back} but a different placement. Specifically, we assume the arrays are distributed on the left edge, right edge and top edge. 
    As shown in Table \ref{tab:Benchmark_3c_KMeans_12_24Beams_left_right_top}, the proposed codebook provides better mean and median gains than the benchmark and 802.15.3c codebooks in all the cases.
    \figref{fig:Comparison_Left_Right_Top_3dB_contour} compares the 3-dB beam contours when $K=12$. We find that the beams of benchmark and 802.15.3c codebook are largely overlapping in the upper half-sphere ($0^\circ \leq \theta \leq 90^\circ$), which explains why their spherical coverage is worse than the proposed codebook which maintains a much better angle separation among 12 beams.
    
    Last, comparing Table \ref{tab:Benchmark_3c_KMeans_12_24Beams_left_right_back} and Table \ref{tab:Benchmark_3c_KMeans_12_24Beams_left_right_top}, we find that the first module placement results in better beamforming gains that the second placement. In other words, it is better to put the third array on the back than on the top edge, if the optimization target is the spherical coverage.
     
	\section{Further Comparisons with Simplified E-field Data} \label{sec:comparison}
    To provide a comprehensive comparison with conventional model-based approach and illustrate the effectiveness of the proposed method in general cases, we perform more simulations based on simplified E-field response data besides HFSS data. 
    
    The uniform antenna array (ULA) with exactly same antenna element is assumed in the comparison. The E-field data are generated for the angular directions uniformly distributions over the spatial frequency, i.e., $\theta  = \arccos(x)$, where $x=-1, -(a-1)/a, \dots, (a-1)/a, 1$. In this section where $1\times L$ ULA is assumed, we choose $a=30 L$. The E-field data at the angle $\theta$ is,
    \begin{align}
        \mathbf{e}_{\Theta}(\theta) = \sqrt{p(\theta)} \exp\left(\j \frac{2\pi\cos (\theta)}{d}  [0, 1, \dots, L-1]^T\right), 
    \end{align}
    and $\mathbf{e}_{\Phi}(\theta)=\mb{0}$, where $\theta$ is angle with respect to the axis of the linear array, $p(\theta)$ is the element radiation pattern.
    
    The K-Means algorithm is used for the comparison, and the K-Means algorithm is initialized by the benchmark codebook. In addition, 5-bit analog beamforming is assumed.
    
    In the ideal case of omni-directional antenna and half-wavelength spacing, the codebook design has been well studied and our proposed method does not bring further improvements. However, our proposed method does bring large gain when designing a codebook for a practical antenna array where the ideal assumptions do not hold.
    
    \subsection{Irregular Antenna Spacing}
    We first consider a scenario where the antenna array is not half-wavelength, which results from form-factor constraints or the multi-frequency bands the array has to support. In particular, a $1\times 4$ ULA with $d=0.65 \lambda$ is simulated. The number 0.65 is chosen by assuming that the antenna array has the antenna spacing of $5$ mm (i.e., half wavelength at 30 GHz), and operates at the 39 GHz band. We want to generate a codebook of 4 beams.
    
\begin{figure*}[t]
    \centering
    \subfigure[Benchmark codebook, Median: $4.76$ dB]{
        \includegraphics[width=0.32 \linewidth]{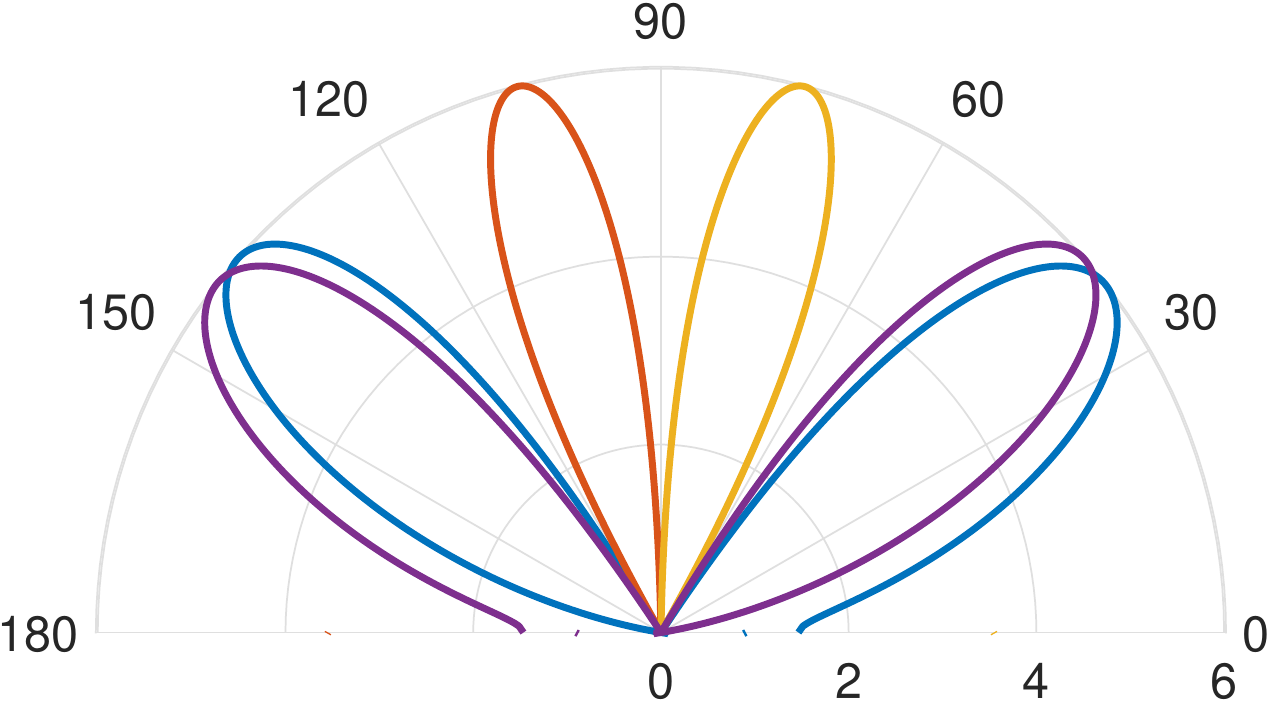}
        \label{fig:D065_pattern_benchmark}}
    \subfigure[802.15.3c codebook, Median: $5.09$ dB]{
        \includegraphics[width=0.32 \linewidth]{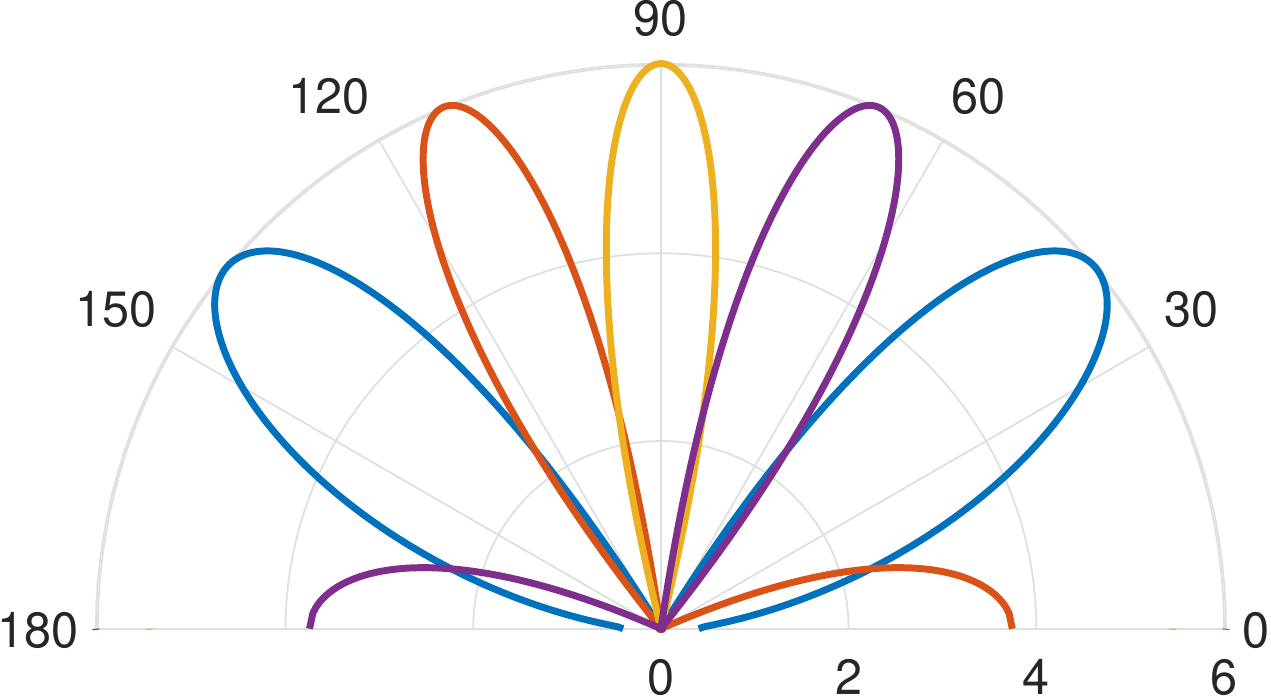}
        \label{fig:D065_pattern_3c}}
    \subfigure[Proposed codebook, Median: $5.38$ dB]{
        \includegraphics[width=0.32 \linewidth]{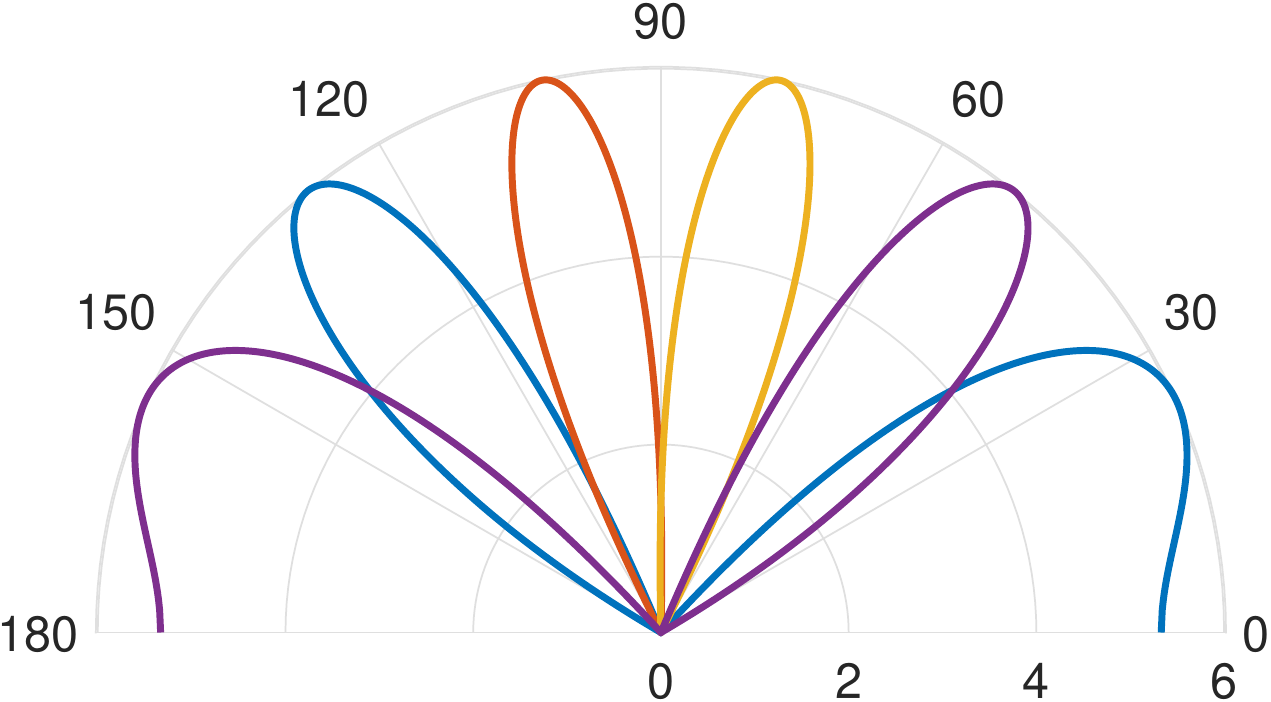}
        \label{fig:D065_pattern_proposed}}
    \caption{The beam patterns of a $1\times 4$ uniform linear array where $d=0.65 \lambda$, $K=4$.}
    \label{fig:Comparison_D065}
    \end{figure*}

\begin{figure}[t]
    \centering
    \includegraphics[width=\linewidth]{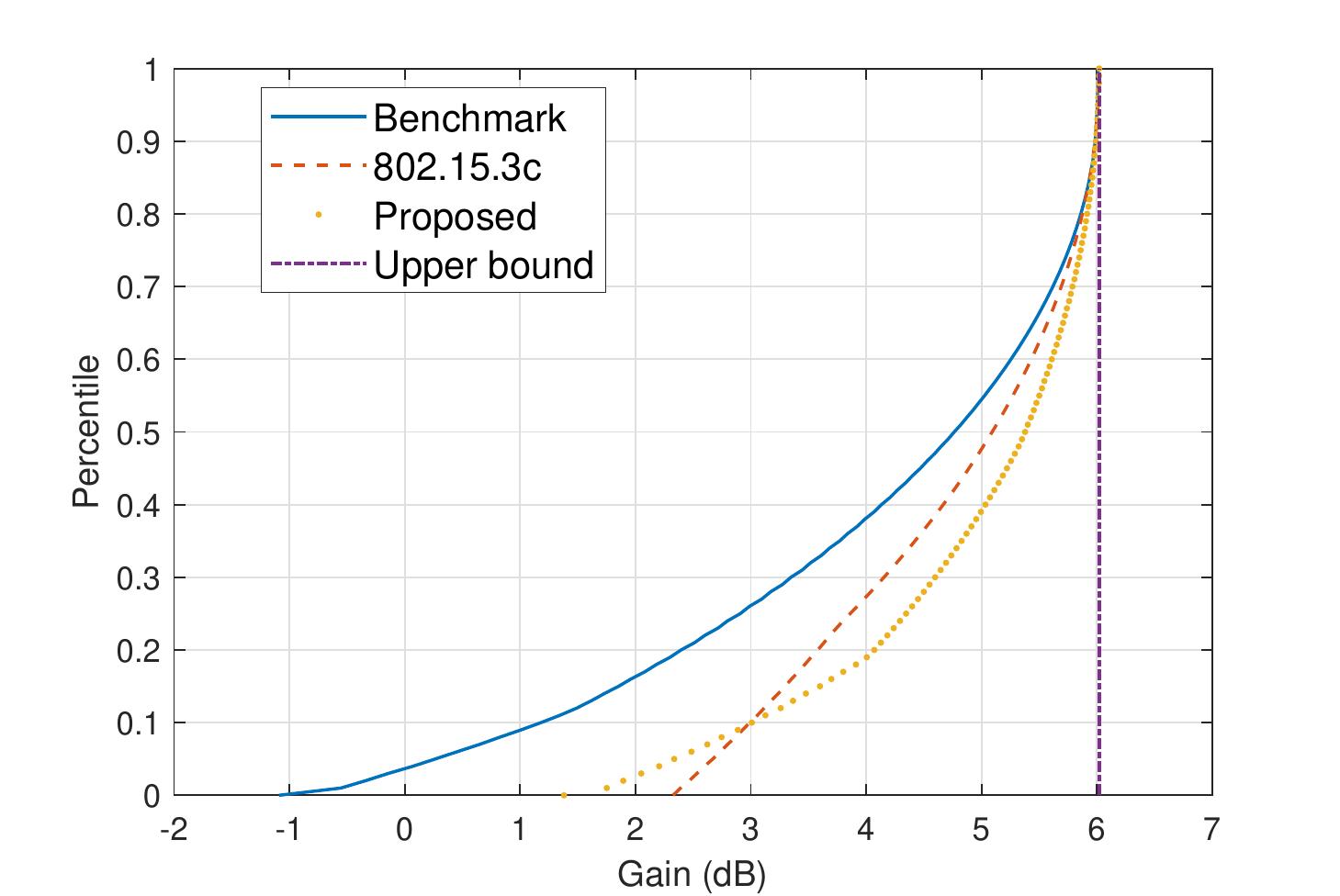}
    \caption{CDF curves of the benchmark, 802.15.3c and the proposed codebooks. It simulated a $1\times 4$ uniform linear array with $d=0.65\lambda$, $K=4$.}
    \label{fig:D065_pattern_compare}
    \end{figure}

    \figref{fig:Comparison_D065} illustrates the radiation pattern of the 4 beams of each codebook. As seen in \figref{fig:Comparison_D065}, there are strong side lobes when the main lobe is pointing away from the broadside direction. The proposed method can adjust the beamforming direction of the beams to fully utilize the side lobes to achieve a better spherical coverage. It is worthy to note that the adjustment is done automatically by the proposed algorithm based on E-field response data. By contrast, the benchmark and 802.15.3c codebooks do not adapt well to the spacing change. 
    The CDF curves of these three codebook is illustrated in \figref{fig:D065_pattern_compare}. 
    It is clear that the proposed codebook shows much better spherical coverage than the other two codebooks. The median gain values are $4.76$, $5.09$, $5.38$ dB, respectively.
    
    Last, the side lobes are also used in \cite{Chen_Li_ICST11, Feng_Wei_WCNC13} to achieve a good coverage. However, their handcrafted approach requires careful design of the beamforming weights and cannot apply to arbitrary antenna spacing.
    
    
    
    \subsection{The directional antenna radiation pattern}
    Now we consider another case where the antenna is directional.
    A simple model of directional radiation pattern is as follows.
    \begin{equation}
    p(\theta) = \sin^{q} (\theta),
    \end{equation}
    where $q$ controls the directionality of the radiation pattern.
    A $1\times 4$ ULA with half-wavelength spacing is considered here.
    
\begin{figure*}[t]
    \centering
    \subfigure[Benchmark codebook, Median: $4.06$ dB]{
        \includegraphics[width=0.32 \linewidth]{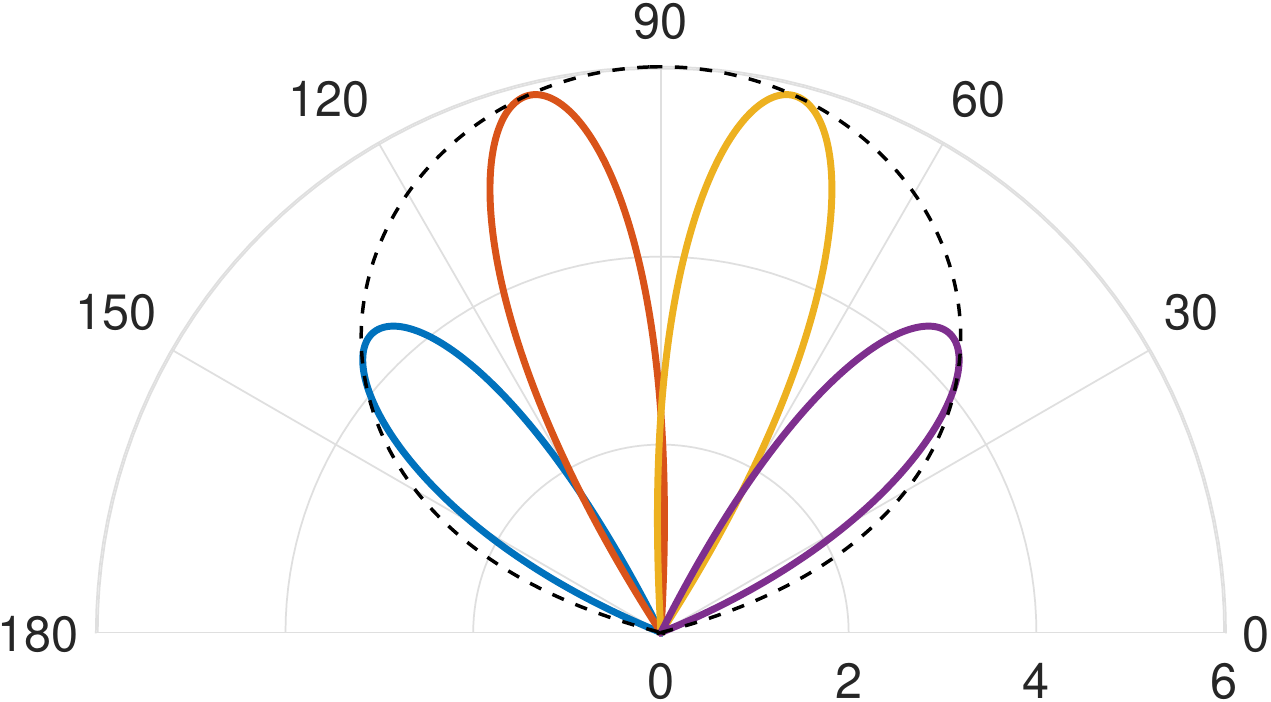}
        \label{fig:Sine1_pattern_benchmark}}
    \subfigure[802.15.3c codebook, Median: $3.96$ dB]{
        \includegraphics[width=0.32 \linewidth]{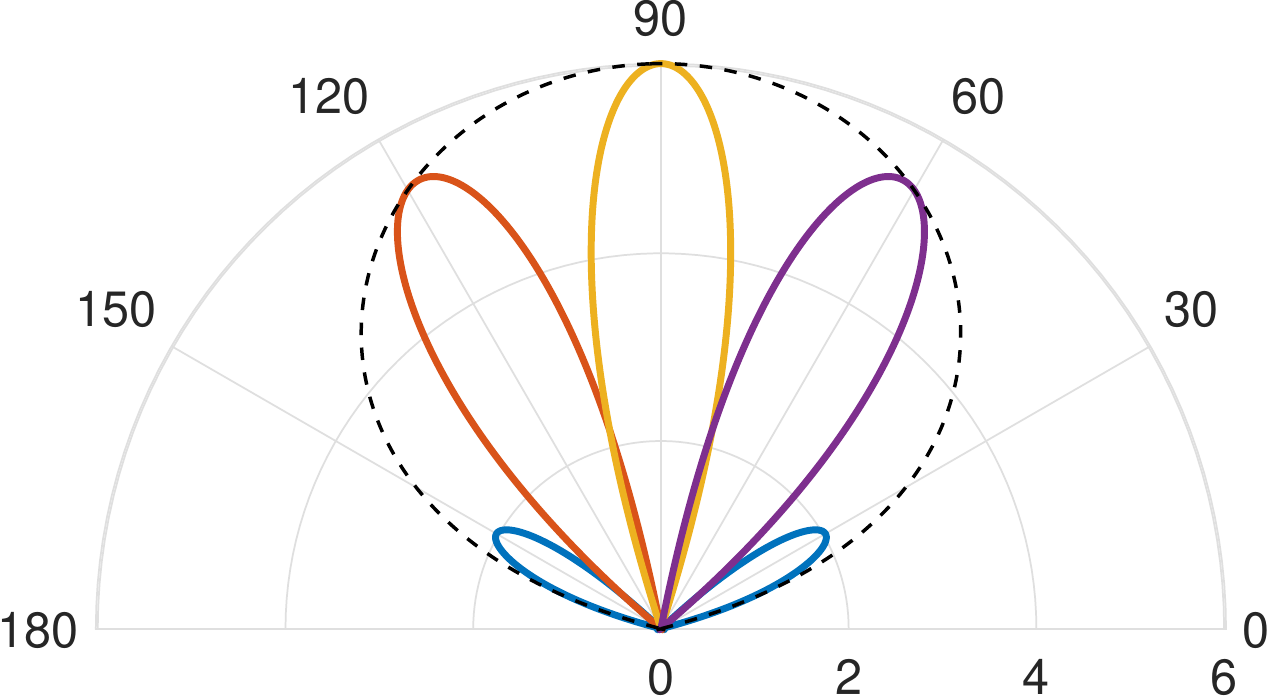}
        \label{fig:Sine1_pattern_3c}}
    \subfigure[Proposed codebook, Median: $4.39$ dB]{
        \includegraphics[width=0.32 \linewidth]{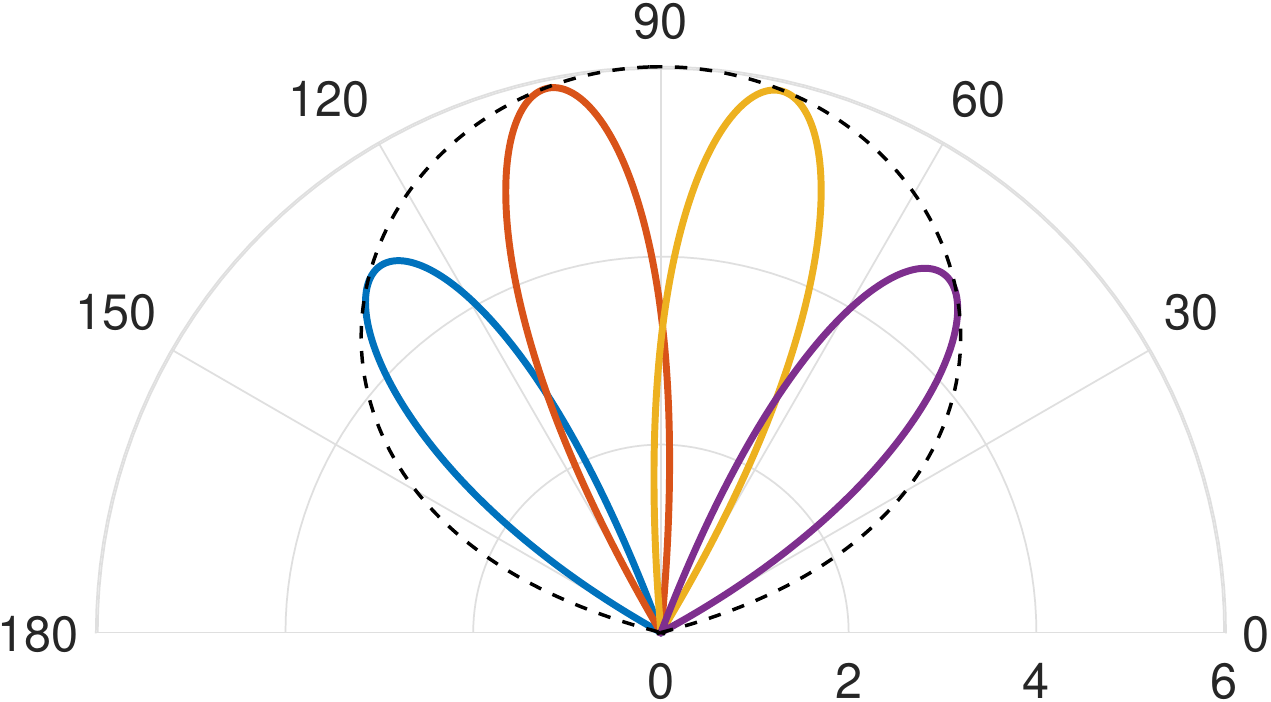}
        \label{fig:Sine1_pattern_proposed}}
    \subfigure[Benchmark codebook, Median: $1.91$ dB]{
        \includegraphics[width=0.32 \linewidth]{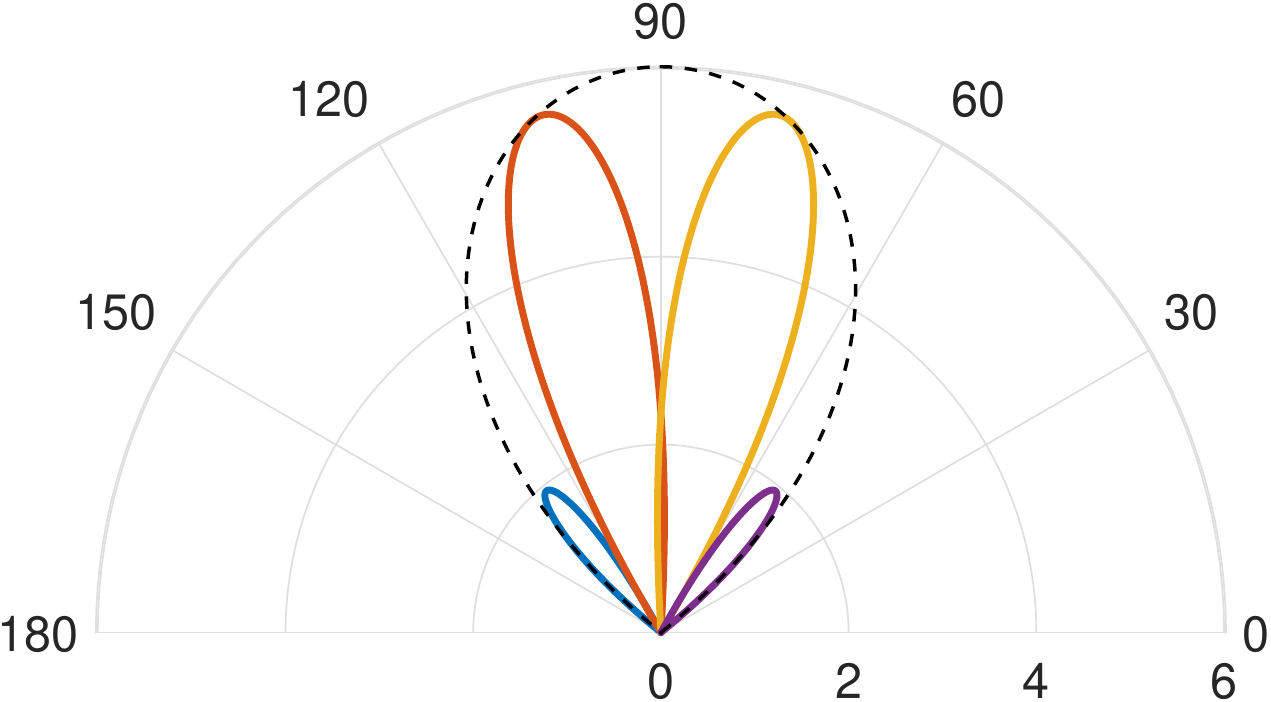}
        \label{fig:Sine3_pattern_benchmark}}
    \subfigure[802.15.3c codebook, Median: $3.02$ dB]{
        \includegraphics[width=0.32 \linewidth]{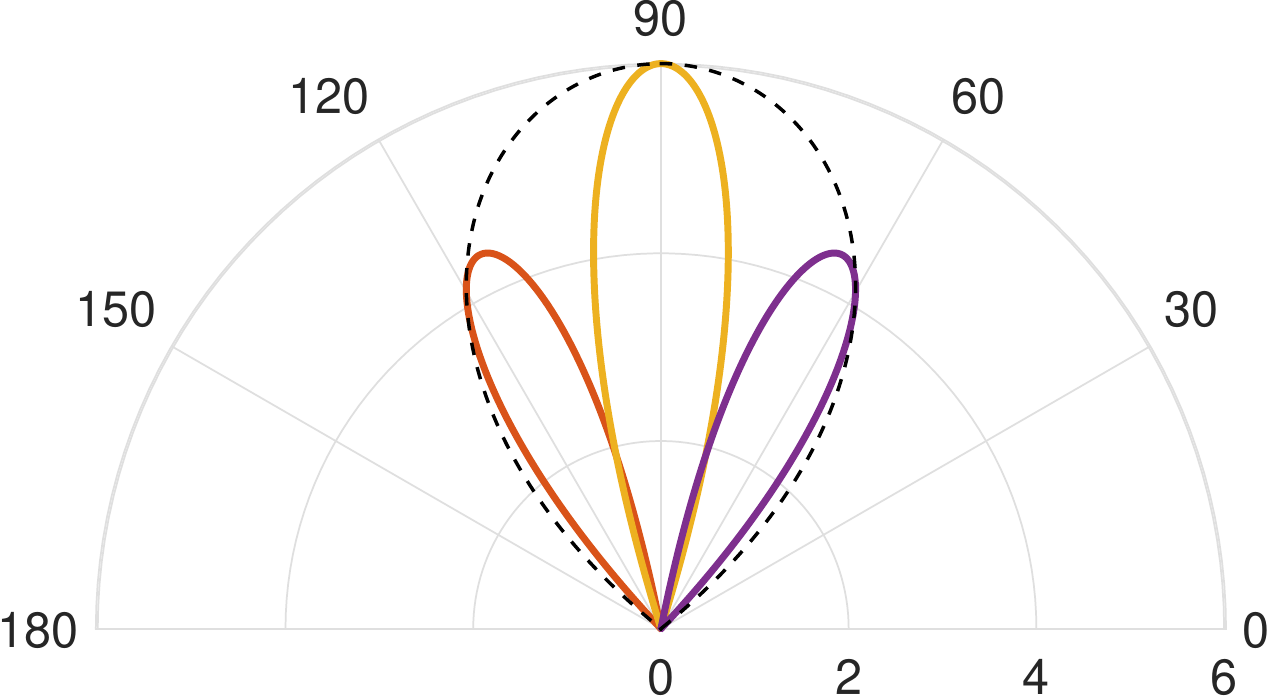}
        \label{fig:Sine3_pattern_3c}}
    \subfigure[Proposed codebook, Median: $3.58$ db]{
        \includegraphics[width=0.32 \linewidth]{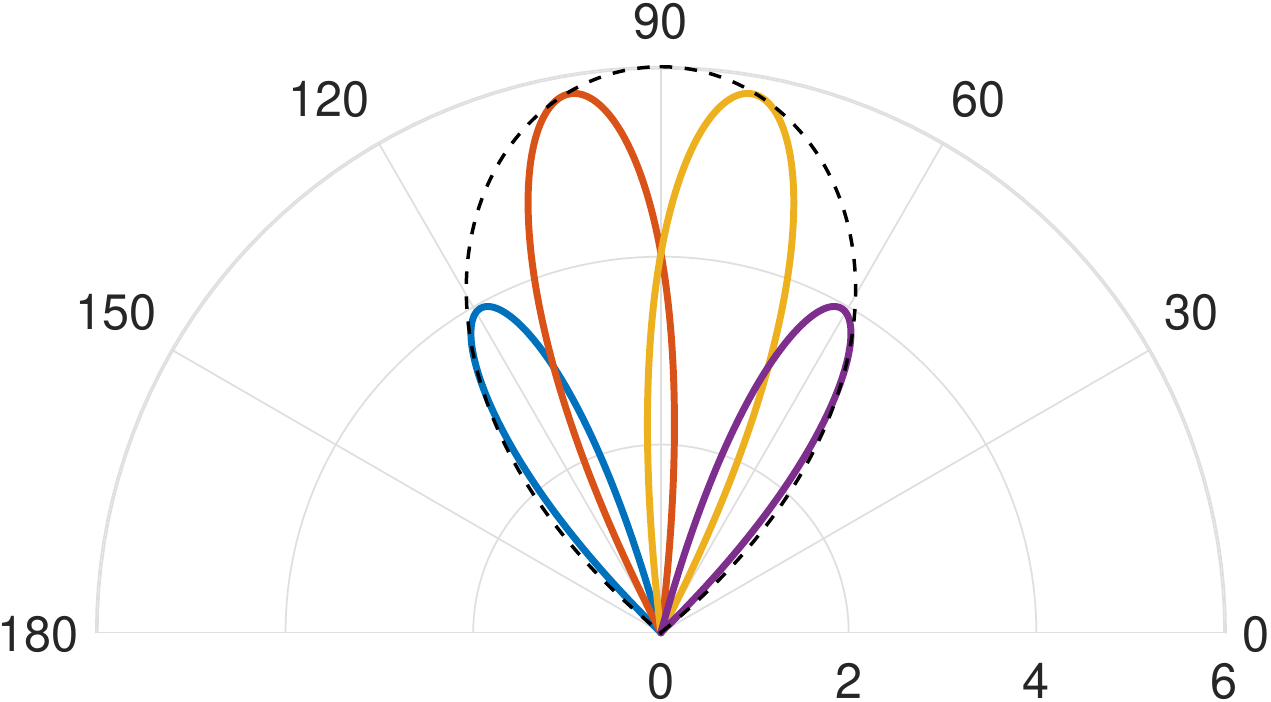}
        \label{fig:Sine3_pattern_proposed}}
    \caption{The beam patterns of a $1\times 4$ uniform linear array where $d=0.5 \lambda$, $K=4$. The element pattern is $p(\theta) = \sin (\theta)$ in (a)-(c), while the element pattern is $p(\theta) = \sin^{3} (\theta)$ in (d)-(f). The dashed envelope stands for the upper bound.}
    \label{fig:Comparison_Sine}
    \end{figure*}

    \figref{fig:Comparison_Sine} shows the comparison of the beam patterns when $q=1$ in (a)-(c) and $q=3$ in (d)-(f). The dashed envelope represents the upper bound. As the parameter $q$ becomes larger, the element pattern as well as the upper bound becomes more directional.
    As seen in the figure, as $q$ increases, two out of the four beams in the benchmark codebook have diminishing gains and provide negligible contribution to the spherical coverage. Similarly, one of the 802.15.3c codewords has relatively small gain. By contrast, the proposed codebook is capable of tweaking the beam direction automatically. The beams are moving towards the broadside direction as the antenna element becomes more directional. The median gain of each codebook is listed below the figures. The proposed codebooks have the largest median gains.

	\section{Conclusion} \label{sec:conclusion}
	In this paper, we have formulated the beam codebook design problem to enhance the spherical coverage of the mmWave terminals. The codebooks designed based on the isotropic antenna assumptions will not work well for mmWave terminals, due to the inherent directional radiation pattern of the mmWave antenna element and the impact from housing components of the terminals, such as coupling, blockage, absorption, reflection, etc. We proposed a novel approach to automatically design the beam codebook solely based on the E-field response of each element. First, a flexible Greedy algorithm is proposed to choose a subset of the candidate codeword pool to form the final codebook according to any given criterion. Second, a machine learning based iterative algorithm is proposed to generate the codebook. Through simulations, we find out that the composite radiation pattern of the proposed codebook is better than the benchmark \revision{and 802.15.3c} codebook. Actually, the performance of the proposed beam codebook is shown to be able to approach the upper bound as codebook size increases. \revision{Furthermore, the proposed data-driven method can be used for any kind of array layouts, placement and antenna type. It is a very generic method capable of designing good codebooks for a wide range of practical scenarios where the conventional model-based method does not work well.}
    
    Note that the proposed method depends closely on the E-field data. There are several possible factors distorting the far-filed E-field response. For example, the protection case of the phone and the hand grip of the users \cite{Hong_Wonbin_TAP17,Raghavan_TMTT19}. We model the hand grip impact on the E-field response and propose an adaptive beam codebook generation method in \cite{AlAmmouri_Access19}.
    
    Last, even without the distortions by hand grip, the E-field response data from simulations or measurements may be different from the true response. For instance, there may be deviations between the antenna and phone model used in the simulations and the manufactured ones. The measurement data may also be inaccurate because of the heating of the phone in the measurement process. A future direction to improve the proposed method is to design a robust beam codebook taking into account these deviations.
	
	\bibliographystyle{IEEEtran}
	\bibliography{IEEEabrv,BeamBook}
	\EOD
\end{document}

%% file: input.tex
\usepackage{acronym}
\usepackage{amsfonts}
\usepackage{bbm}
\usepackage{times}
\usepackage{cite}
\usepackage{amsmath}
\usepackage{array}
\usepackage{amssymb}
\usepackage{stfloats}
\usepackage{diagbox}
\usepackage{graphicx}
\usepackage{footnote}
\usepackage{amsthm}
\usepackage{booktabs}
\usepackage{array}
\usepackage{algorithmic}
\usepackage{algorithm}
\usepackage{subeqnarray}
\usepackage{cases}
\usepackage{threeparttable}
\usepackage{color}
\usepackage{epstopdf}
\usepackage{multirow}
\usepackage{tabularx}
\usepackage{enumerate}
\usepackage{multicol}
\usepackage{subfigure}

\newcommand{\fig}[1]{Fig.~\ref{#1}}
\newcommand{\secref}[1]{{Section}~\ref{#1}}
\newcommand{\figref}[1]{{Fig.}~\ref{#1}}
\newcommand{\tabref}[1]{{Table}~\ref{#1}}

\def\bb0{{\mathbb{0}}}

\def\ba{{\mathbf{a}}}
\def\bb{{\mathbf{b}}}

\def\bm{{\mathbf{m}}}
\def\bn{{\mathbf{n}}}

\def\bw{{\mathbf{w}}}
\def\bx{{\mathbf{x}}}

\def\b0{{\mathbf{0}}}

\def\bA{{\mathbf{A}}}

\def\bD{{\mathbf{D}}}

\def\bI{{\mathbf{I}}}

\def\bM{{\mathbf{M}}}

\def\bU{{\mathbf{U}}}

\def\bW{{\mathbf{W}}}

\def\sf0{{\mathsf{0}}}

\def\rm0{{\mathrm{0}}}

\def\nn{\nonumber}

\newcommand{\mb}{\mathbf}
\newcommand{\mr}{\mathrm}

\def\l{\left}
\def\r{\right}
\def\j{\mathrm{j}}

\acrodef{CSI}[CSI]{channel state information}
\acrodef{CSIT}[CSIT]{channel state information at the transmitter}
\acrodef{CSIR}[CSIR]{channel state information at the receiver}
\acrodef{MIMO}[MIMO]{multiple-input multiple-output}
\acrodef{SISO}[SISO]{single-input single-output}
\acrodef{MISO}[MISO]{multiple-input single-output}
\acrodef{SIMO}[SIMO]{single-input multiple-output}
\acrodef{ADCs}[ADCs]{analog-to-digital convertors}
\acrodef{SNR}[SNR]{signal-to-noise ratio}
\acrodef{AWGN}[AWGN]{additive white Gaussian noise}
\acrodef{MRT}[MRT]{maximal ratio transmission}
\acrodef{DFT}[DFT]{Discrete Fourier Transform}
\acrodef{ULA}[ULA]{uniform linear array}
\acrodef{UPA}[UPA]{uniform planar array}
\acrodef{LS}[LS]{least squares}
\acrodef{ML}[ML]{maximum likelihood}
\acrodef{AML}[AML]{approximate ML}
\acrodef{LMMSE}[LMMSE]{linear MMSE}
\acrodef{ALMMSE}[ALMMSE]{approximate LMMSE}
\acrodef{QIHT}[QIHT]{quantized iterative hard thresholding}
\acrodef{QIST}[QIST]{quantized iterative soft thresholding}
\acrodef{PAPR}[PAPR]{peak-to-average power ratio}
\acrodef{SVD}[SVD]{singular value decomposition}
\acrodef{CDF}[CDF]{cumulative distribution function}
\acrodef{EIRP}[EIRP]{equivalent isotropic radiated power}
\acrodef{HFSS}[HFSS]{high-frequency structure simulator}
\acrodef{QCQP}[QCQP]{quadratically constrained quadratic program}
\acrodef{SDR}[SDR]{semi-definite relaxation}
\acrodef{SDP}[SDP]{semi-definite programming}
\acrodef{KKT}[KKT]{Karush-Kuhn-Tucker}

